\RequirePackage{fix-cm}
\documentclass[twocolumn,epjc3]{svjour3}  
\smartqed  
\journalname{Eur. Phys. J. C}
\usepackage{graphicx}
\usepackage{psfrag}
\usepackage{color}
\usepackage{amssymb}
\usepackage{amsmath}
\usepackage{epstopdf}

\newcommand{\be}{\begin{eqnarray}}
\newcommand{\ee}{\end{eqnarray}}

\newcommand{\bfb}{{\bf b}_{\perp}}

\newcommand{\bfp}{{\bf p}_{\perp}}

\newcommand{\Dp}{{\bf \Delta}_{\perp}}

\begin{document}
\title{
Wigner distributions and orbital angular momentum of a proton}
\author{D. Chakrabarti$^1$, T. Maji$^1$,  C. Mondal$^1$ \and A. Mukherjee$^{2}$ }
\institute{$ $Department of Physics, 
Indian Institute of Technology Kanpur,
Kanpur 208016, India\\
\and Department of Physics, 
Indian Institute of Technology Bombay, Mumbai 400076, India}
\date{Received: date / Revised version: date}
\maketitle

\begin{abstract}
The Wigner distributions for $u$ and $d$ quarks in a proton are calculated using the light front wave functions (LFWFs) of the scalar quark-diquark model for nucleon constructed from the soft-wall AdS/QCD correspondence. We present a detail study of the quark orbital angular momentum(OAM) and its correlation with quark spin and  proton spin. The quark density distributions, considering the different polarizations of quarks and proton, in transverse momentum plane as well as in transverse impact parameter plane are presented for both  $u$ and $d$ quarks. 

\end{abstract}

\section{Introduction\label{intro}}
A complete understanding of partonic structure of nucleon is one of the  challenging tasks in the particle physics. Both theoretical and experimental efforts are going on to unravel the  three dimensional distributions of the partons and their contributions to the nucleon spin and angular momentum.  Because of the nonperturbative nature of QCD, it is very difficult to perform first principle calculations of  the  hadron properties. 
 However a perturbative approach in light cone framework allows us to calculate the parton distribution function(PDF), $f(x)$, which gives the probability of having a parton with light-cone longitudinal momentum fraction $x$ inside a nucleon but it contains no information about the transverse structure or angular momentum distributions. The spin correlation of partons are described by the helicity distribution, $g_{1}(x)$, and transversity distributions, $h_1(x)$. The generalized parton distributions(GPDs) and the transverse momentum dependent distributions(TMDs) encode informations about the three dimensional structure of the nucleons. 
In the deeply virtual Compton scattering(DVCS), deeply virtual meson electroproduction (DVMPs), a more general views of parton distributions, in 
the collinear frame, is studied by GPDs\cite{Ji,Radu,Goeke,Diehl} which are functions of longitudinal momentum and two transverse impact parameter coordinates. TMDs\cite{Collins,Sivers,Mulders,Boer} are functions of the transverse momentum of the parton and appear in the semi inclusive deep inelastic scattering(SIDIS)  where the collinear picture is no longer enough to explain the single spin asymmetry(SSA). 

Wigner distributions are six dimensional  distributions containing more general informations about the nucleon structure. Wigner distributions do not have probabilistic interpretation but  in certain limits, reduce to GPDs and TMDs.  
The Wigner distributions are defined as  functions of three momentums and three positions of a parton inside a nucleon. The concept of Wigner distributions was first introduced in \cite{ji03_04}. In \cite{pasquini11},  five dimensional Wigner distributions were proposed in the light-front formalism with three momentum and two position components of a parton. 
Wigner distributions integrated over transverse momentum give the  GPDs at zero skewness, the TMDs are  obtained by integrating over transverse impact parameters with zero momentum transfer and the integration over transverse momentum and transverse positions provide the PDFs.
The Wigner distributions after integrating over the light cone energy  of the parton are interpreted as a Fourier transform of corresponding generalized transverse momentum dependent distributions (GTMDs) which are functions of the light cone three-momentum of the parton as well as  the momentum transfer to the nucleon. 
 Angular momentum of a quark is extracted from Wigner distributions taking the phase space average. The spin-spin and spin-orbital angular momentum(OAM) correlations between a nucleon and a quark inside the nucleon can also be described from phase space average of Wigner distributions.  Wigner distributions have been studied in different models e.g., in lightcone constituent quark model\cite{LCCQM,Lorce12}, in  chiral soliton model\cite{chi_QSM,QSM2}, light front dressed quark model\cite{MNO}, lightcone spectator model\cite{liu_ma_WD}. In this work, we investigate the Wigner distributions for unpolarized and polarized proton and the orbital angular momentum(OAM) and spin-spin and spin-OAM correlations in a scalar diquark model of proton \cite{Gut} with the light front wavefunctions modeled  from AdS/QCD prediction. 
 
The paper is organized as follows.  We first  introduce the lightfront scalar diquark model  in Sec.\ref{model} and the Wigner distributions in Sec.\ref{WD}. The different definitions of orbital angular momentum are discussed in Sec.\ref{oam}.  Then in Sec.\ref{results}, both  analytical  and numerical   results in our model are discussed in detail. The correlation between the quark and proton spins and  quark spin and OAM correlations are discussed in Sec.\ref{corr}.  The results are also compared with other models. The GTMDs in this model are briefly discussed in Sec.\ref{gtmds} and finally we conclude in Sec.\ref{concl}.

\section{Light-front diquark model}\label{model}
In the diquark spectator model, one of the three valence quarks interacts with external photon and other two valence quarks are considered as a diquark state  of spin-0(scalar diquark) or spin-1 (vector diquark). Therefore the proton state $|P~;S\rangle$ can be treated as a two particle state in the Fock-state expansion. In this paper we consider the scalar diquark model developed in \cite{Gut,CM_EPJ}. The average light-front momentum of the scalar diquark is $P_X=\big((1-x)P^+,P^-_X,-\bfp \big)$, where $x$ is the longitudinal momentum fraction carried by the struck quark.

The two-particle fock-state expansion for $J^z=\pm\frac{1}{2}$ are given by
\be
|P;\pm\rangle & =& \sum_q \int \frac{dx~ d^2\bfp}{2(2\pi)^3\sqrt{x(1-x)}}\nonumber\\
&& \bigg[ \psi^{\pm}_{q+}(x,\bfp)|+\frac{1}{2},0; xP^+,\bfp\rangle \nonumber \\
 &+&  \psi^{\pm}_{q-}(x,\bfp)|-\frac{1}{2},0; xP^+,\bfp\rangle\bigg],\label{fock_PS}
\ee
Where the $|\lambda_q,\lambda_s; xP^+,\bfp \rangle $ represents a 2-particle state with a quark of helicity $\lambda_q$, and a diquark(spectator) of helicity $\lambda_s$. The $xP^+$ and $\bfp$ are the longitudinal momentum and transverse momentum  of the active quark respectively. The $\psi^{\lambda_N}_{q\lambda_q}$ are the light-front wave functions corresponding to the nucleon helicity $\lambda_N=\pm$ and quark helicity $\lambda_q=\pm$. We adopt the generic ansatz for the quark-diquark model of the valence Fock state of the nucleon LFWFs \cite{Gut}, assuming vanishing quark mass
\be 
\psi^{+}_{q+}(x,\bfp)&=&\phi^{(1)}_q(x,\bfp),\nonumber\\
\psi^{+}_{q-}(x,\bfp)&=&-\frac{p^1+ip^2}{xM}\phi^{(2)}_q(x,\bfp),\nonumber\\
\psi^{-}_{q+}(x,\bfp)&=&\frac{p^1-ip^2}{xM}\phi^{(2)}_q(x,\bfp),\label{LFWF}\\
\psi^{-}_{q-}(x,\bfp)&=& \phi^{(1)}_q(x,\bfp),\nonumber
\ee 
where $\varphi_q^{(1)}(x,\bfp) $ and $\varphi_q^{(2)}(x,\bfp) $ are the wave functions predicted by soft-wall AdS/QCD in \cite{BT} with the AdS/QCD scale parameter $\kappa=0.4 ~GeV $.
\be
\varphi_q^{(i)}(x,\bfp)&=&N_q^{(i)}\frac{4\pi}{\kappa}\sqrt{\frac{\log(1/x)}{1-x}}x^{a_q^{(i)}}(1-x)^{b_q^{(i)}}\nonumber\\
&&\exp\bigg[-\frac{\bfp^2}{2\kappa^2}\frac{\log(1/x)}{(1-x)^2}\bigg],
\label{phi}
\ee
The values of the parameters $a^{(i)}_q$, $b^{(i)}_q$ and $N^{(i)}_q$ are fixed in \cite{chi_odd,tm_rel} by fitting the nucleon form-factor data.  For completeness, we list the parameters in Table \ref{table1}.
\begin{table}[ht]
\centering 
\begin{tabular}{| c | c | c | c| c | c| c |}
    \hline
     parameters&  $a^{(1)}$ & $a^{(2)}$ & $b^{(1)}$ & $b^{(2)}$ & $N^{(1)}$ & $N^{(2)}$ \\ \hline
    $u$ quark & 0.02 & 1.05 & 0.022 & -0.15 & 2.055 & 1.322 \\ \hline
    $d$ quark & 0.1 & 1.07 & 0.38  & -0.2 & 1.7618 & -2.4827 \\ \hline
    \end{tabular} 
\caption{The parameters in the  light front diquark model for $\kappa=0.4$ GeV.} 
\label{table1} 
\end{table} 
This is a very simplistic model of the proton. It describes the proton by a  scalar diquark and a quark and does not assume the $SU(4)$ symmetry of the usual diquark models where both scalar and axial vector diquarks are considered.

\section{Wigner distribution}\label{WD}
In the light-front framework, the 5-dimensional Wigner distribution is defined as \cite{meissner0809} :  
\be
\rho^{q[\Gamma]}(\bfb,\bfp,x;S)\!\!=\!\!\!\int \!\! \frac{d^2\Dp}{(2\pi)^2} e^{-i\Dp.b_\perp} W^{q[\Gamma]}(\Dp,\bfp,x;S)
\label{wig_rho}
\ee
Where the correlator $W^{[\Gamma]}$  at 
at $\Delta^+=0$ and fixed light-cone time $ z^+=0$, is given by\cite{pasquini11}:
\be
W^{q[\Gamma]}(\Dp,\bfp,x;S)&=&\frac{1}{2}\int \frac{dz^-}{(2\pi)} \frac{d^2z_T}{(2\pi)^2} e^{ip.z}\nonumber\\
 \langle P^{\prime\prime}; S| &&\!\!\!\!\!\!\! \bar{\psi}^q(-z/2)\Gamma \mathcal{W}_{[-z/2,z/2]} \psi^q(z/2) |P^\prime;S\rangle \bigg|_{z^+=0}
\label{Wigner_W}
\ee

with the Dirac structure $\Gamma$ e.g, $\gamma^+,\gamma^+\gamma^5$.
The $P^\prime=(P^+,P^{\prime -},\frac{\Dp}{2})$ and the $P^{\prime\prime}=(P^+,P^{\prime\prime -}-\frac{\Dp}{2})$ are the initial and final momentum of proton. The $W^{[\Gamma]}$ depends on the average momentum  $P=\frac{1}{2}(P^{\prime\prime}+P^\prime)$ of proton, average quark momentum $\bfp=\frac{1}{2}(\bfp^{\prime\prime}+\bfp^\prime)$, the proton helicity $S$ and the transverse momentum transfer to the proton$\Dp=(P^{\prime\prime}_\perp-P^\prime_\perp)$. The Wilson line $\mathcal{W}_{[-z/2,z/2]}$ ensures the gauge invariance of the operator. 
We choose the symmetric frame where the components of 4-momentums, with skewness $\xi=0$, are 
\be
P&=&[P^+,P^-,\textbf{0}_\perp],\\
p&=&[xP^+,p^-,\bfp],\\
\Delta&=&[0,0,\Dp] ,
\ee
with $P^-=\frac{1}{2}(P^{\prime\prime-}+P^{\prime-})=\frac{4M^2+\Dp^2}{4P^+}$ (we use the notation $v^\pm=v^0\pm v^3$).
We calculate the matrix element of Eq.(\ref{Wigner_W}) in scalar diquark model using the wave functions predicted by soft-wall AdS/QCD.
The Wigner distributions, with the proton helicity $\Lambda$ and the quark helicity $\lambda$, for unpolarized and longitudinally polarized proton is defined as:
\be 
\rho^q_{\Lambda \lambda}(\bfb,\bfp,x)&=&\frac{1}{2}[\rho^{q[\gamma^+]}(\bfb,\bfp,x;\Lambda \hat{S}_z) +\nonumber\\
&&\lambda \rho^{q[\gamma^+\gamma^5]}(\bfb,\bfp,x;\Lambda \hat{S}_z)],
\label{rho_Lamlam}
\ee
which can be decomposed as:
\be 
\rho^q_{\Lambda \lambda}(\bfb,\bfp,x)&=&\frac{1}{2}[\rho^{q}_{UU}(\bfb,\bfp,x) +\nonumber\\
&& \Lambda \rho^{q}_{LU}(\bfb,\bfp,x)+ \lambda\rho^{q}_{UL}(\bfb,\bfp,x) \nonumber\\
&&+\Lambda\lambda \rho^{q}_{LL}(\bfb,\bfp,x)].
\label{rho_long}
\ee
corresponding to the proton spin $\Lambda=\uparrow, \downarrow $ and quark spin $\lambda=\uparrow,\downarrow$ (where $\uparrow$ and $\downarrow$ are corresponding to  $+1$ and $-1$ respectively).
Where  the Wigner distribution  $\rho^q_{UU}(\bfb,\bfp,x)$ of unpolarized quarks in an unpolarized proton,  and the distortions $\rho^q_{LU}(\bfb,\bfp,x)$ 
due to unpolarized quarks in a longitudinally polarized proton, $\rho^q_{UL}(\bfb,\bfp,x)$ due to longitudinally polarized quarks in an unpolarized proton and $\rho^q_{LL}(\bfb,\bfp,x)$ due to longitudinally polarized quark in a longitudinally polarized proton,  are defined as 
\be 
\rho^{q}_{UU}(\bfb,\bfp,x)&=&\frac{1}{2}[\rho^{q[\gamma^+]}(\bfb,\bfp,x; +\hat{S}_z)\nonumber\\
&& + \rho^{q[\gamma^+]}(\bfb,\bfp,x; -\hat{S}_z)],\label{rho_UU_def}\\
\rho^{q}_{LU}(\bfb,\bfp,x)&=&\frac{1}{2}[\rho^{q[\gamma^+]}(\bfb,\bfp,x; +\hat{S}_z) \nonumber\\
&& - \rho^{q[\gamma^+]}(\bfb,\bfp,x; -\hat{S}_z)],\label{rho_LU_def}\\
\rho^{q}_{UL}(\bfb,\bfp,x)&=&\frac{1}{2}[\rho^{q[\gamma^+\gamma^5]}(\bfb,\bfp,x; +\hat{S}_z) \nonumber\\
&&+\rho^{q[\gamma^+\gamma^5]}(\bfb,\bfp,x; -\hat{S}_z)],\label{rho_UL_def}\\
\rho^{q}_{LL}(\bfb,\bfp,x)&=&\frac{1}{2}[\rho^{q[\gamma^+\gamma^5]}(\bfb,\bfp,x; +\hat{S}_z) \nonumber\\
&&-\rho^{q[\gamma^+\gamma^5]}(\bfb,\bfp,x; -\hat{S}_z)].\label{rho_LL_def}
\ee 
 These four distributions are related with the Fourier transforms of the GTMDs as:
\be
\rho^q_{UU}(\bfb,\bfp,x)&=& \mathcal{F}^q_{1,1}(x,0,\bfp^2,\bfp.\bfb,\bfb^2), \label{rhoUU_F11}\\
\rho^q_{LU}(\bfb,\bfp,x)&=&-\frac{1}{M^2}\epsilon^{ij}_\perp p^i_\perp \nonumber\\
&& \frac{\partial}{\partial b^j_\perp} \mathcal{F}^q_{1,4}(x,0,\bfp^2,\bfp.\bfb,\bfb^2),\label{rhoLU_F14} \\
\rho^q_{UL}(\bfb,\bfp,x)&=&\frac{1}{M^2}\epsilon^{ij}_\perp p^i_\perp \nonumber\\
&&\frac{\partial}{\partial b^j_\perp} \mathcal{G}^q_{1,1}(x,0,\bfp^2,\bfp.\bfb,\bfb^2),\label{rhoUL_G11}\\
\rho^q_{LL}(\bfb,\bfp,x)&=&  \mathcal{G}^q_{1,4}(x,0,\bfp^2,\bfp.\bfb,\bfb^2).\label{rhoLL_G14}
\ee
Where the $\chi^q = \mathcal{F}^q_{1,1}, \mathcal{F}^q_{1,4}, \mathcal{G}^q_{1,1}, \mathcal{G}^q_{1,4}$ can be expressed as Fourier transform of corresponding GTMDs $X^q= F^q_{1,1}, F^q_{1,4}, G^q_{1,1}, G^q_{1,4}$.
\be 
\chi^q(x,0,\bfp^2,\bfp.\bfb,\bfb^2) &=& \int\frac{d^2\Dp}{(2\pi)^2} e^{-i\Dp.\bfb} \nonumber\\
&& \!\!\!\!\! X^q(x,0,\bfp^2,\bfp.\Dp,\Dp^2),\label{chi_GTMDs}
\ee
 
Integrating over all the variables, the Wigner distributions give
\be 
\int dx d^2\bfp d^2\bfb \rho^q_{UU}(\bfb,\bfp,x)&=&n_q, \label{norm_UU}\\
\int dx d^2\bfp d^2\bfb \rho^q_{LU}(\bfb,\bfp,x)&=&0, \label{norm_LU} \\
\int dx d^2\bfp d^2\bfb \rho^q_{UL}(\bfb,\bfp,x)&=&0, \label{norm_UL}\\
\int dx d^2\bfp d^2\bfb \rho^q_{LL}(\bfb,\bfp,x)&=& \Delta q. \label{norm_LL}
\ee
Where the $n_q$ is the flavor factors, $n_u=2$, $n_d=1$ and the $\Delta q$ is the axial charge. 

 Wigner distributions cannot have a direct probabilistic interpretation, however integrating over momentum and position, Wigner distributions can be reduced to probability distributions. Integrating over $\bfb$ with $\Dp=0$, the Wigner distributions reduce to the transverse momentum dependent parton distributions(TMDs). At $z_\perp=0$, the $\bfp$ integration of Wigner distributions give generalized parton distributions(GPDs).
The unpolarized TMD $f^q_1(x,\bfp^2)$ and GPD $H^q(x,0,\Dp^2)$ can be extracted as
\be 
f^q_1(x,\bfp^2)&=&F^q_{1,1}(x,0,\bfp^2,0,0),\\
H^q(x,0,\Dp^2)&=& \int d^2\bfp F^q_{1,1}(x,0,\bfp^2,\bfp.\Dp,\Dp^2),\label{H_F11}
\ee  
and the TMD $g^q_{1L}(x,\bfp^2)$ and GPD $\tilde{H}^q(x,0,\Dp^2)$ can be expressed as:
\be 
g^q_{1L}(x,\bfp^2)&=& G^q_{1,4}(x,0,\bfp^2,0,0),\\
\tilde{H}^q(x,0,\Dp^2)&=&\int d^2\bfp G^q_{1,4}(x,0,\bfp^2,\bfp.\Dp,\Dp^2).\label{Ht_G14}
\ee
The $\bfp$ and $\bfb$ integration of the $\rho^q_{LU}$ and $\rho^q_{UL}$ give zero. So, there are no TMD and GPD corresponding to $F_{1,4}$ and $G_{1,1}$ GTMDs.

The Wigner distributions can also be reduced to three dimensional quark densities by integrating over two mutually orthogonal components of transverse position and momentum, e,g. $b_y$ and $p_x$ ($b_x$ and $p_y$), which are not constraint by Heisenberg uncertainty principle as:
\be
\int db_y dp_x \rho^{q[\Gamma]}(\bfb,\bfp,x;S)=\tilde{\rho}^{q[\Gamma]}(b_x,p_y,x;S),
\label{rho_bxpy}
\ee
with $\Delta_y=z_x=0$.
Note that the integration over other mixed transverse components $b_x$ and $p_y$ gives the same quark density as Eq.(\ref{rho_bxpy}), with a opposite momentum i.e, $\tilde{\rho}^{q[\Gamma]}(b_y,p_x,x;S)=\tilde{\rho}^{q[\Gamma]}(b_x,-p_y,x;S)$. 
 These relations are true only when there is axial symmtery i.e., for unpolarized or longitudinally polarized proton.

\section{Orbital angular momentum}\label{oam}
Jaffe and Manohar showed in the light-cone gauge that the spin of the nucleon can be decomposed into the quark spin, quark OAM, gluon spin and gluon OAM\cite{Jaffe_Mon}.
\be
S^q+\ell^q + S^g + \ell^g = \frac{1}{2}. \label{decom_Jaf}
\ee 
For the diquark model, the above sum rule can  be written as
\be
S^q+\ell^q + S^D + \ell^D = \frac{1}{2},
\ee
where the super-script $D$ is for diquark, and for scalar diquark $S^D=0$.  The canonical OAM operator for quark is defined as
\be
\hat{\ell}^q_z(b^-,\bfb,p^+,\bfp)&=&\frac{1}{2}\int \frac{dz^-d^2\textbf{z}_\perp}{(2\pi)^3}e^{ip.z}\nonumber\\
\bar{\psi}^q(b^--\frac{z^-}{2},\textbf{b}_\perp)\gamma^+&&\!\!\!\! (\textbf{b}_\perp \times(-i\stackrel{\leftrightarrow}{\partial}_\perp))\psi^q(b^-+\frac{z^-}{2},\textbf{b}_\perp). \nonumber\\
\ee 
From the definition of Wigner operator (Eq.(\ref{Wigner_W})), the OAM density operator can be expressed as
\be
 \hat{\ell}^q_z=2(\textbf{b}_\perp \times \bfp)\hat{W}^{q[\gamma^+]}.
\ee
Thus in light-front gauge the average canonical OAM for quark is written in terms of Wigner distribution as.
\be 
\ell^q_z &=& \int\frac{d\Delta^+ d^2\Dp}{2P^+(2\pi)^3}\langle P^{\prime\prime};S|\hat{\ell}^q_z|P^\prime;S\rangle \nonumber\\
&=&\!\!\!\! \int dx d^2\bfp d^2\bfb (\bfb\times\bfp)_z\rho^{q[\gamma^+]}(\bfb,\bfp,x,\hat{S}_z).\label{OAM_ell_def}
\ee

Where, the distribution $\rho^{q[\gamma^+]}(\bfb,\bfp,x,\hat{S}_z)$ can be written from Eqs.(\ref{rho_UU_def},\ref{rho_LU_def}) as:
\be
\rho^{q[\gamma^+]}(\bfb,\bfp,x,+\hat{S}_z)&=&\rho^q_{UU}(\bfb,\bfp,x)\nonumber\\
&&+\rho^q_{LU}(\bfb,\bfp,x)
\ee
From Eq.(\ref{rhoUU_F11}) we see that
\be 
\int dx d^2\bfp d^2\bfb (\bfb\times\bfp)_z \rho^q_{UU}(\bfb,\bfp,x)=0,
\label{OAM_rhoUU}
\ee
 which satisfies the angular momentum sum rule for unpolarized proton, the total angular momentum of constituents sum up to zero. Using Eq.(\ref{rhoLU_F14}) and Eq.(\ref{chi_GTMDs}), the twist-2 canonical quark OAM in the light-front gauge is
\be 
\ell^q_z &=&-\int dx d^2\bfp \frac{\bfp^2}{M^2}F^q_{1,4}(x,0,\bfp^2,0,0). \label{OAM_ell}
\ee 

The Jaffe-Manohar decomposition ( Eq.(\ref{decom_Jaf})) is not gauge invariant. 
Ji proposed a gauge invariant decomposition of nucleon spin as\cite{Ji_97}
\be
S^q + L^q + J^g = \frac{1}{2},\label{Decom_Ji} 
\ee 
where $L^q$ is the kinetic OAM for the quark $q$.
However, Chen et al.\cite{chen} proposed an idea to decompose the gauge field $A_\mu $ into a pure gauge part, $A^{pure}_\mu$,  and a physical part, $A^{phy}_\mu$ to give a gauge invariant definition of the Jaffe-Manohar decomposition.  

The kinetic OAM of quark appearing in the Ji sum rule is defined in terms of GPDs as\cite{Ji_97}:
\be 
L^q_z=\frac{1}{2}\int dx \bigg[x\bigg(H^q(x,0,0)+E^q(x,0,0)\bigg) -\tilde{H}^q(x,0,0)\bigg] ,\label{OAM_L}
\ee
where $H^q(x,\xi,t)$ and $E^q(x,\xi,t)$  are unpolarized GPDs and $\tilde{H}^q(x,\xi,t)$ is the helicity dependent GPD. In our model calculation, the explicit expressions are given in Sec.\ref{results}. A comparative study between longitudinal component of  canonical OAM and kinetic OAM are shown in the Fig. \ref{fig_OAM_x} and the values are given in Table \ref{tab_OAM}.
 Note that the above relation (Eq.\ref{OAM_L}) does not hold for density level  interpretation in the transverse plane \cite{Liu}.

The spin-orbit correlation  is given by the operator
\be
C^q_z(b^-,\bfb,p^+,\bfp)&=&\frac{1}{2}\int \frac{dz^-d^2\textbf{z}_\perp}{(2\pi)^3}e^{ip.z}\nonumber\\
\!\!\!\bar{\psi}^q(b^--\frac{z^-}{2},\textbf{b}_\perp)&&\!\!\!\!\!\!\! \gamma^+\gamma^5(\textbf{b}_\perp \times(-i\stackrel{\leftrightarrow}{\partial}_\perp))\psi^q(b^-+\frac{z^-}{2},\textbf{b}_\perp).\nonumber\\
\ee 
The correlation between quark spin and quark OAM can be expressed  with Wigner distributions $\rho^q_{UL}$ and equivalently in terms of GTMD as:
\be 
C^q_z&=&\int dx d^2\bfp d^2\bfb (\bfb\times\bfp)_z \rho^q_{UL}(\bfb,\bfp,x) \nonumber \\
&=& \int dx d^2\bfp \frac{\bfp^2}{M^2}G^q_{1,1}(x,0,\bfp^2,0,0). \label{Cqz}
\ee 
Where $C^q_z>0$ implies the quark spin and OAM tend to be aligned  and $C^q_z<0$ implies they are anti-aligned. In our model, the quark spin and OAM tend to be anti-aligned for both $u$ and $d$ quarks.

One can see from Eq.(\ref{rhoLL_G14}),  a similar correlator   with $\rho^q_{LL}$  vanishes
\be 
\int dx d^2\bfp d^2\bfb (\bfb\times\bfp)_z \rho^q_{LL}(\bfb,\bfp,x)=0. \label{OAM_rhoLL}
\ee

\section{Results}\label{results}
We calculate the Wigner distributions of proton in light-front AdS/QCD quark-diquark model.  Using Eq.(\ref{fock_PS}) in Eq.(\ref{Wigner_W}) the quark-quark correlator, $W^{q[\Gamma]}(\Dp,\bfp,x;S)$, can be expressed in terms of LFWFs as.
\be
W^{q[\gamma^+]}(\Dp,\bfp,x;\pm\hat{S}_z)&=&\frac{1}{16\pi^3}\bigg[\psi^{\pm\dagger}_{q+}(x,\bfp^{\prime\prime})\psi^{\pm}_{q+}(x,\bfp^{\prime}) \nonumber\\
&&\!\!\!\!\! +\psi^{\pm\dagger}_{q-}(x,\bfp^{\prime\prime})\psi^{\pm}_{q-}(x,\bfp^{\prime}) \bigg],\label{W_V}\\
W^{q[\gamma^+\gamma^5]}(\Dp,\bfp,x;\pm\hat{S}_z)&=&\frac{1}{16\pi^3}\bigg[\psi^{\pm\dagger}_{q+}(x,\bfp^{\prime\prime})\psi^{\pm}_{q+}(x,\bfp^{\prime})\nonumber\\
&&\!\!\! \!\! -\psi^{\pm\dagger}_{q-}(x,\bfp^{\prime\prime})\psi^{\pm}_{q-}(x,\bfp^{\prime}) \bigg].\label{W_A} 
\ee
for the Dirac structures $\Gamma=\gamma^+,\gamma^+\gamma^5$. In the symmetric frame the initial and final momentums of the struck quark are 
\be 
\bfp^{\prime}=\bfp-(1-x)\frac{\Dp}{2},\\
\bfp^{\prime\prime}=\bfp+(1-x)\frac{\Dp}{2}
\ee
respectively. Using the wave functions from Eq.(\ref{LFWF},\ref{phi}) in Eqs.(\ref{W_V},\ref{W_A}), the explicit expressions for Wigner distributions are
\be 
\rho^q_{UU}(\bfb,\bfp,x)&=&\frac{1}{16\pi^3}\int\frac{d\Delta_\perp}{2\pi} \Delta_\perp \rm{J}_0(|\Delta_\perp||b_\perp|)\nonumber\\
 &&\!\!\!\! \exp\big(-2\tilde{a}(x)\tilde{\textbf{p}}^2_\perp \big) \bigg[|A^{(1)}_q(x)|^2 +\nonumber\\
&&\!\!\!\!\!\!\!\!\!\!\!\!\!\!\!\!\!\!\! \bigg(\bfp^2 - \frac{\Dp^2}{4}(1-x)^2\bigg)\frac{1}{M^2x^2}|A^{(2)}_q(x)|^2 \bigg],\label{rho_UU}\\
\rho^q_{LU}(\bfb,\bfp,x)&=&-\frac{1}{M^2}\epsilon^{ij}_\perp p^i_\perp \frac{\partial}{\partial b^j_\perp}\bigg[-\frac{1}{16\pi^3}\int\frac{d\Delta_\perp}{2\pi} \nonumber\\
\Delta_\perp \rm{J}_0(|\Delta_\perp||b_\perp|)&&\!\!\!\!\!\!\! \exp\big(-2\tilde{a}(x)\tilde{\textbf{p}}^2_\perp \big) 
\frac{(1-x)}{x^2}|A^{(2)}_q(x)|^2\bigg],\label{rho_LU}\\
\rho^q_{UL}(\bfb,\bfp,x)&=&\frac{1}{M^2}\epsilon^{ij}_\perp p^i_\perp \frac{\partial}{\partial b^j_\perp}\bigg[-\frac{1}{16\pi^3}\int\frac{d\Delta_\perp}{2\pi} \nonumber\\
 \Delta_\perp \rm{J}_0(|\Delta_\perp||b_\perp|)&&\!\!\!\!\!\!\!\!\! \exp\big(-2\tilde{a}(x)\tilde{\textbf{p}}^2_\perp \big) 
\frac{(1-x)}{x^2}|A^{(2)}_q(x)|^2\bigg],\label{rho_UL}\\
\rho^q_{LL}(\bfb,\bfp,x)&=&\frac{1}{16\pi^3}\int\frac{d\Delta_\perp}{2\pi} \Delta_\perp \rm{J}_0(|\Delta_\perp||b_\perp|)\nonumber\\
&&\!\!\!\!\! \exp\big(-2\tilde{a}(x)\tilde{\textbf{p}}^2_\perp \big) 
 \bigg[|A^{(1)}_q(x)|^2 -\nonumber\\
&&\!\!\!\!\!\!\!\!\!\!\!\!\!\!\!\!\!\!
\bigg(\bfp^2 - \frac{\Dp^2}{4}(1-x)^2\bigg)\frac{1}{M^2x^2}|A^{(2)}_q(x)|^2 \bigg].\label{rho_LL}
\ee
Where 
\be 
A^{(i)}_q(x)&=& N^{(i)}_q \frac{4\pi}{\kappa}\sqrt{\frac{\log(1/x)}{(1-x)}}x^{a^{(i)}_q}(1-x)^{b^{(i)}_q},\\
\tilde{a}(x)&=& \frac{\log(1/x)}{2\kappa^2(1-x)^2},\\
\tilde{\textbf{p}}^2_\perp &=&\bfp^2+\frac{\Dp^2}{4}(1-x)^2.
\ee
At the limit $\xi=0$, the GTMDs are 
\be 
F^q_{1,1}(x,\Dp^2,\bfp^2) &=& \frac{1}{16\pi^3}\bigg[|A^{(1)}_q(x)|^2 +
\bigg(\bfp^2 - \frac{\Dp^2}{4}(1-x)^2\bigg)\nonumber\\
&&\frac{1}{M^2x^2}|A^{(2)}_q(x)|^2 \bigg]\exp\big[-2\tilde{a}(x)\tilde{\textbf{p}}^2_\perp \big],\label{F11} \\
F^q_{1,4}(x,\Dp,\bfp^2) &=& -\frac{1}{16\pi^3}\bigg[\frac{(1-x)}{x^2}|A^{(2)}_q(x)|^2\bigg] \nonumber\\
&&\exp\big[-2\tilde{a}(x)\tilde{\textbf{p}}^2_\perp \big],\label{F14}\\
G^q_{1,1}(x,\Dp^2,\bfp^2) &=& -\frac{1}{16\pi^3}\bigg[\frac{(1-x)}{x^2}|A^{(2)}_q(x)|^2\bigg] \nonumber\\
&&\exp\big[-2\tilde{a}(x)\tilde{\textbf{p}}^2_\perp \big] ,\label{G11}\\
G^q_{1,4}(x,\Dp,\bfp^2) &=& \frac{1}{16\pi^3}\bigg[|A^{(1)}_q(x)|^2 -
\bigg(\bfp^2 - \frac{\Dp^2}{4}(1-x)^2\bigg)\nonumber\\
&&\frac{1}{M^2x^2}|A^{(2)}_q(x)|^2 \bigg]\exp\big[-2\tilde{a}(x)\tilde{\textbf{p}}^2_\perp \big].\label{G14}
\ee
We find $F^q_{1,4}=G^q_{1,1}$ to the leading order as found in \cite{Lorce14} for scalar diquark model. Thus the distributions $\rho^q_{LU}=-\rho^q_{UL}$.
\begin{figure*}[htbp]
\begin{minipage}[c]{0.98\textwidth}
\small{(a)}\includegraphics[width=7cm,clip]{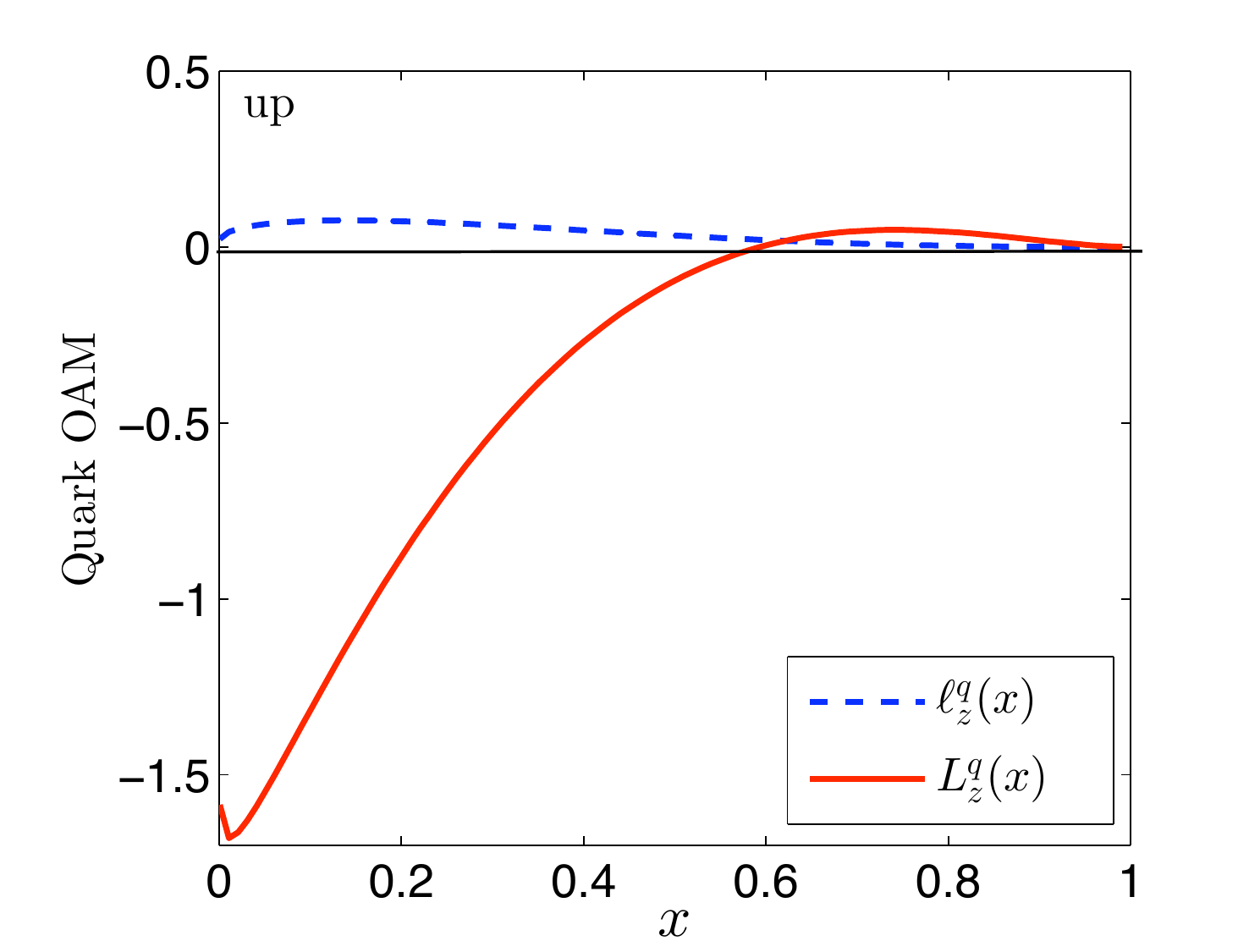}
\hspace{0.1cm}%
\small{(b)}\includegraphics[width=7cm,clip]{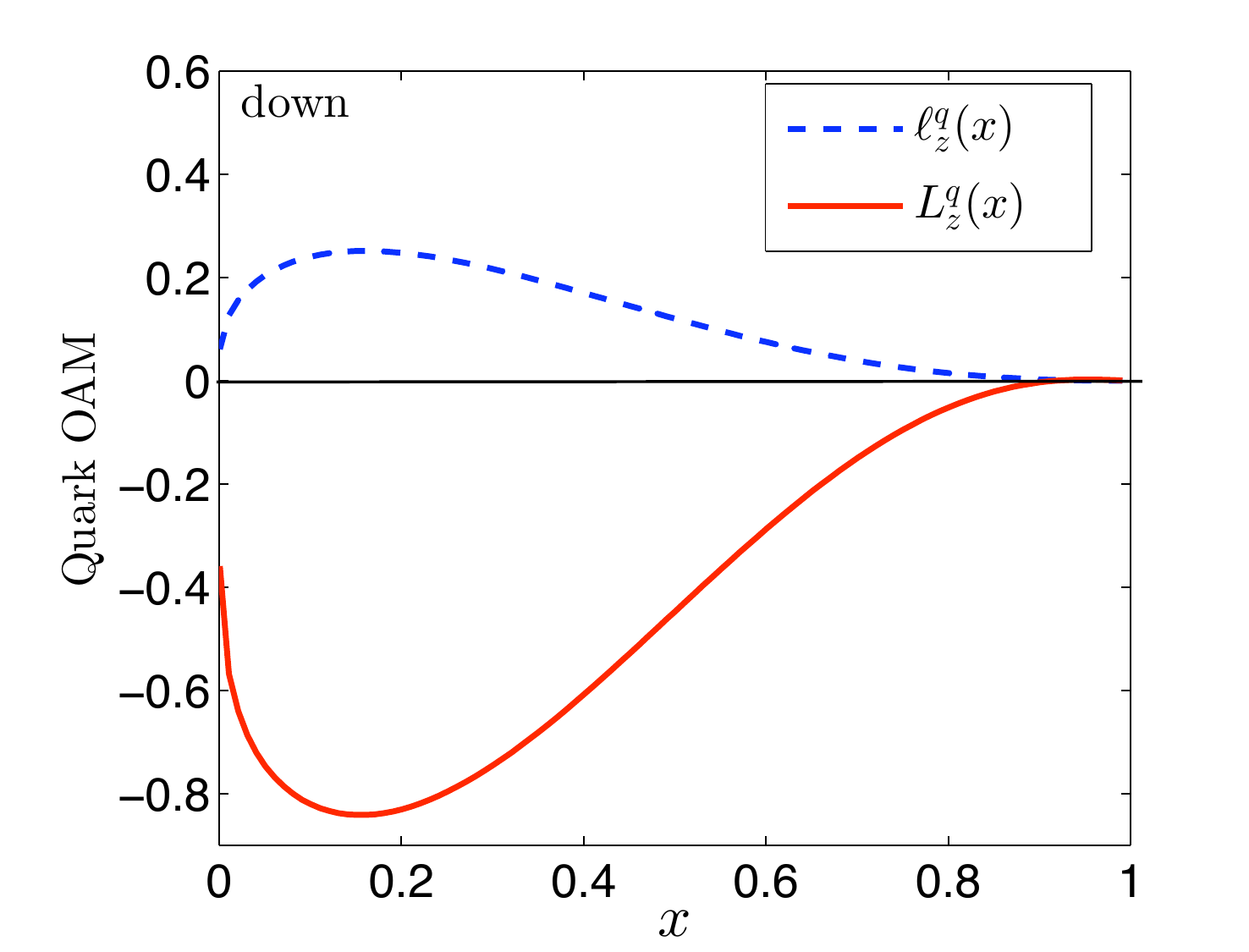}
\end{minipage}
\caption{\label{fig_OAM_x} The variation of canonical OAM $\ell^q_z(x)$ and kinetic OAM $L^q_z(x)$ with longitudinal momentum fraction $x$, for (a)  $u$ quark and (b)  $d$ quark.}
\end{figure*}

From Eq.(\ref{OAM_ell}), the canonical OAM can be written as
\be 
\ell^q_z=\int dx \ell^q_z(x).\label{OAM_ell_res}
\ee
Using Eq.(\ref{F14}), $\ell^q_z(x)$ can be written as 
\be 
\ell^q_z(x)= \frac{\kappa^2}{M^2 \log(1/x)}F^q_2(x)(1-x)^5. \label{OAM_ell_x}
\ee
In this model, $\ell^q_z$ can also be related with pretzelosity $h^\perp_{1T}$ as 
\be \ell^q_z=-\int dx d^2\bfp \frac{\bfp^2}{2M^2}(1-x)h^{q\perp}_{1T}(x,\bfp^2). \label{OAM_TMD}
\ee
 $h^{q\perp}_{1T}(x,\bfp^2)$ is one of the eight leading twist TMDs. In this light-front scalar diquark model  $h^{q\perp}_{1T}(x,\bfp^2)$ is written as\cite{tm_rel}
\be 
\!\! {h}^{q\perp  }_{1T}(x,\textbf{p}^2_{\perp})&\!\!=&\!\! - \frac{2\log(1/x)}{\pi\kappa^2}\exp\bigg[-\frac{\textbf{p}_{\perp}^2\log(1/x)}{\kappa^2(1-x)^2}\bigg]F^q_2(x).\label{h1Tp}
\ee
where $F^q_2(x)$ is given in Eq(\ref{F2}).
Using Eq.(\ref{F11}) and Eq.(\ref{G14}) in Eq.(\ref{H_F11}) and (\ref{Ht_G14}), the GPDs $H$ and  $\tilde{H}$  can be expressed  as
\be
H^q(x,0,t)&=&\!\!\bigg[F^q_1(x)(1-x)^2+F^q_2(x)(1-x)^4 \frac{\kappa^2}{M^2\log(1/x)}\bigg] \nonumber\\
&&\!\!\!\!\!\!\!(1-\frac{|t|}{4\kappa^2} \log(1/x))\exp\bigg[-\frac{|t|}{4\kappa^2} \log(1/x)\bigg],\label{H}\\
\tilde{H}^{q}(x,0,t)&=&\!\!\! \bigg[F^q_1(x)(1-x)^2 - F^q_2(x)(1-x)^4\frac{\kappa^2}{M^2\log(1/x)}\bigg] \nonumber\\
&&\!\!\!\!\!\!\! (1+\frac{|t|}{4\kappa^2} \log(1/x))\exp\bigg[-\frac{|t|}{4\kappa^2} \log(1/x)\bigg].\label{Ht}
\ee
In the AdS/QCD light-front scalar diquark model the helicity flip GPD $E$ is given\cite{CM_EPJ} as
\be  
E^q(x,0,t)&=& 2F^q_3(x)(1-x)^3 \exp\bigg[-\frac{|t|}{4\kappa^2} \log(1/x)\bigg].\label{E}
\ee
Where  $Q^2=-q^2=-t$, the square of the momentum transferred in the process and  is taken to be zero for OAM calculation. 

The kinetic OAM of quarks(Eq.(\ref{OAM_L})) can be written as
\be 
L^q_z=\int dx L^q_z(x).\label{OAM_L_res}
\ee
Where in this model, using Eqs.(\ref{H}), (\ref{Ht}) and Eq.(\ref{E}) at $t=0$ limit, the $L^q_z(x)$ reads
\be 
L^q_z(x)&=&\frac{1}{2}\bigg[-F^q_1(x)(1-x)^3 +F^q_2(x)(1-x)^4(1+x)\nonumber\\
&&\frac{\kappa^2}{M^2\log(1/x)}+2F^q_3(x)x(1-x)^3 \bigg].\label{OAM_L_x}
\ee
Where
\be 
F^q_1(x)&=&|N^{(1)}_q|^2 x^{2a^{(1)}_q}(1-x)^{2b^{(1)}_q-1},\\
F^q_2(x)&=&|N^{(2)}_q|^2 x^{2a^{(2)}_q-2}(1-x)^{2b^{(2)}_q-1}, \label{F2}\\
F^q_3(x)&=&N^{(1)}_q N^{(2)}_q x^{a^{(1)}+a^{(2)}_q-1}(1-x)^{b^{(1)}_q+b^{(2)}_q-1}.
\ee
The variation of the quark OAMs $\ell^q_z(x)$ and $L^q_z(x)$ with longitudinal momentum fraction $x$ is sown in Fig.\ref{fig_OAM_x} for $u$ and $d$ quark.

\subsection{Unpolarized proton}
\begin{figure*}[htbp]
\begin{minipage}[c]{0.98\textwidth}
\small{(a)}\includegraphics[width=7cm,clip]{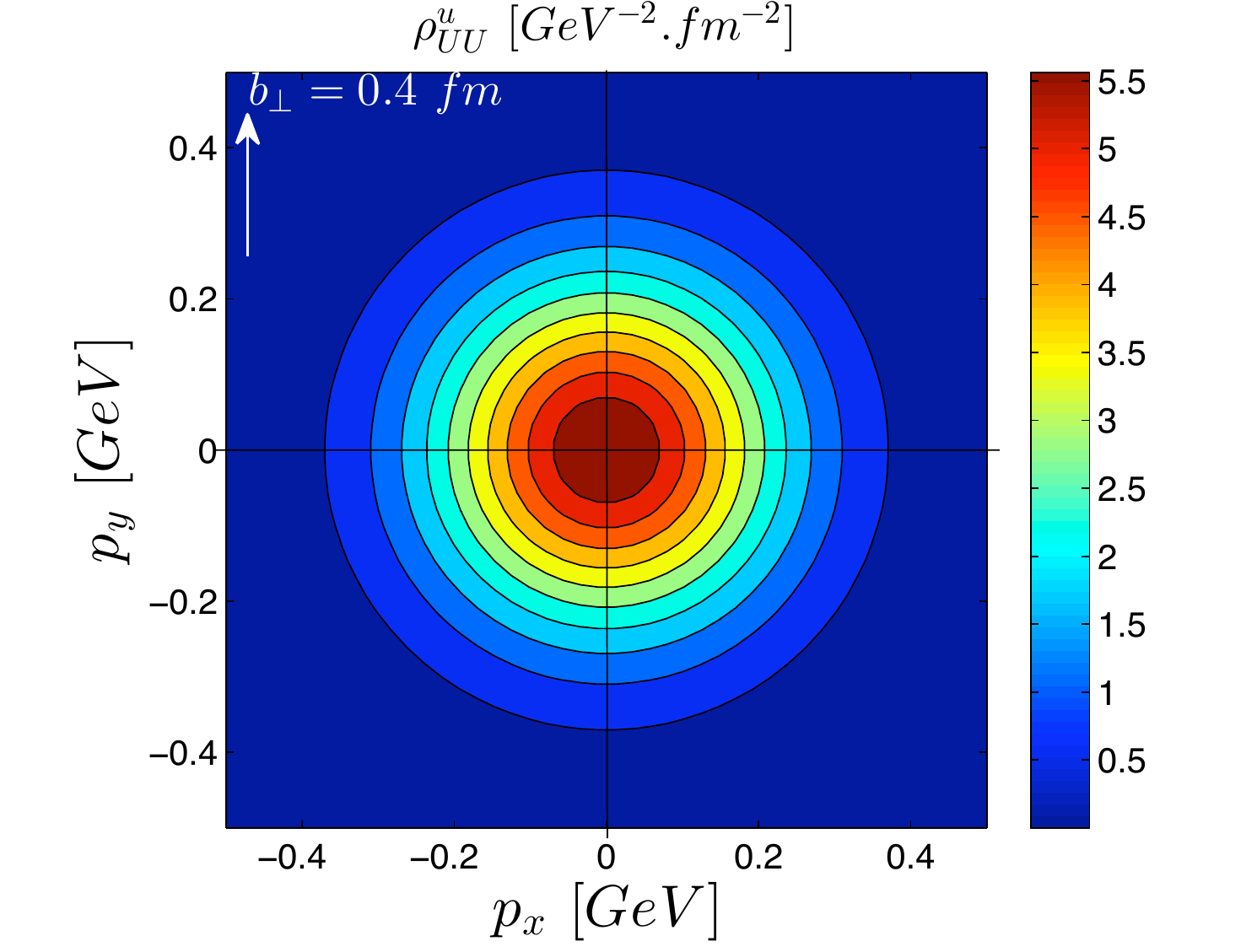}
\hspace{0.1cm}%
\small{(b)}\includegraphics[width=7cm,clip]{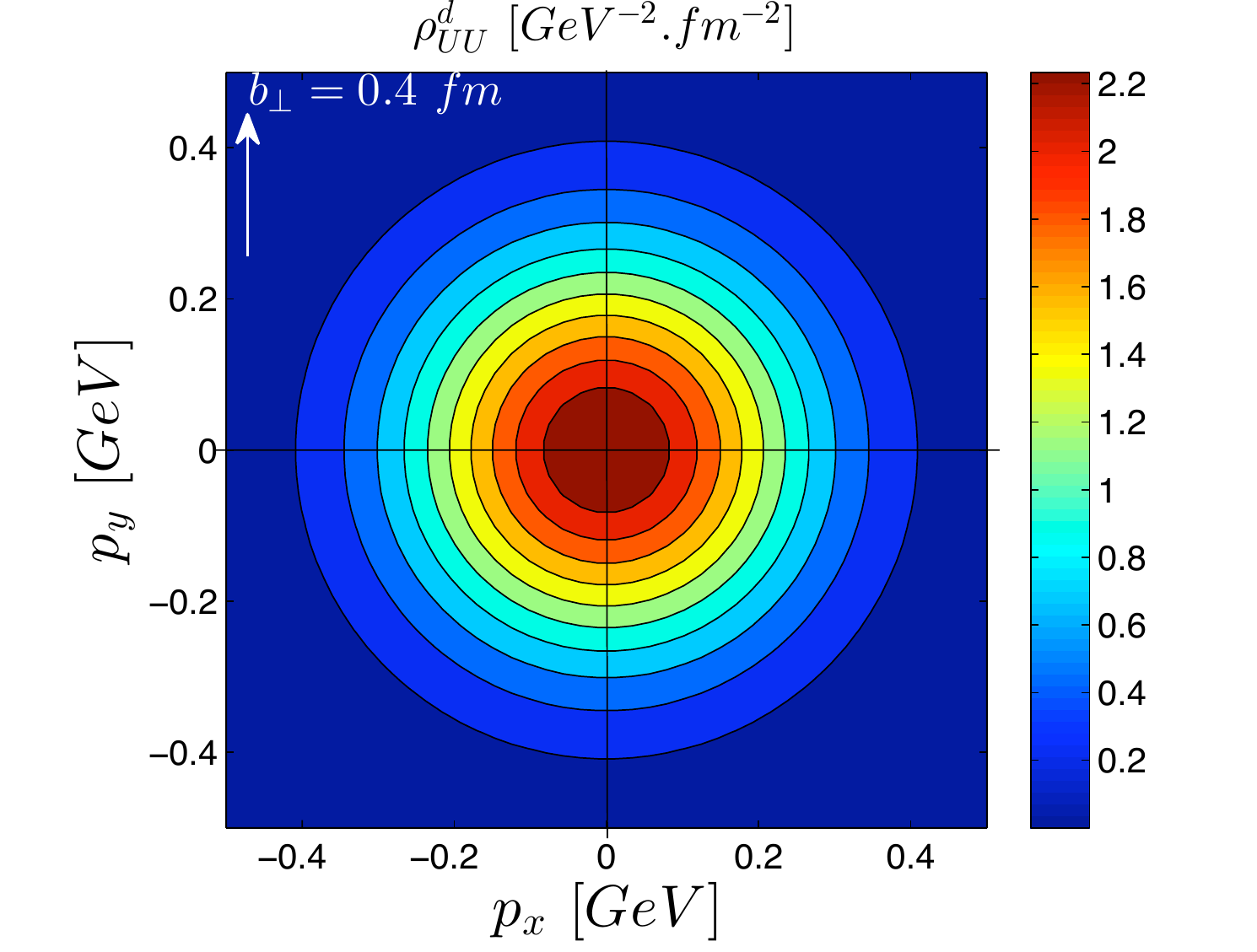}
\end{minipage}
\begin{minipage}[c]{0.98\textwidth}
\small{(c)}\includegraphics[width=7cm,clip]{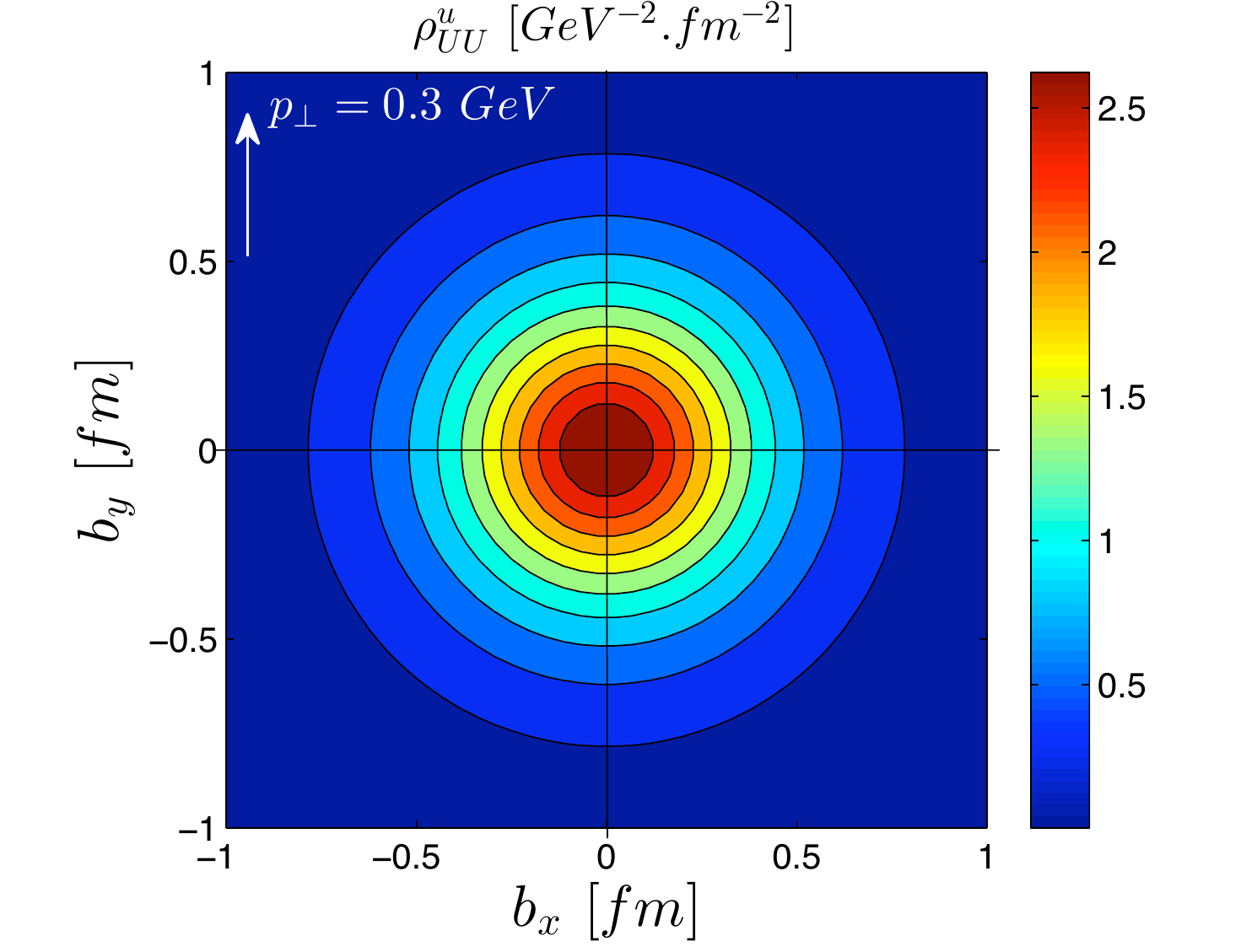}
\hspace{0.1cm}%
\small{(d)}\includegraphics[width=7cm,clip]{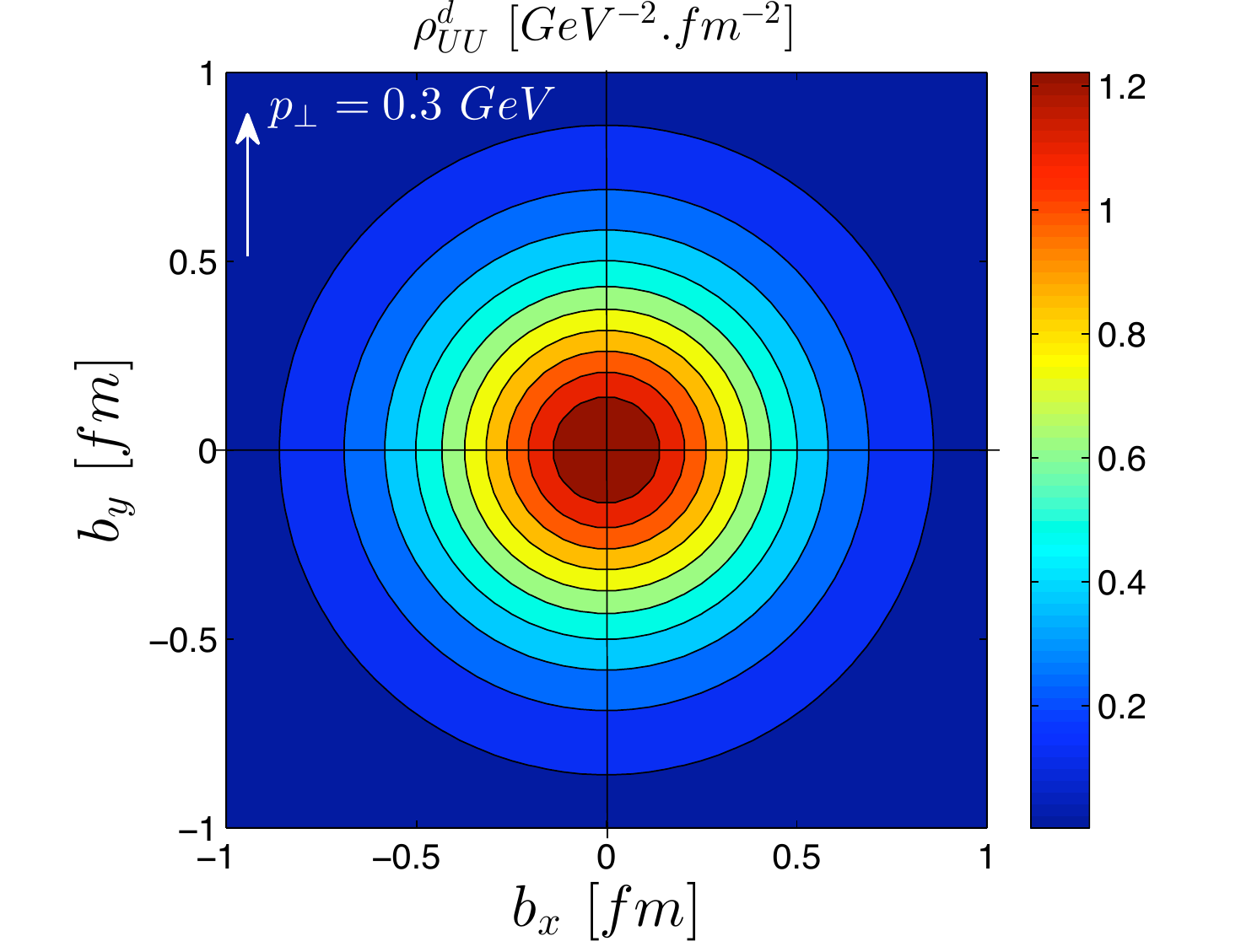}
\end{minipage}
\caption{\label{plot_rhoUU}The Wigner distributions of unpolarized quarks in an unpolarized proton in the transverse momentum plane(a,b) with $\bfb=0.4\hat{y}~ fm$ and in the transverse impact parameter plane(c,d) with $\bfp=0.3\hat{y}~ GeV$ for $u$ quarks(left column) and $d$ quarks(right column).}
\end{figure*}

In our numerical study, we have considered the active quark to be either a $u$ or $d$ quark, the spectator always being a diquark. In other words, when we calculate the Wigner distribution for the $u$ quark, we have not incorporated any contribution from the $u$ quark that is part of the diquark.
The first Mellin moment of $\rho^q_{UU}(\bfb,\bfp,x)$ is shown in Fig.\ref{plot_rhoUU}.  Fig.\ref{plot_rhoUU}(a) and Fig.\ref{plot_rhoUU}(b) represent the distributions in transverse momentum plane for $u$ quark and $d$ quark respectively. The fixed impact parameter $\bfb$ is taken along $\hat{y}$ and $b_y=0.4~fm$. 
The variation of $\rho^q_{UU}(\bfb,\bfp)$ in the transverse impact parameter plane are shown in Fig.\ref{plot_rhoUU}(c) and Fig.\ref{plot_rhoUU}(d) for $u$ and $d$ quark respectively, with fixed transverse momentum $\bfp$ along $\hat{y}$ for $p_y=0.3~GeV$. The distributions $\rho^u_{UU}$ and $\rho^d_{UU}$ are circularly symmetric, in transverse momentum plane as well as transverse impact parameter plane, with a positive maxima at the centre $(p_x=p_y=0)$, $(b_x=b_y=0)$ and gradually decrease towards periphery, for both $u$ and $d$ quarks. The peak of the distribution for $u$ quark is large compare to $d$ quark in both the planes.  
\begin{figure*}[htbp]
\begin{minipage}[c]{0.98\textwidth}
\small{(a)}\includegraphics[width=7cm,clip]{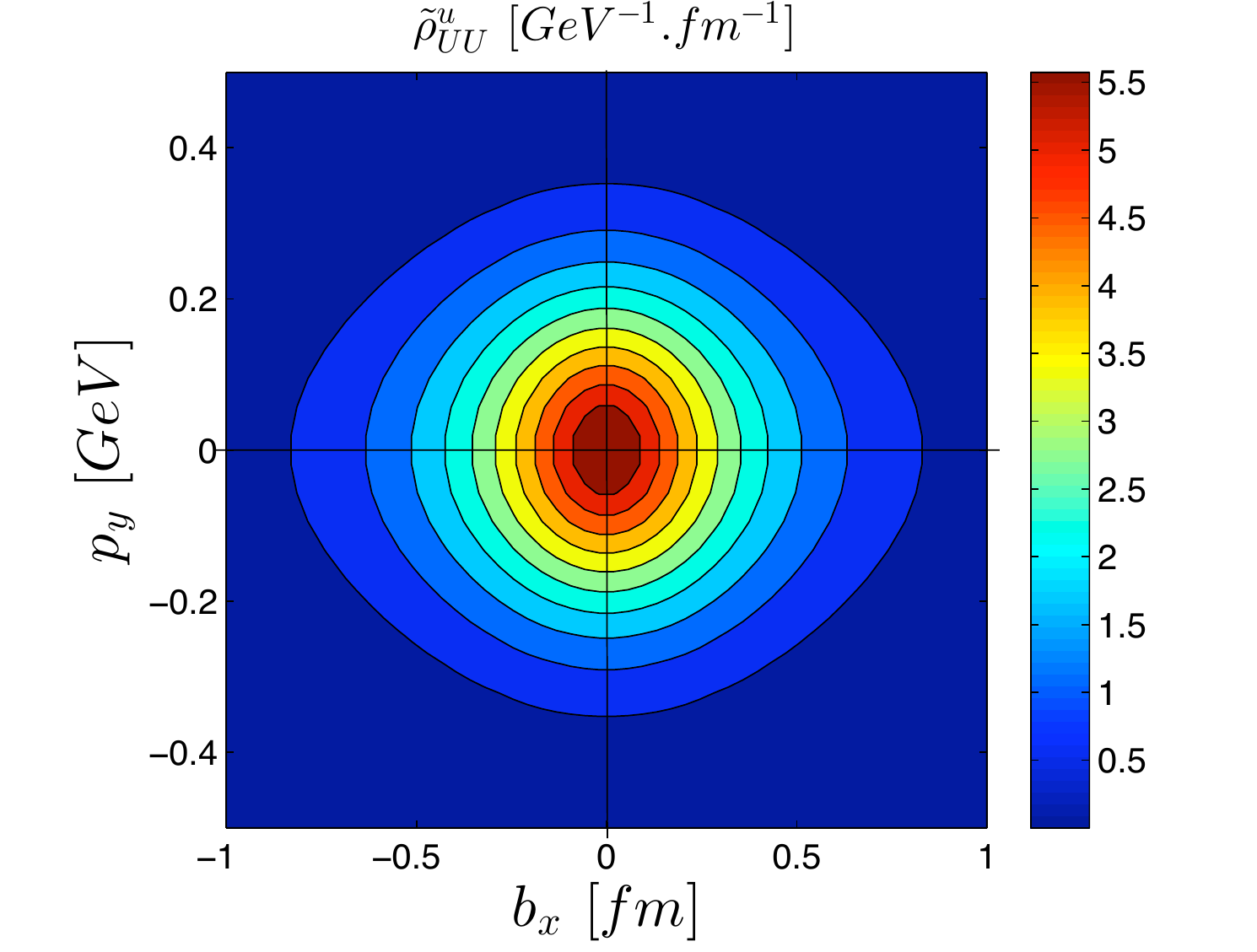}
\hspace{0.1cm}%
\small{(b)}\includegraphics[width=7cm,clip]{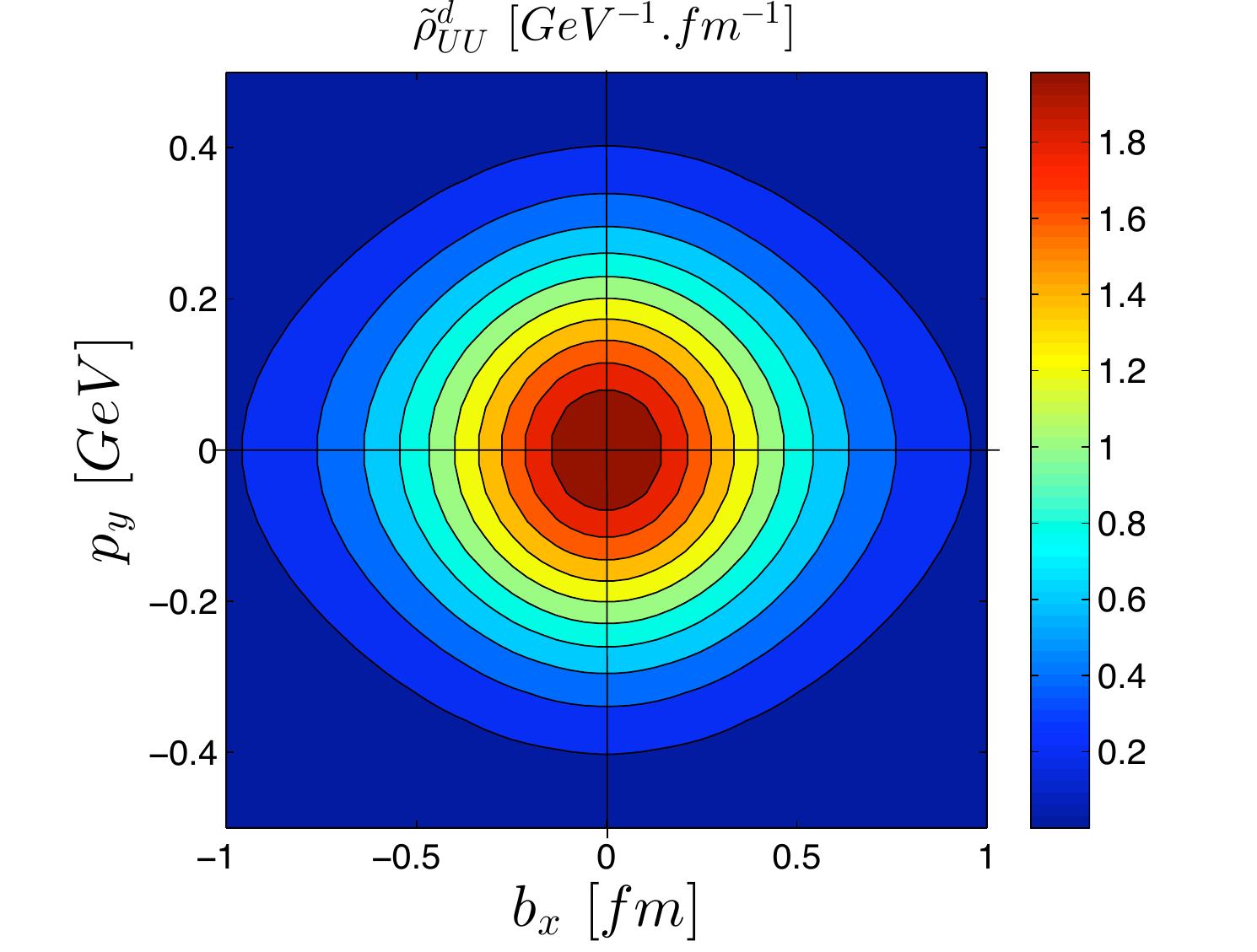}
\end{minipage}
\caption{\label{plot_rhoUU_bxpy} $\tilde{\rho}^q_{UU}(b_x,p_y)$ in mixed transverse plane for $u$ quark(a) and $d$ quark(b).}
\end{figure*}

The average quadrupole distortions $Q^{ij}_b(\bfp)$ and $Q^{ij}_p(\bfb)$ are defined as\cite{pasquini11}
\be 
Q^{ij}_b(\bfp)= \frac{\int d^2\bfb (2b^i_\perp b^j_\perp-\delta^{ij}\bfb^2)\rho_{UU}(\bfb,\bfp)}{\int d^2\bfb \bfb^2 \rho_{UU}(\bfb,\bfp)}\\
Q^{ij}_p(\bfb)= \frac{\int d^2\bfp (2p^i_\perp p^j_\perp-\delta^{ij}\bfp^2)\rho_{UU}(\bfb,\bfp)}{\int d^2\bfp \bfp^2 \rho_{UU}(\bfb,\bfp)}.\label{Eq_Qdis}
\ee
In this model, the average quadrupole distortion is found to be zero. Since the wave functions in soft-wall AdS/QCD model are of gaussian type, the $\rho_{UU}$ and $\rho_{LL}$ are even in $\bfp$ and $\bfb$ resulting to the zero quadrupole distortion. 

As we discussed before, the three dimensional quark densities can be extracted from the Wigner distributions by integrating over one transverse momentum $p_x$ and one transverse position $b_y$ variables(see Eq.(\ref{rho_bxpy})). $\tilde{\rho}_{UU}(b_x,p_y)$ in mixed transverse plane are shown in Fig.\ref{plot_rhoUU_bxpy}  for $u$ and $d$ quarks. We find that the distributions are axially symmetric. 
Therefore, there is no favored configuration between $\bfb\perp\bfp$ and $\bfb\parallel\bfp$ unlike the  light-cone constituent quark model(LCCQM) \cite{LCCQM} or chiral quark soliton model($\chi$QSM) \cite{chi_QSM}. At $b_x=p_y=0$, the probability density for $u$ and $d$ quark is maximum and decreases as $e^{-\alpha p_y^2}$ and $e^{-\beta b_x^2}$. Where the $\alpha$ and $\beta$ are positive constants and we observe $\alpha>\beta$ for both $u$ and $d$ quarks .

\begin{figure*}[htbp]
\begin{minipage}[c]{0.98\textwidth}
\small{(a)}\includegraphics[width=7cm,clip]{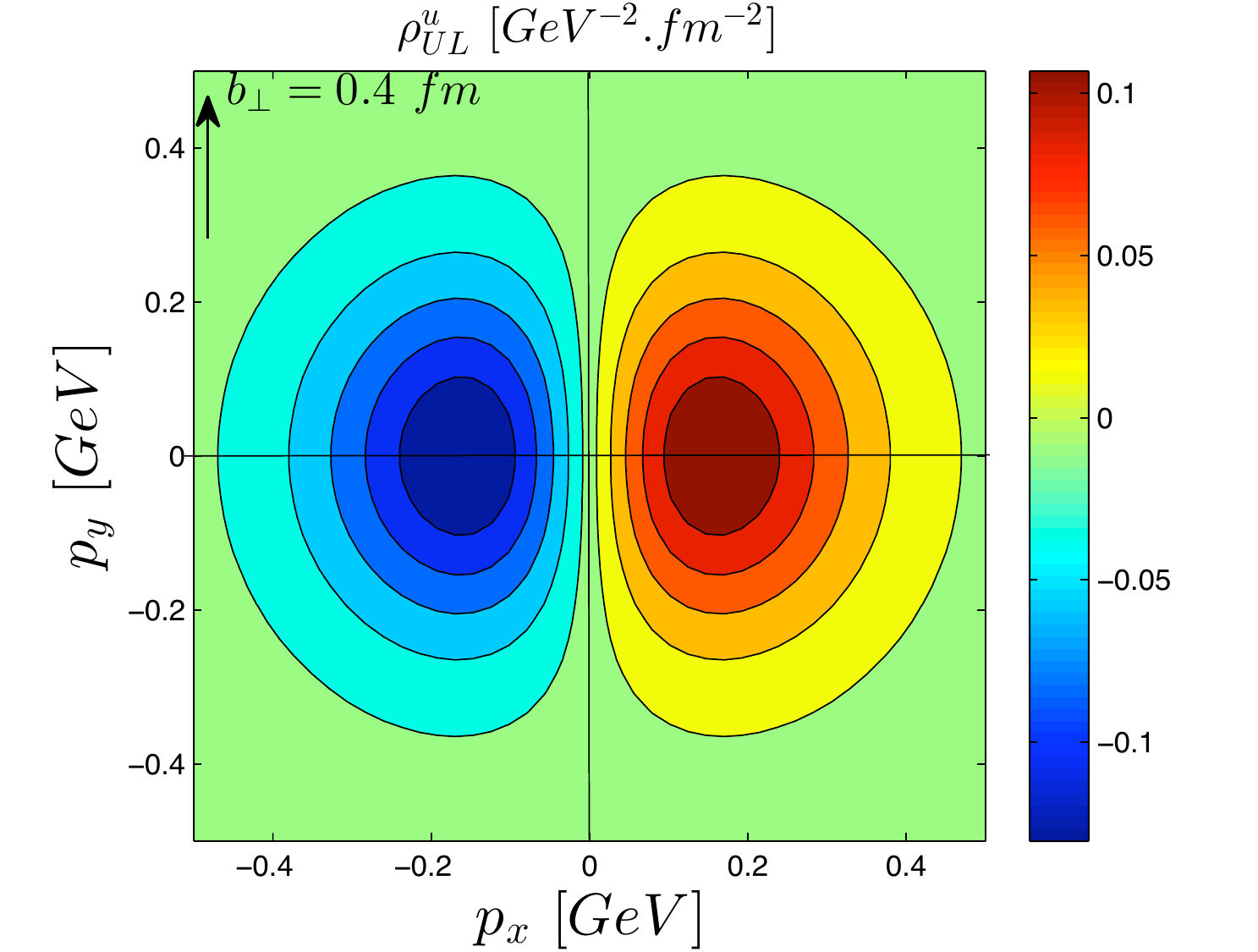}
\hspace{0.1cm}%
\small{(b)}\includegraphics[width=7cm,clip]{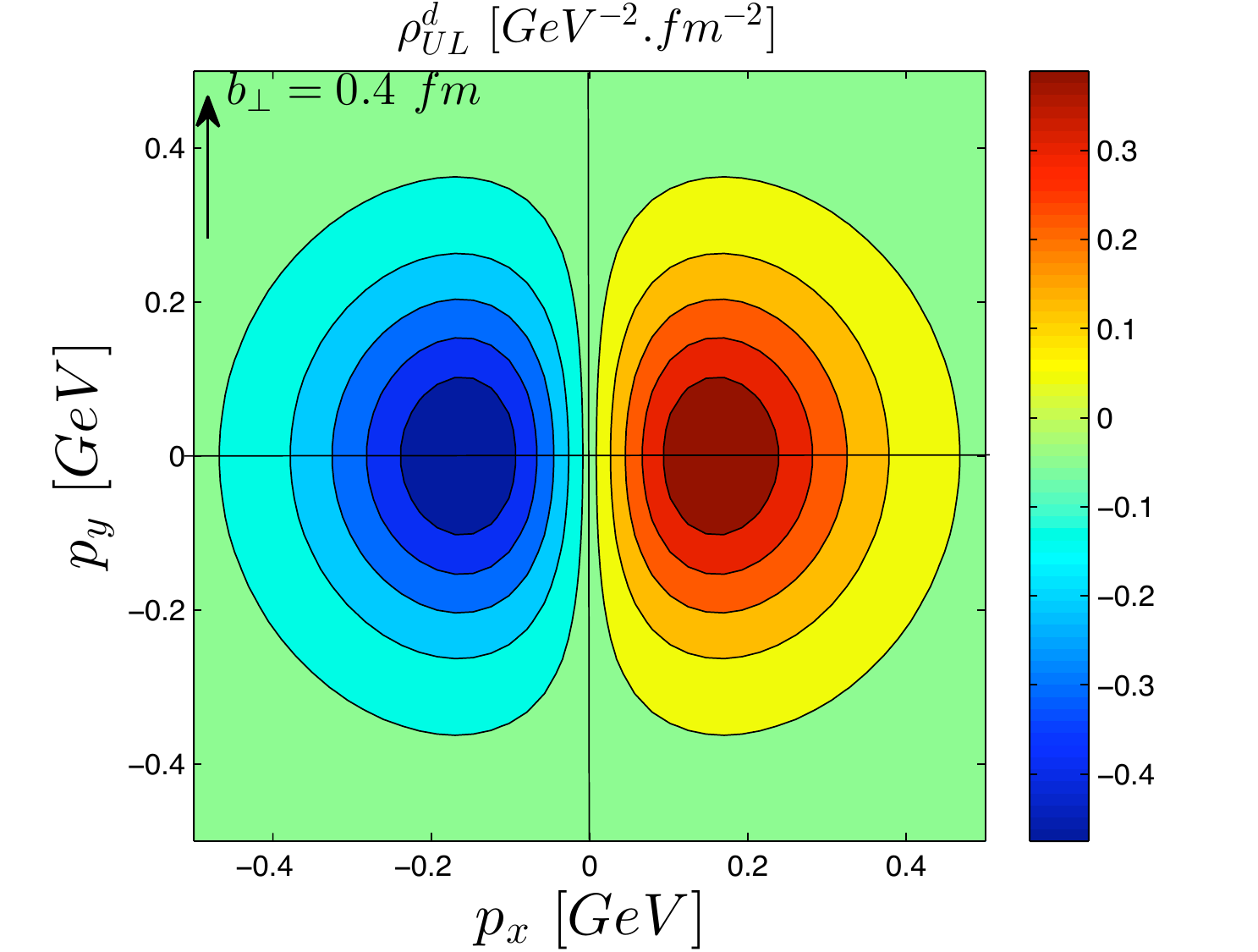}
\end{minipage}
\begin{minipage}[c]{0.98\textwidth}
\small{(c)}\includegraphics[width=7cm,clip]{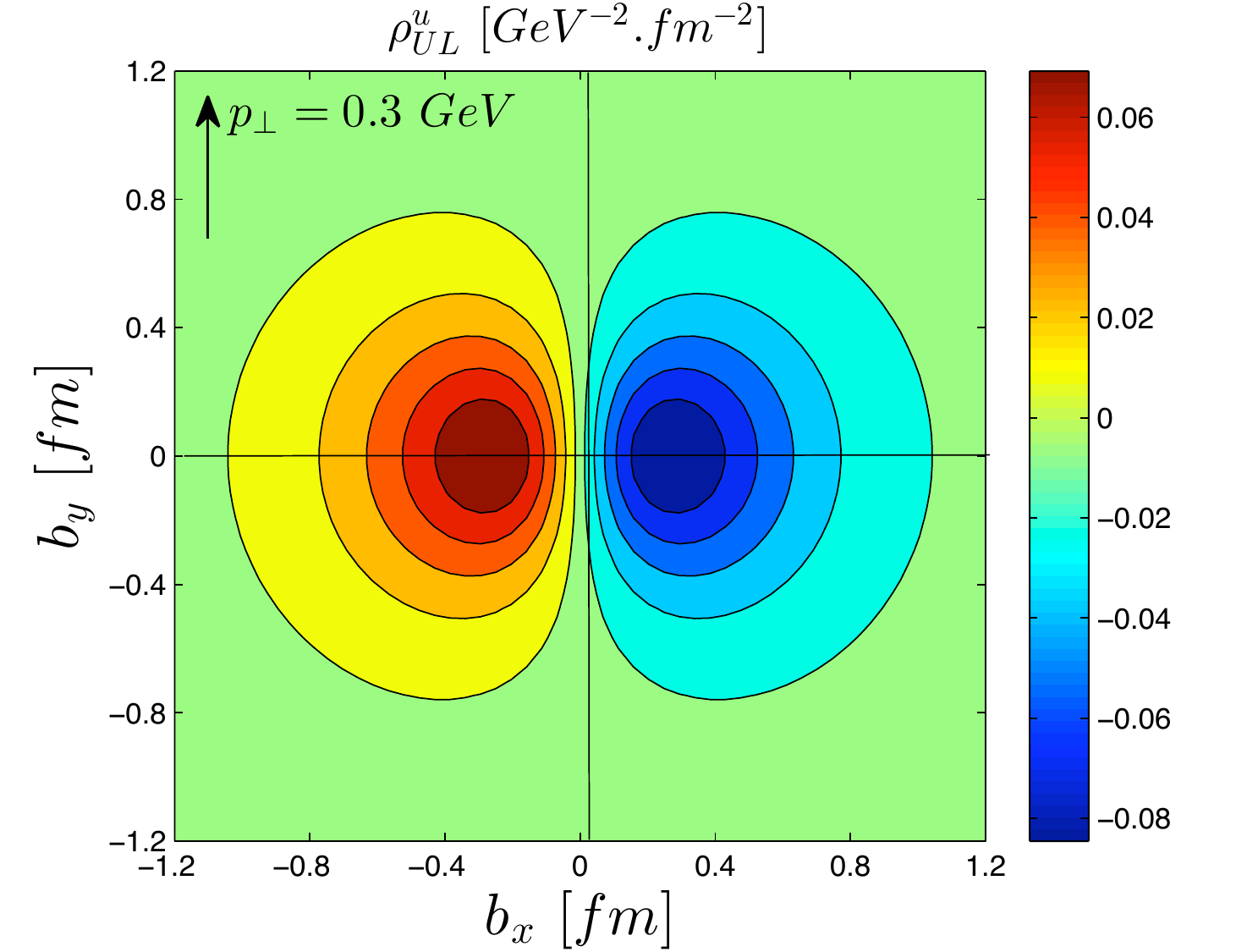}
\hspace{0.1cm}%
\small{(d)}\includegraphics[width=7cm,clip]{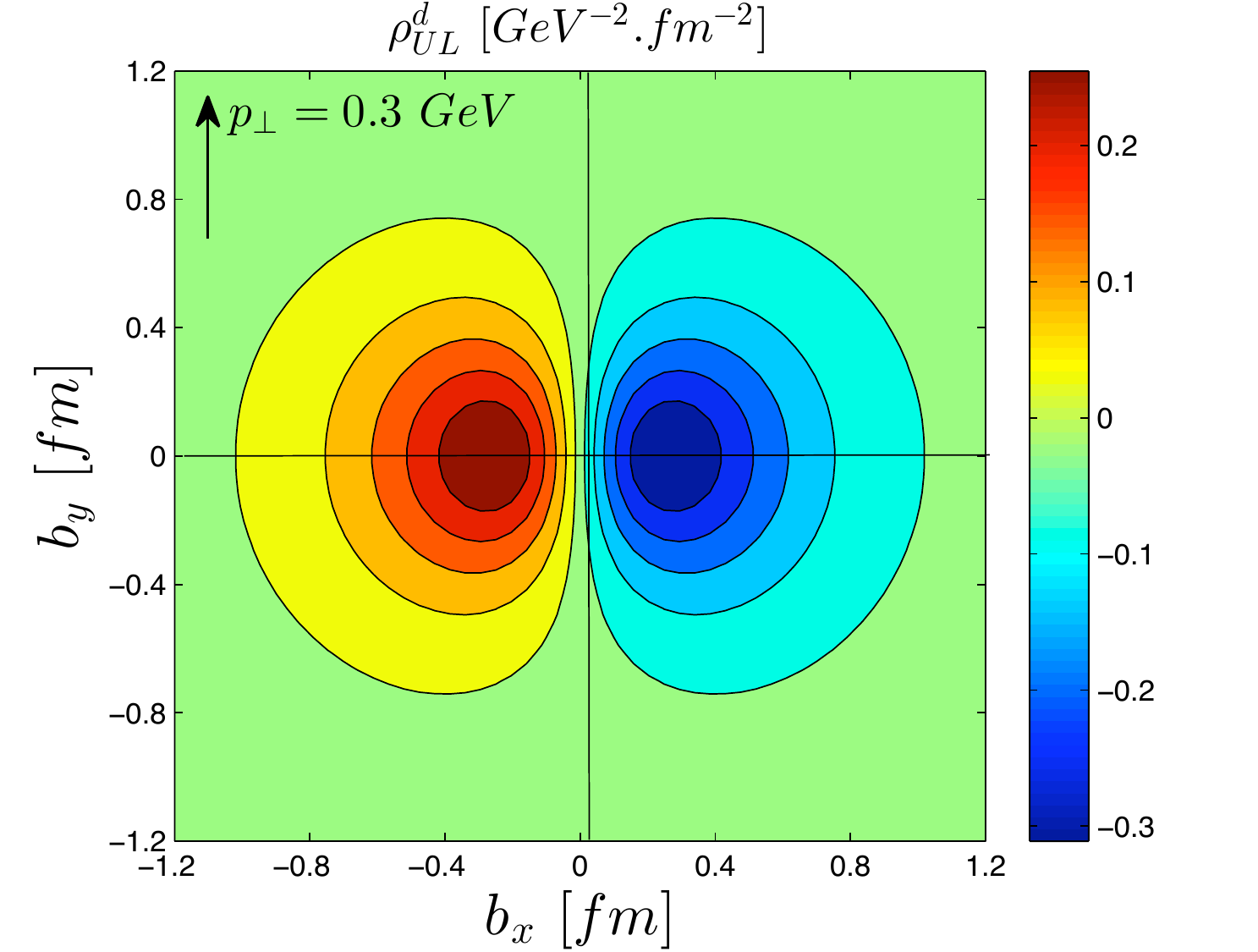}
\end{minipage}
\caption{\label{plot_rhoUL}The dipolar  behavior of $\rho^q_{UL}$ in the transverse momentum plane(a,b) with $\bfb=0.4\hat{y}~ fm$ and in the transverse impact parameter plane(c,d) with $\bfp=0.3\hat{y}~ GeV$ for $u$ quarks(left column) and $d$ quarks(right column).}
\end{figure*}

The Wigner distributions $\rho^q_{UL}(\bfb,\bfp)$, in the transverse momentum plane, are shown in Fig.\ref{plot_rhoUL}(a) and (b) for $u$ and $d$ quarks respectively. The fixed transverse impact parameter $\bfb$ is along $\hat{y}$ with $b_y=0.4~ fm$. The Fig.\ref{plot_rhoUL}(c) and (d) represent the distribution $\rho^q_{UL}(\bfb,\bfp)$ in transverse impact parameter plane, for $u$ and $d$ quark for $\bfp=p_y\hat{y}=0.3 ~GeV$. We observe a dipolar distributions having same polarity for $u$ and $d$ quarks. 
\begin{figure*}[htbp]
\begin{minipage}[c]{0.98\textwidth}
\small{(a)}\includegraphics[width=7cm,clip]{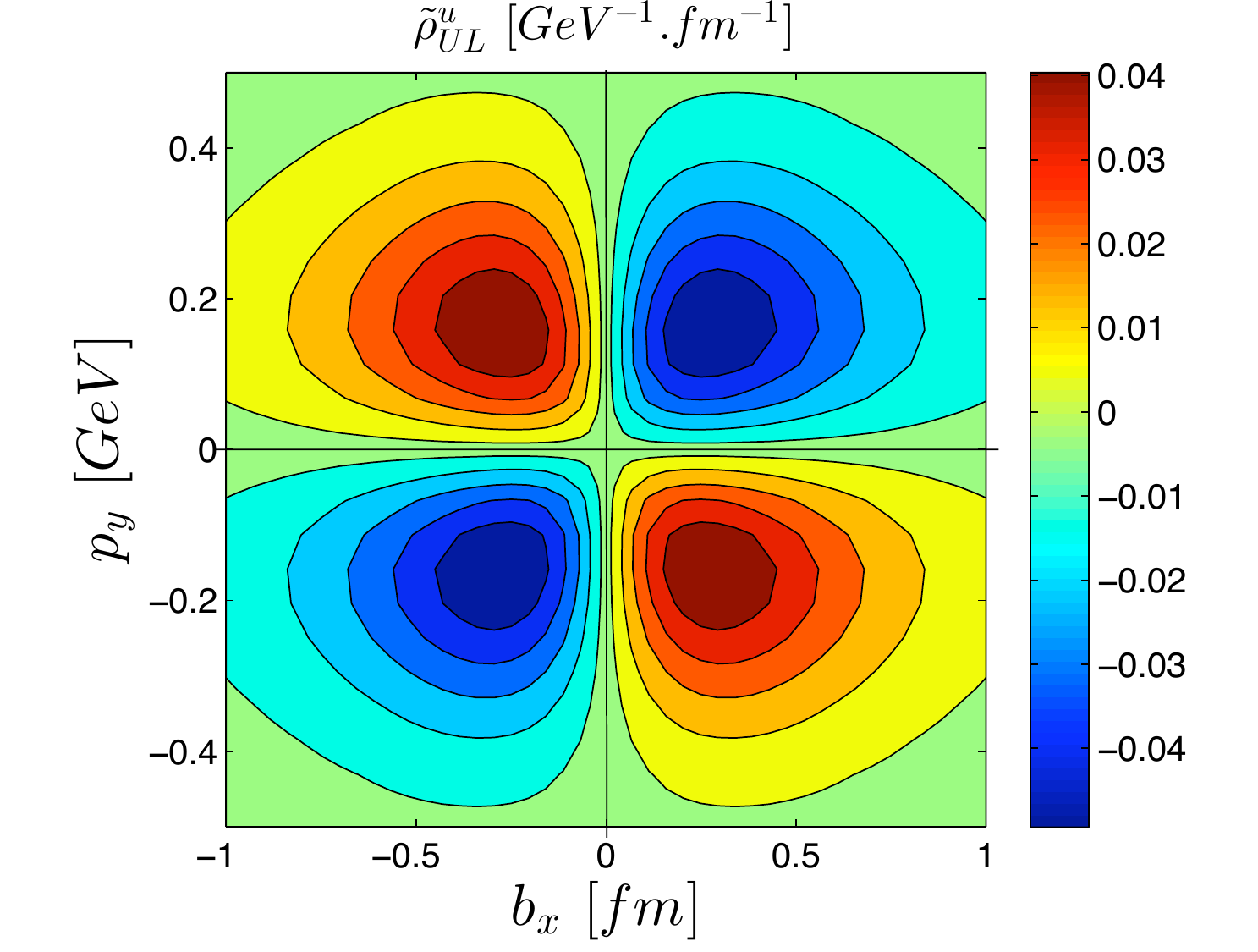}
\hspace{0.1cm}%
\small{(b)}\includegraphics[width=7cm,clip]{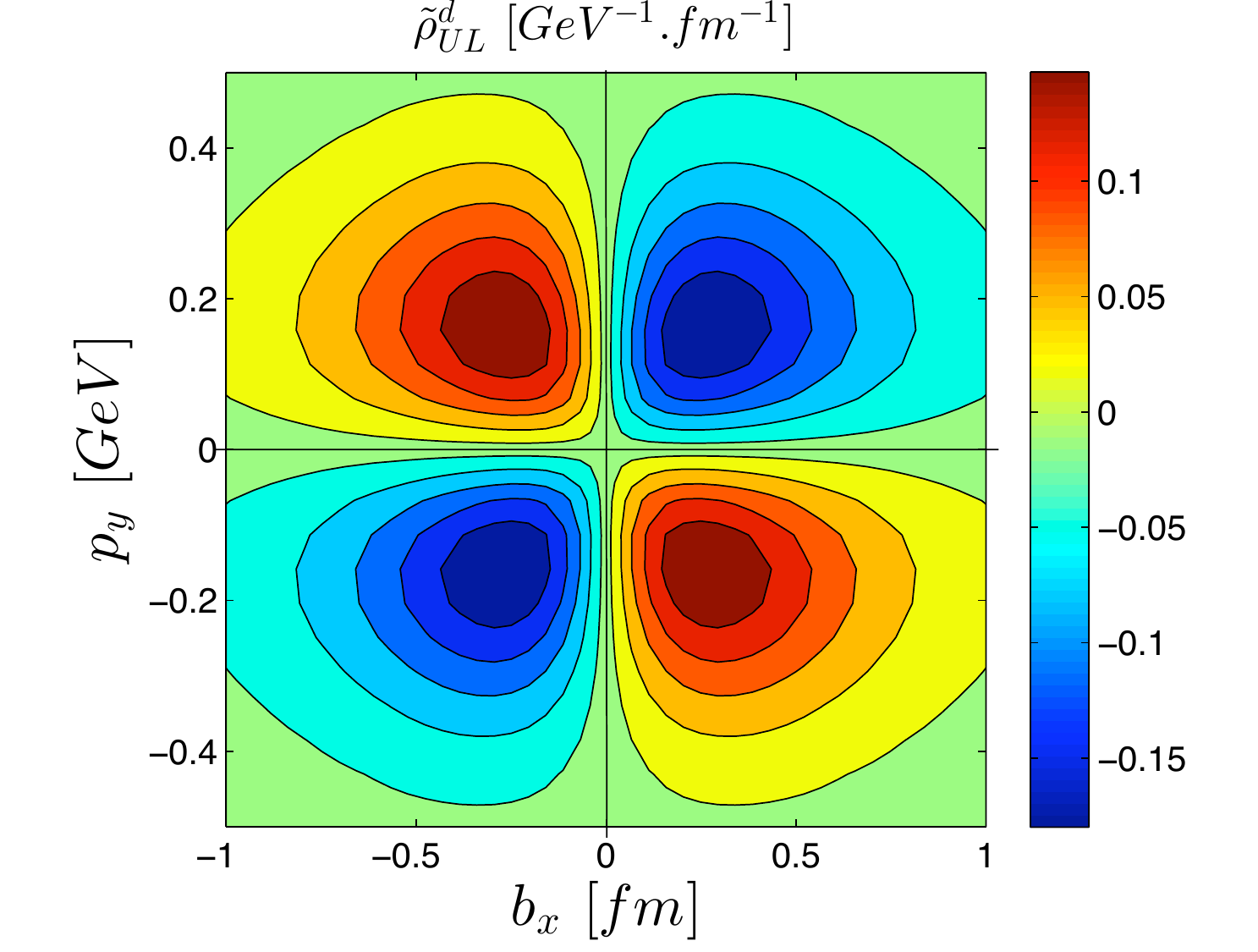}
\end{minipage}
\caption{\label{plot_rhoUL_bxpy}  $\tilde{\rho}^q_{UL}(b_x, p_y)$ in mixed transverse plane corresponding to $u$ quark(a) and $d$ quark(b).}
\end{figure*}
$\tilde{\rho}^q_{UL}(b_x,p_y)$ in the transverse mixed plane are shown in Fig.\ref{plot_rhoUL_bxpy}. We find a quadrupole distribution for $u$ and $d$ quarks. Using Eq.(\ref{G11}) in Eq.(\ref{Cqz}) we calculate the $C^q_z$, the correlation between quark spin and quark OAM. The values are: $C^u_z=-0.0348$ for $u$ quark and $C^d_z=-0.1201$ for $d$ quarks. Therefore in this model, the quark OAM tends to be anti-aligned($C^u_z<0,C^d_z<0 $) to quark spin for both $u$ and $d$ quarks.

%

\subsection{Longitudinally Polarized Proton}
\begin{figure*}[htbp]
\begin{minipage}[c]{0.98\textwidth}
\small{(a)}\includegraphics[width=7cm,clip]{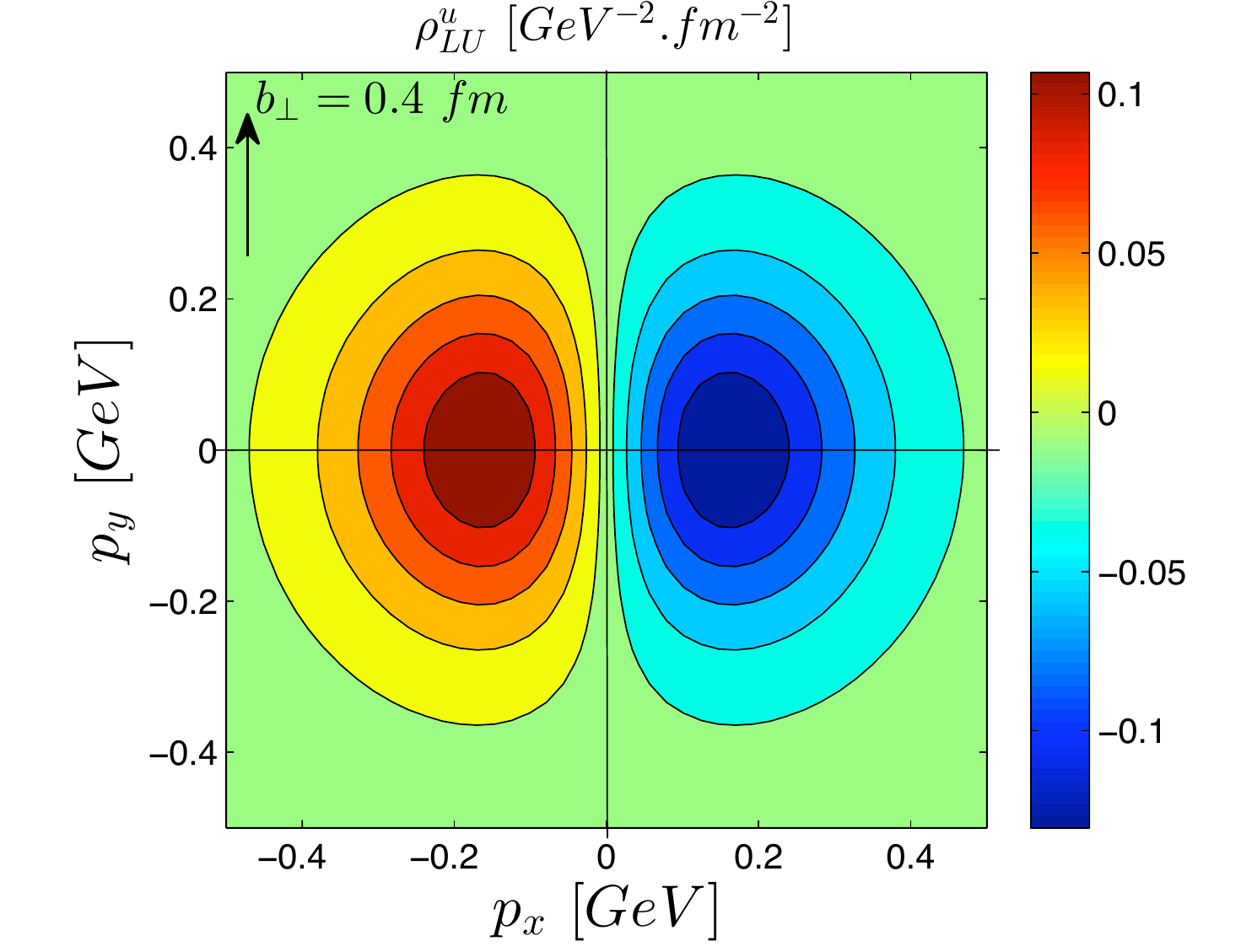}
\hspace{0.1cm}%
\small{(b)}\includegraphics[width=7cm,clip]{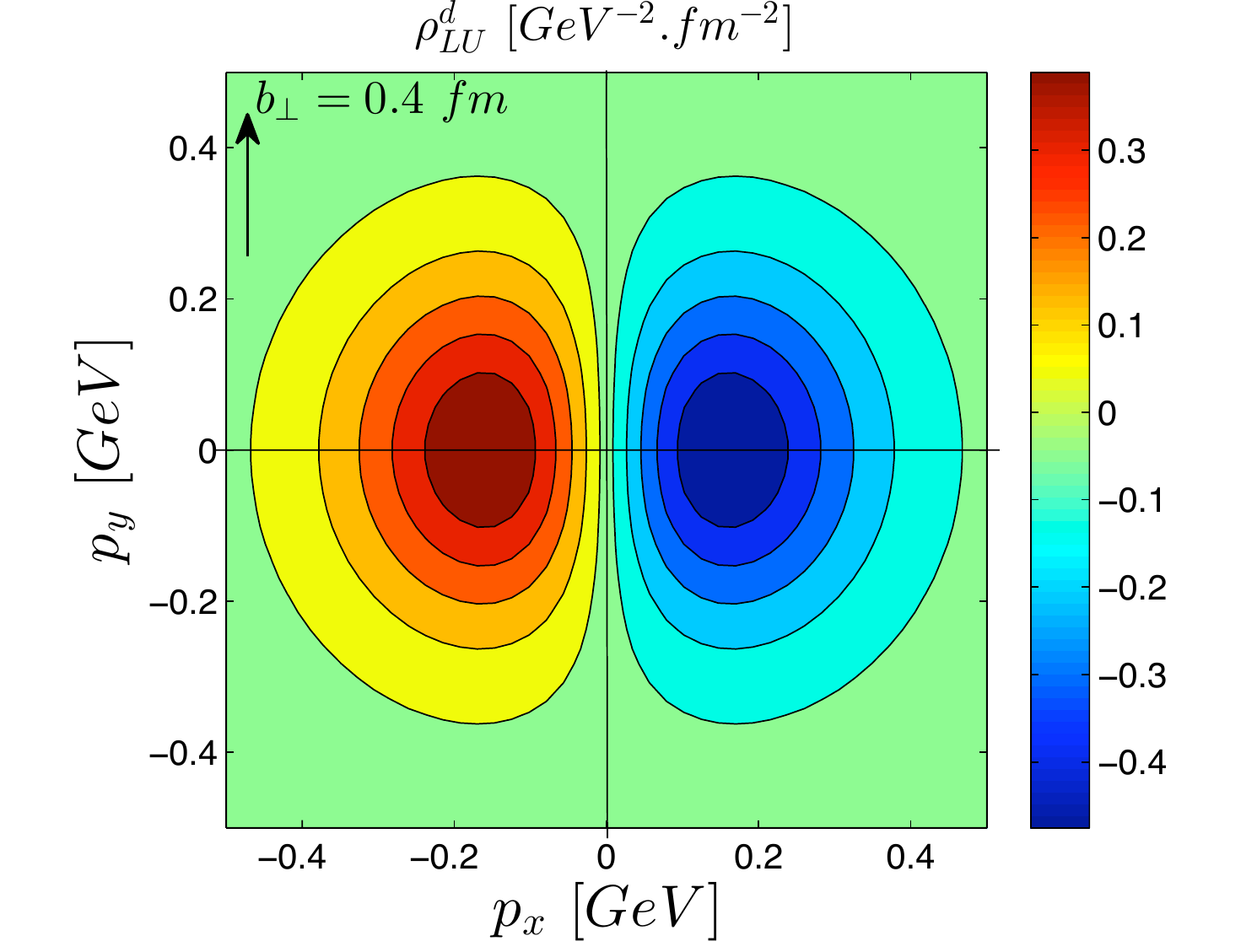}
\end{minipage}
\begin{minipage}[c]{0.98\textwidth}
\small{(c)}\includegraphics[width=7cm,clip]{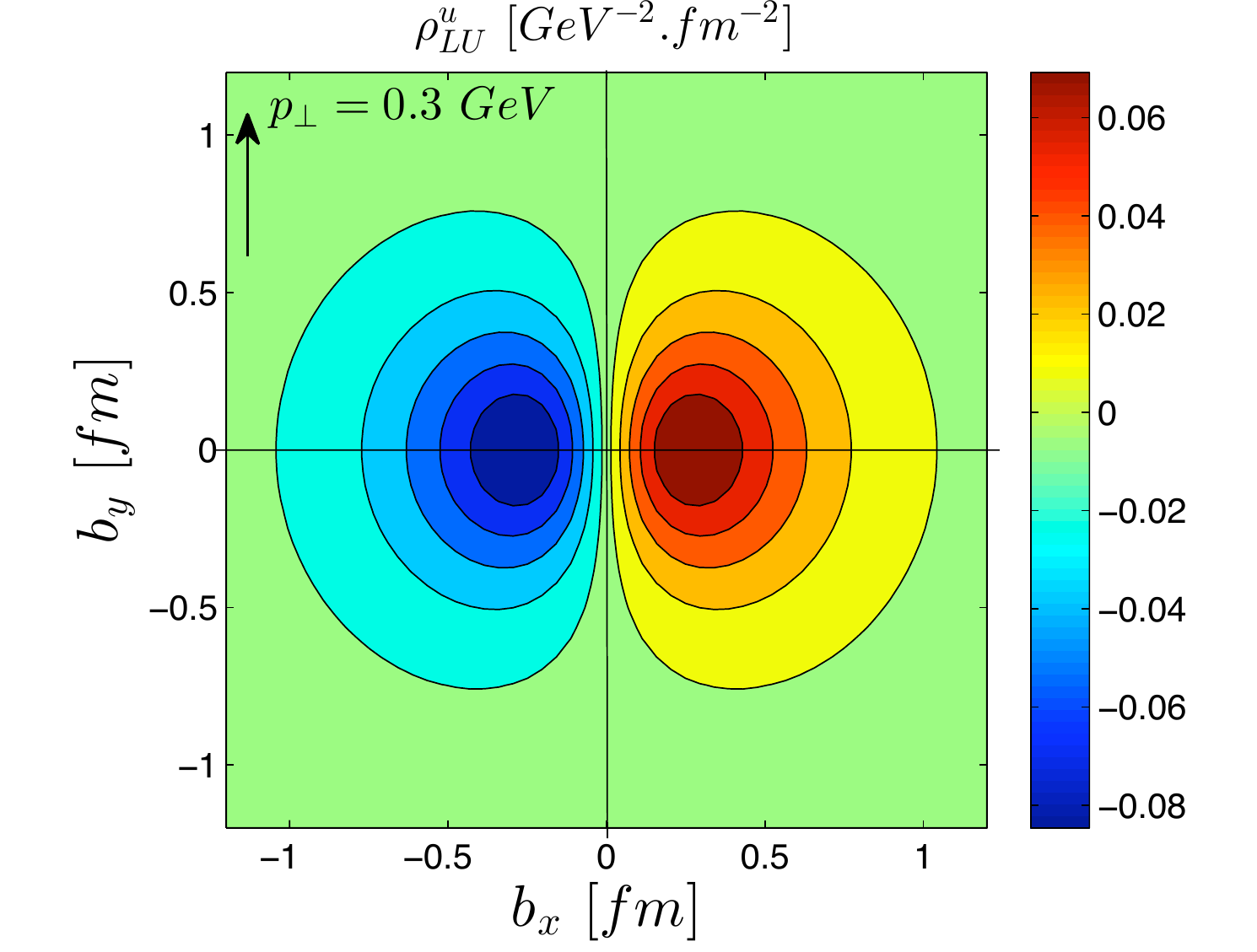}
\hspace{0.1cm}%
\small{(d)}\includegraphics[width=7cm,clip]{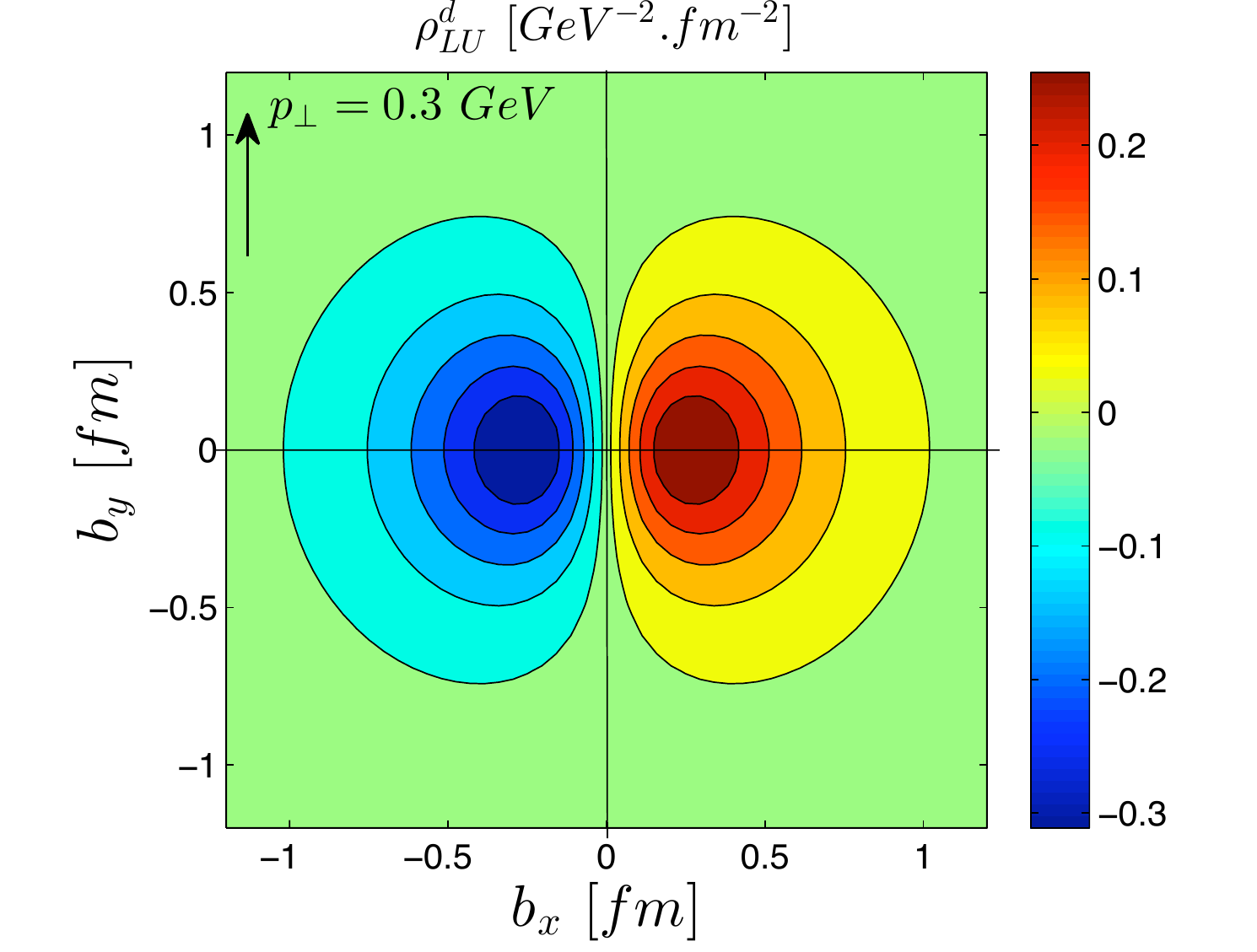}
\end{minipage}
\caption{\label{plot_rhoLU}The dipolar  behavior of $\rho^q_{LU}$ in the transverse momentum plane(a,b) with $\bfb=0.4\hat{y}~ fm$ and in the transverse impact parameter plane(c,d) with $\bfp=0.3\hat{y}~ GeV$ for $u$ quarks(left column) and $d$ quarks(right column).}
\end{figure*}

The Wigner distributions $\rho^q_{LU}(\bfb,\bfp)$ are shown in Fig.\ref{plot_rhoLU} for $u$ and $d$ quarks.  Fig.\ref{plot_rhoLU}(a) and (b) show the variation of  $\rho^q_{LU}(\bfb,\bfp)$ in transverse momentum plane for $u$ and $d$ quarks respectively with $\bfb$ is along $\hat{y}$ and $b_y=0.4~fm$. The variation of $\rho^q_{LU}(\bfb,\bfp)$ in transverse impact parameter plane is shown in Fig.\ref{plot_rhoLU}(c) and (d) with fixed $\bfp$ along $\hat{y}$, $p_y=0.3~ GeV$. We find dipolar distributions for $u$ and $d$ quarks. The polarity of the dipolar distribution $\rho^q_{LU}$ is opposite to the polarity of $\rho^q_{UL}$ . The maximum value of $\rho^q_{LU}(\bfb,\bfp)$ for $u$ quark is less than that for $d$ quarks in both the planes.
\begin{figure*}[htbp]
\begin{minipage}[c]{0.98\textwidth}
\small{(a)}\includegraphics[width=7cm,clip]{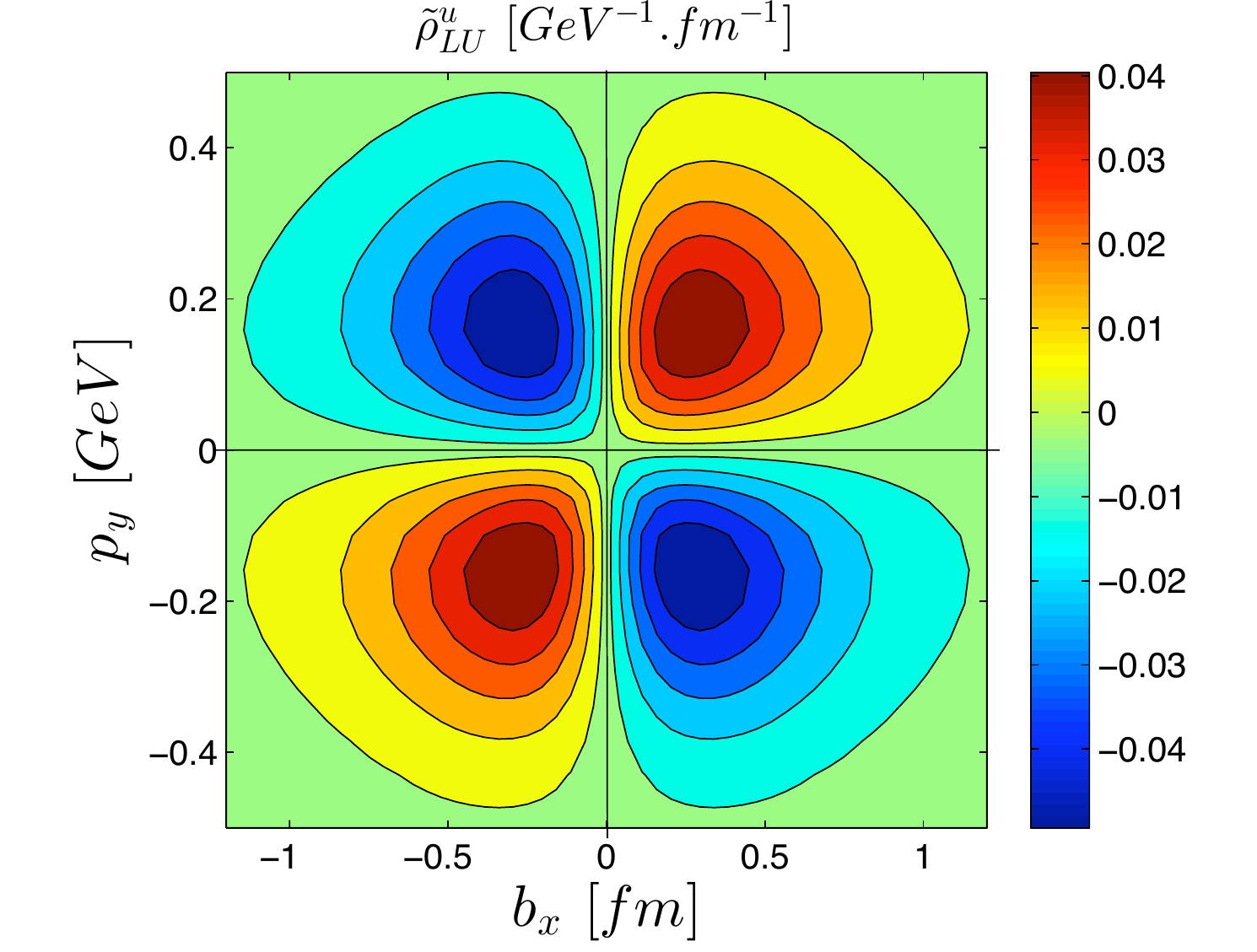}
\hspace{0.1cm}%
\small{(b)}\includegraphics[width=7cm,clip]{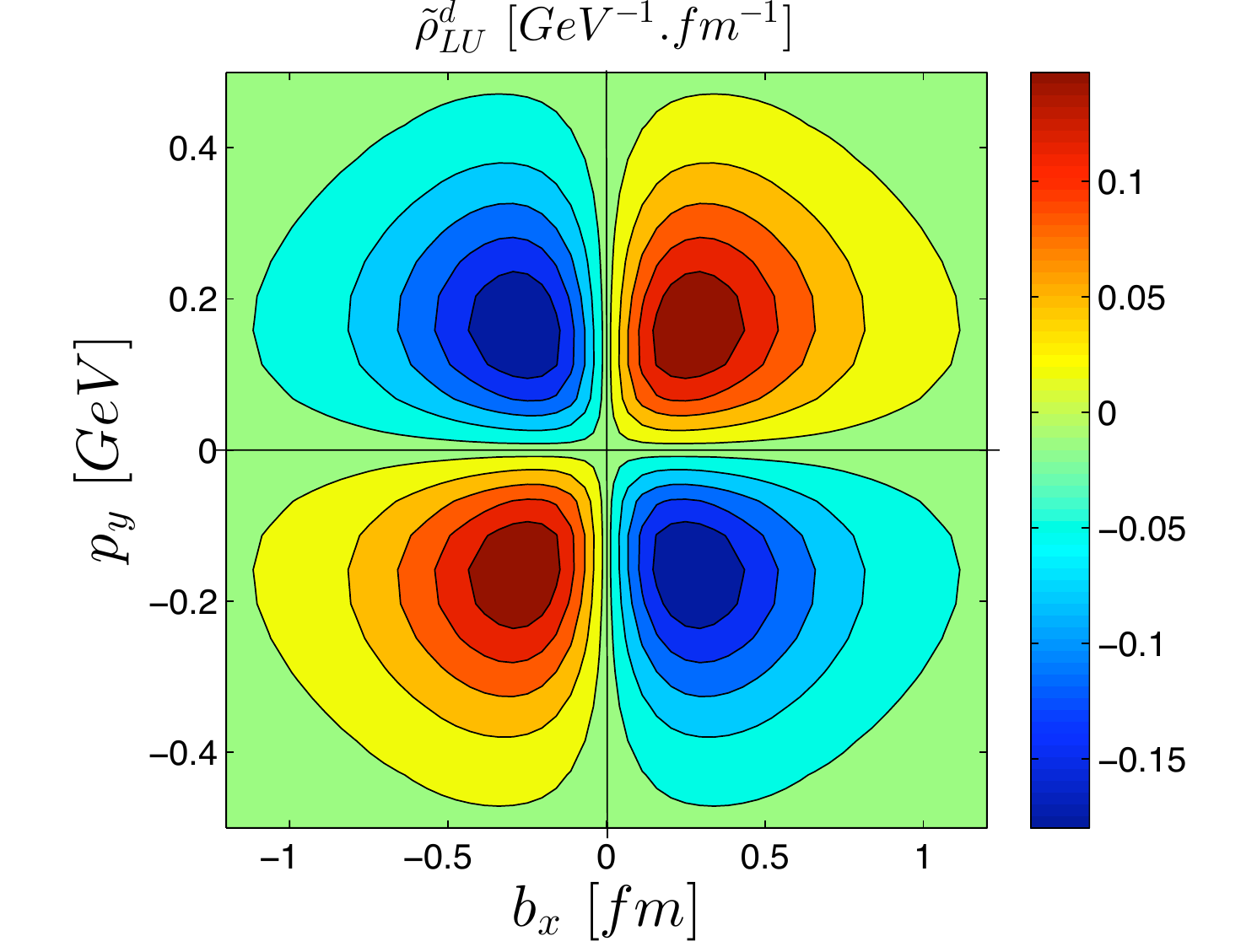}
\end{minipage}
\caption{\label{plot_rhoLU_bxpy} The $\tilde{\rho}^q_{LU}(b_x,p_y)$ in mixed transverse plane corresponding to $u$ quark(a) and $d$ quark(b).}
\end{figure*}
  Fig.\ref{plot_rhoLU_bxpy}(a) and (b) represent the distribution $\tilde{\rho}^q_{LU}(b_x,p_y)$ in the mixed transverse plane for $u$ and $d$ quarks respectively. We observe quadrupole distributions for both $u$ and $d$ quarks. 
The quadrupole structures in $\rho^q_{LU}(\bfb,\bfp)$ and $\rho^q_{UL}(\bfb,\bfp)$ are found due to the presence of the derivative terms in Eq.(\ref{rhoLU_F14}) and Eq.(\ref{rhoUL_G11}).
\begin{table}[ht]
\centering 
\begin{tabular}{ p{2.4cm}  p{2.0cm}  p{1.7cm}  p{1.7cm}}
     & $q$-OAM &  $u$ & $d$ \\ \hline
     $~~Canonical$ & $\ell_z^q$ & 0.0348 & 0.1201 \\
     $~~Kinetic$ & $~L_z^q$ & -0.3812 & -0.4258 \\
\end{tabular} 
\caption{In the light-front AdS/QCD scalar diquark  model, the values of canonical OAM $\ell^q_z$ and the kinetic OAM $L^q_z$ for $u$ and $d$ quark.}
\label{tab_OAM} 
\end{table}

From Eqs.(\ref{OAM_ell_res}) and (\ref{OAM_L_res}), we calculate the canonical OAM and kinetic OAM of quarks in this model. The values of quark OAM are given in Table.\ref{tab_OAM}.   Note that in quark-diquark model, the total proton OAM is given by the sum of quark and diquark angular momenta, so unlike the quark models $u$ and $d$ quark contributions do not add upto the total proton OAM and hence  the sum of kinetic OAM of $u$ and $d$ in Table.\ref{tab_OAM} is  not the same as total canonical OAM of the $u$ and $d$.
The correlation between the canonical OAM of quark and proton spin can be understood from the sign of the $\ell^q_z$. 
In our model calculation, the positive values of $\ell^q_z$ for both $u$ and $d$  imply that the proton spin tends to be aligned to quark OAM for both $u$ and $d$ quarks.
The spin contribution of the quark to the proton spin is given by\cite{pasquini11}
\be s^q=\frac{1}{2}\Delta q &=&\frac{1}{2}\int dx \tilde{H}^q(x,0,0)\nonumber\\
&=&\frac{1}{2}\int dx~d^2p_\perp G_{1,4}^q(x,0,\bfp^2,0,0)
\ee
where $\Delta q$ is the axial charge.  In our model, we get $s^u=0.946$ and $s^d=0.396$. 
 It is well known that the spectator diquark model has its own limitations\cite{Becchetta}. Though the functional behaviors of the GPDs and GTMDs are well reproduced in our model, the axial charges for both $u$ and $d$ quarks are over estimated.  
The model is defined at a very low scale $Q_0^2 \approx 0.09~ GeV^2$. The axial charge is scale dependent and known to be negative at larger scales.  In \cite{Gut}, the authors have extended the result to an arbitrary scale $Q^2$    
and studied the evolution of unpolarized pdfs in this model. Our result
agrees closely with theirs, in spite of the fact that  the fit parameters 
are slightly different. In their model \cite{1411}, the pdfs are slightly smaller in
magnitude. When polarized pdfs or helicity distributions are computed, the $d$ 
quark helicity distribution comes out to be positive, although it is expected
to be negative from the recent fit of the data \cite{0904}. In \cite{nnpdf}, it has been shown
that NNPDF allows for a  positive total ${\Delta d(x)/d(x)}$ (where $\Delta d(x)$ stands for helicity distribution)
for larger values of $x$, this 
is also obtained in some other models, for example in \cite{0705}, the above ratio was
calculated in perturbative QCD taking into account the valence Fock
components with non-vanishing orbital angular momentum and it was found that 
$ \Delta d(x)/d(x)$  is positive as $x \approx 0.75$ and approaches 1 as $x
\rightarrow 1$. Positive values of this ratio was also found in an SU(6)
breaking quark model calculation in \cite{close}. Another way to parametrize the model
would be to fit the data of the helicity distributions with the model
parameters, instead of the form factors and the GPDs.
Since in the scalar diquark model,  $\ell_z^q+s^q+\ell_z^D=1/2$ (as $s^D=0$), the diquark contribution  to the canonical OAM is $\ell_z^D=-0.484$ for $u$ struck-quark and $\ell_z=-0.016$ for $d$ struck-quark.
 The contributions of different partial waves to the quark OAM in LCCQM have been studied in \cite{Lorce12}.
\begin{figure*}[htbp]
\begin{minipage}[c]{0.98\textwidth}
\small{(a)}\includegraphics[width=7cm,clip]{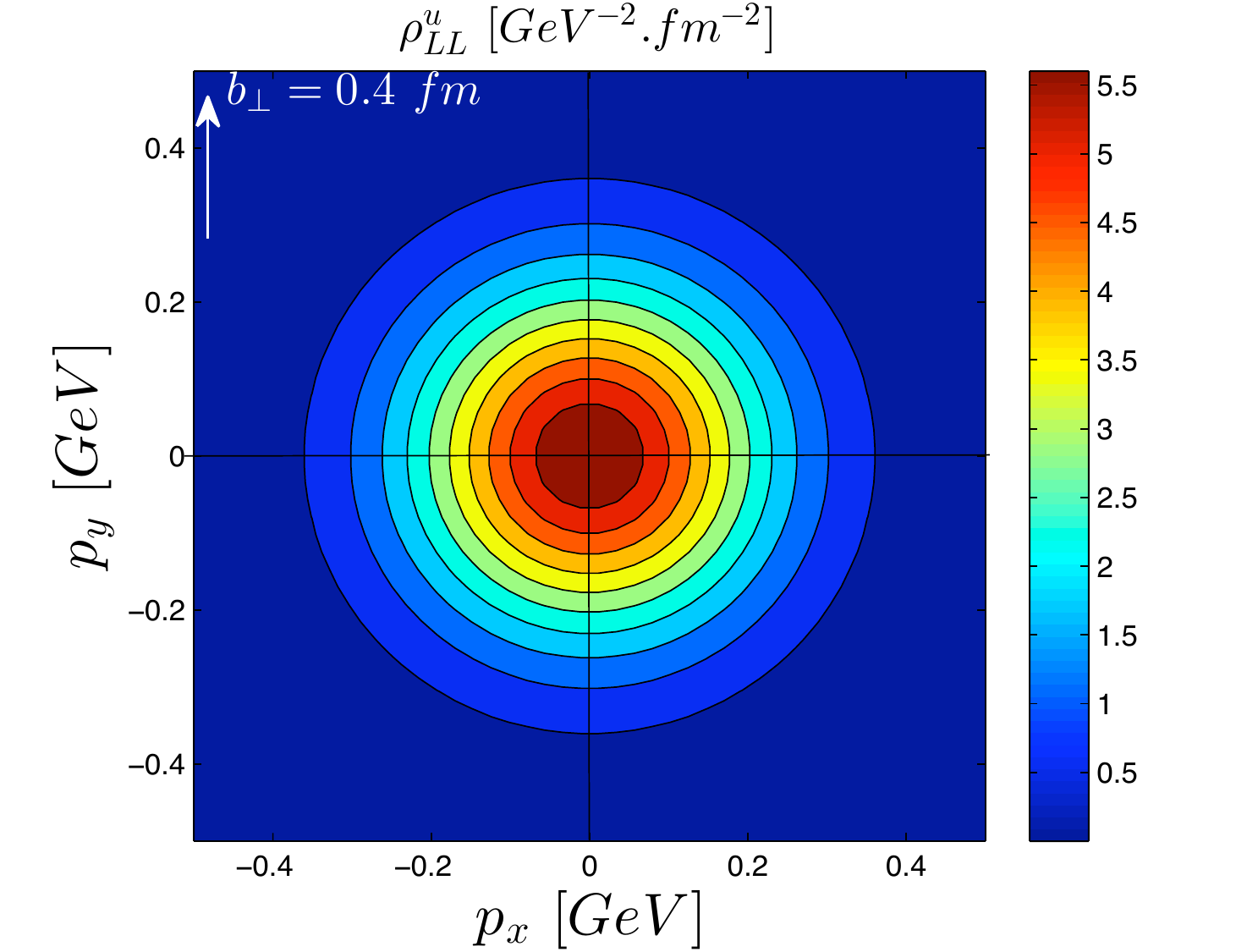}
\hspace{0.1cm}%
\small{(b)}\includegraphics[width=7cm,clip]{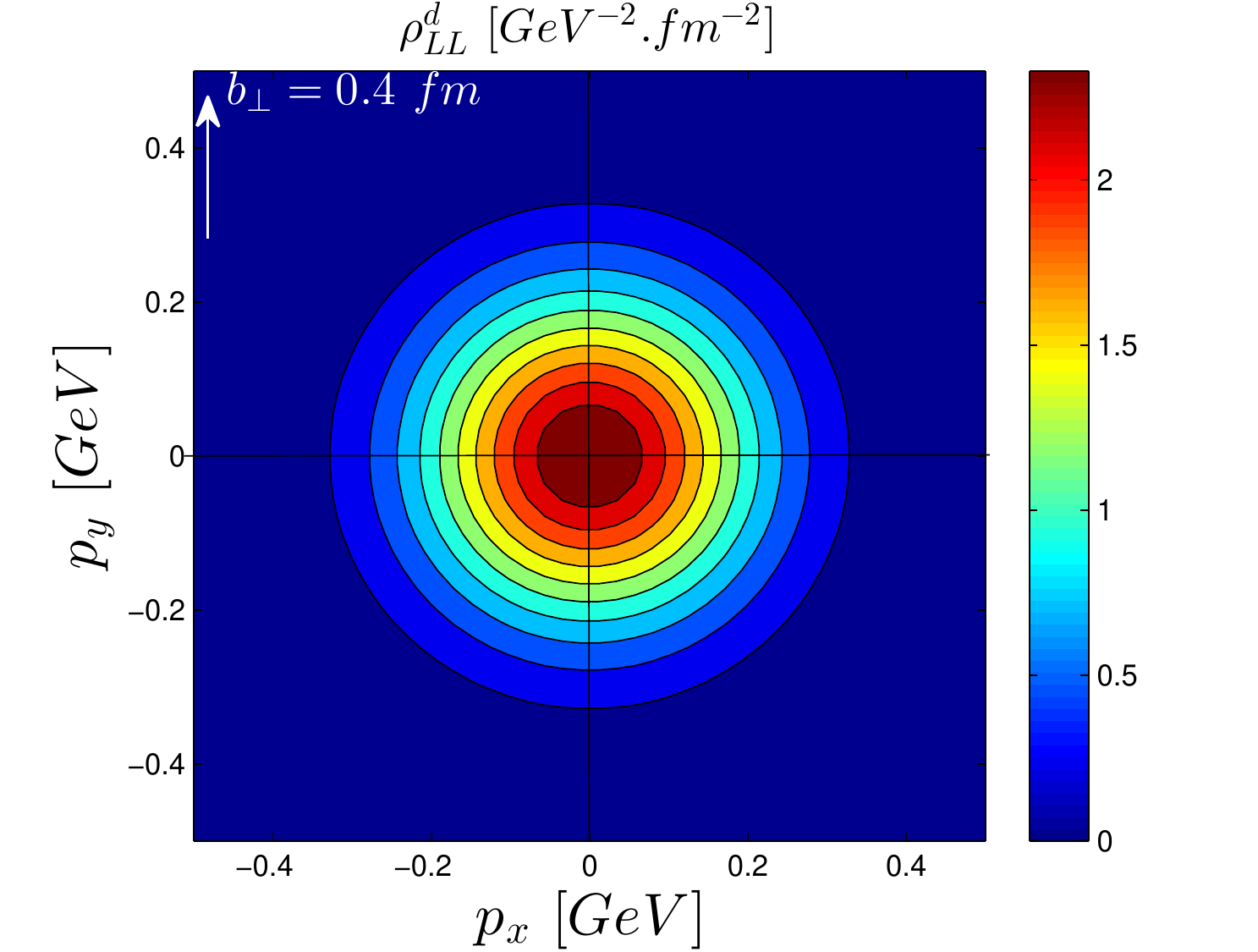}
\end{minipage}
\begin{minipage}[c]{0.98\textwidth}
\small{(c)}\includegraphics[width=7cm,clip]{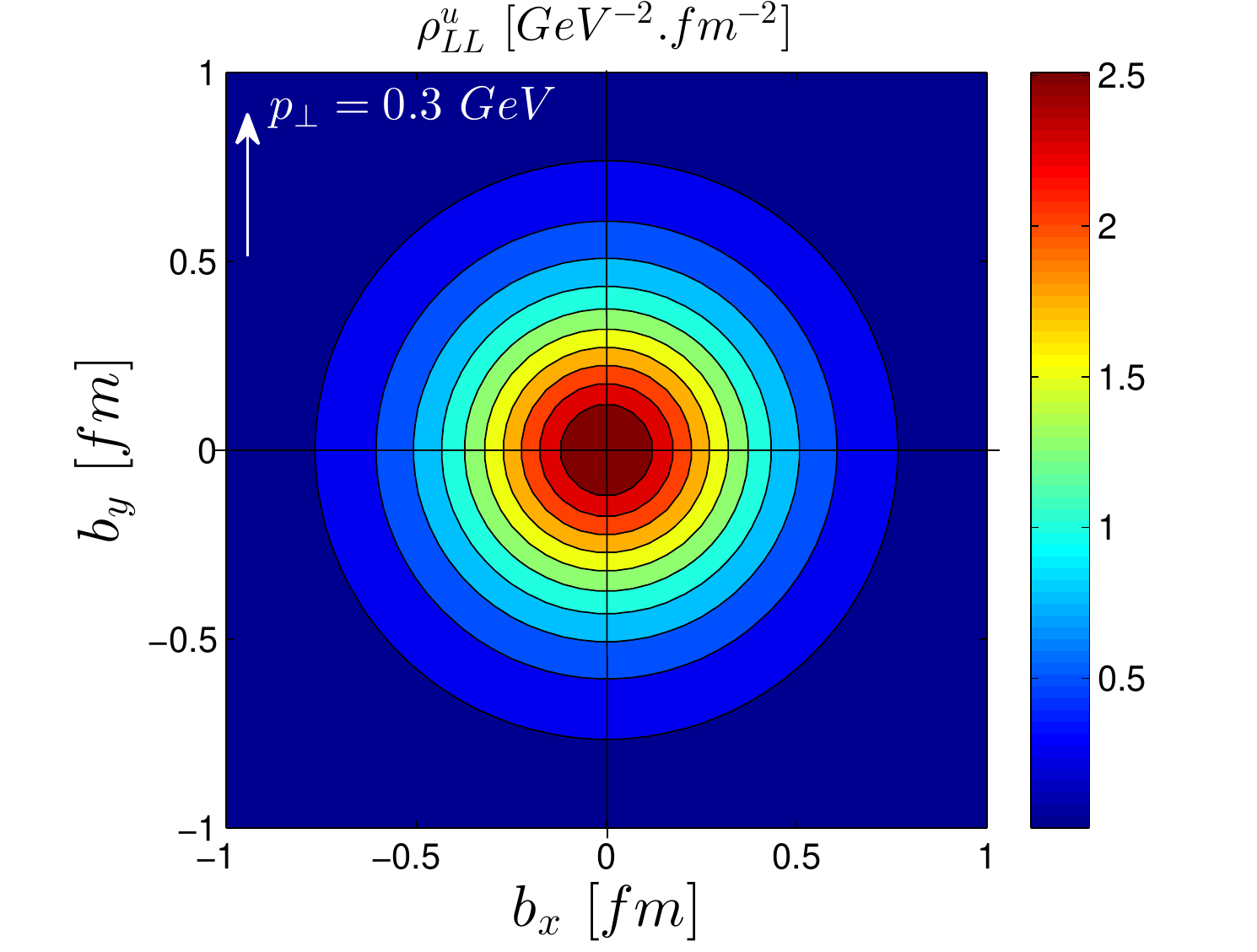}
\hspace{0.1cm}%
\small{(d)}\includegraphics[width=7cm,clip]{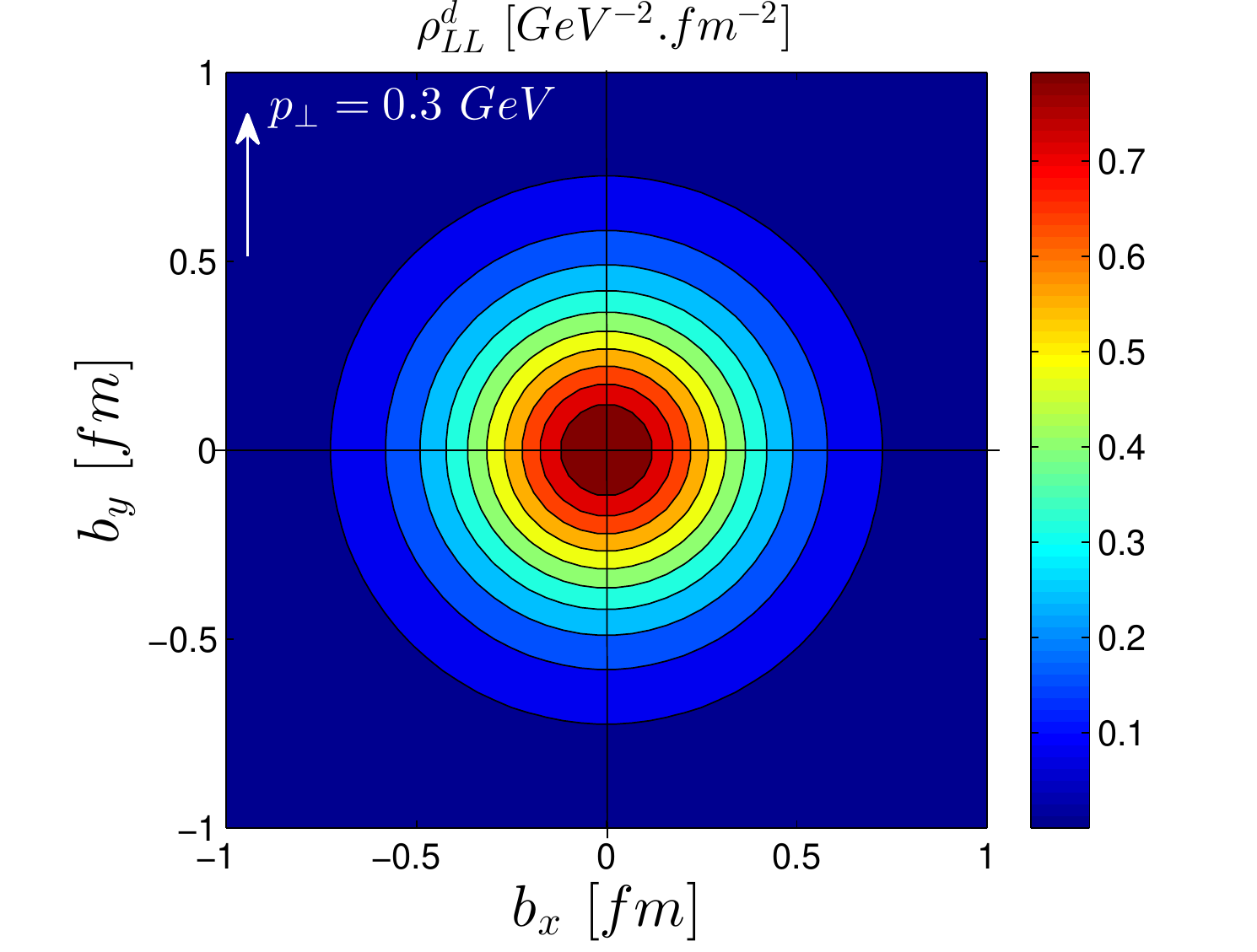}
\end{minipage}
\caption{\label{plot_rhoLL}The $\rho^q_{LL}(\bfb,\bfp)$ in the transverse momentum plane (a,b) with $\bfb=0.4\hat{y}~ fm$ and in the transverse impact parameter plane(c,d) with $\bfp=0.3\hat{y}~ GeV$ for $u$ and $d$ quarks respectively. }
\end{figure*}
\begin{figure*}[htbp]
\begin{minipage}[c]{0.98\textwidth}
\small{(a)}\includegraphics[width=7.cm,clip]{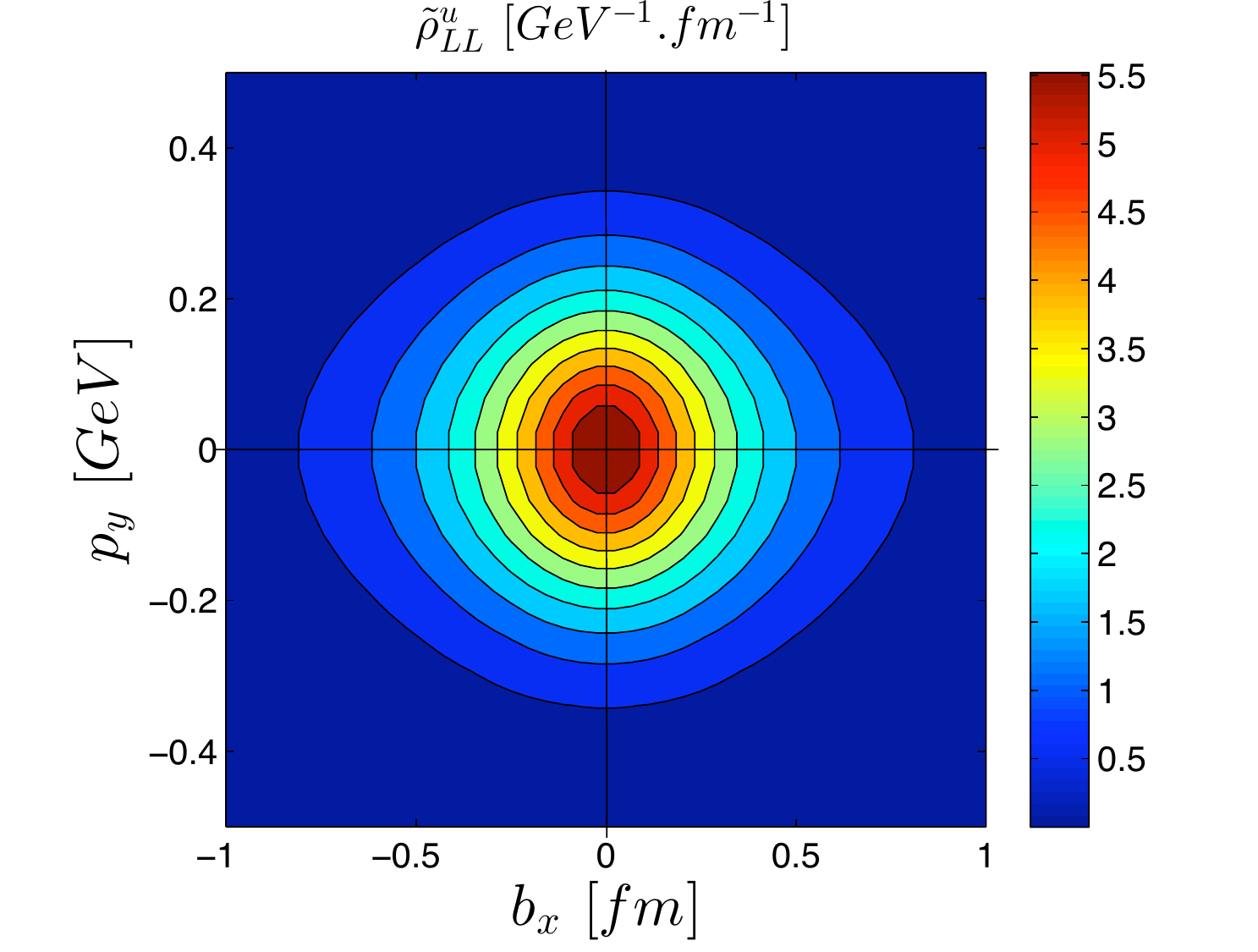}
\hspace{0.1cm}%
\small{(b)}\includegraphics[width=7.cm,clip]{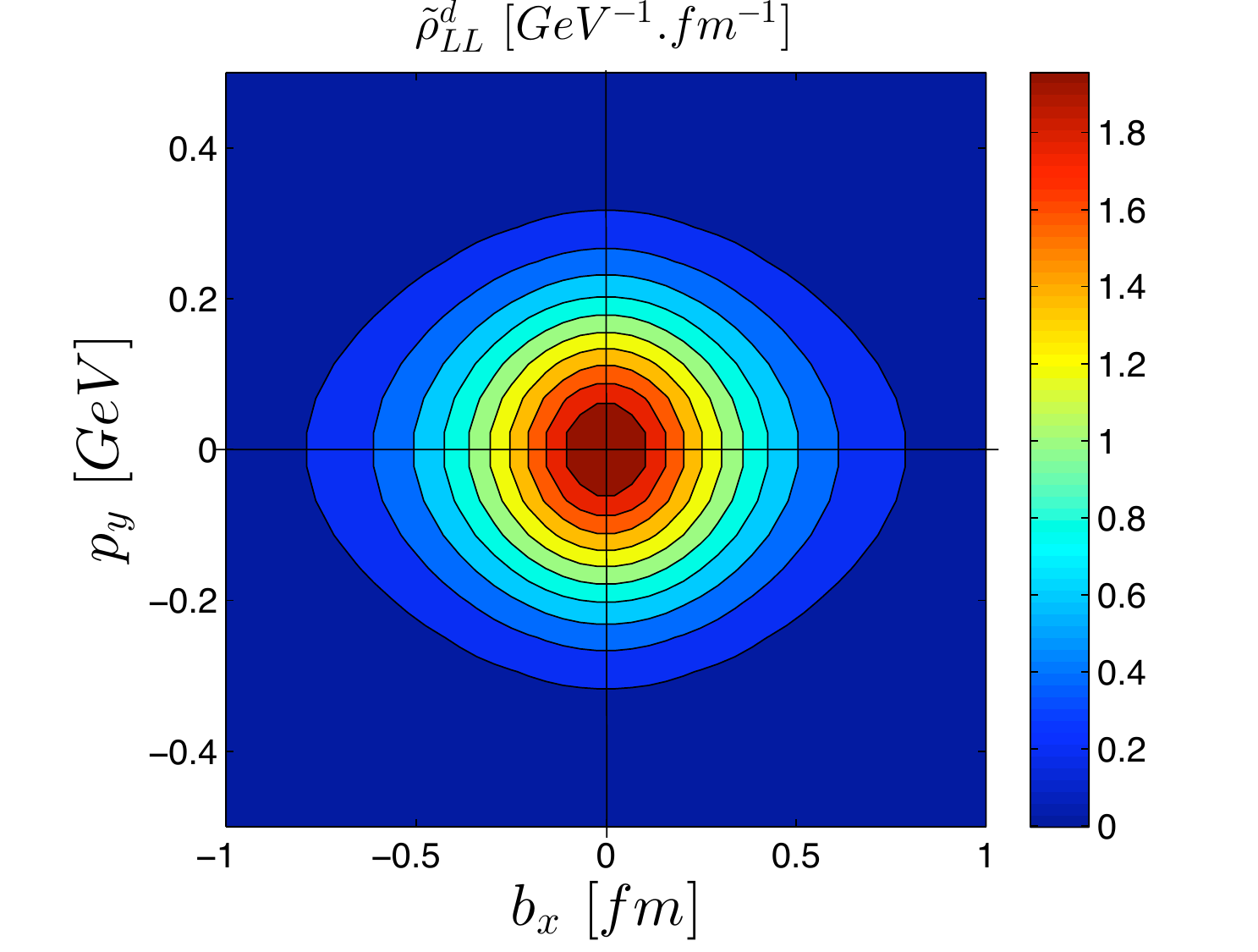}
\end{minipage}
\caption{\label{plot_rhoLL_bxpy} The $\tilde{\rho}^q_{LL}(b_x,p_y)$ in mixed transverse plane for $u$ and $d$ quarks.}
\end{figure*}

The Wigner distributions for longitudinally polarized quark in a longitudinally polarized proton, $\rho^q_{LL}(\bfb,\bfp)$, are shown in Fig.\ref{plot_rhoLL}. The Fig.\ref{plot_rhoLL}(a) and (b) represent  $\rho^q_{LL}(\bfb,\bfp)$ in transverse momentum plane with fixed $\bfb=0.4~fm~\hat{y}$ and  Fig.\ref{plot_rhoLL}(c) and (d) show the plots in the transverse impact parameter plane with $\bfp=0.3~GeV~\hat{y}$ . The distributions are circularly symmetric for $u$ and $d$ quarks in both the planes. The circular symmetry implies that the $\rho_{LL}$ can not contribute to the quark OAM as shown in Eq.(\ref{OAM_rhoLL}).  The picks of the distributions are at the centre (0,0) in both the planes. 
Therefore the quark polarization and the proton polarization tend to be parallel for $u$ and $d$ quarks.  Fig.\ref{plot_rhoLL_bxpy} represents the distribution $\tilde{\rho}^q_{LL}(b_x,p_y)$ in a mixed transverse plane. The distributions are axially symmetric for both $u$ and $d$ quarks.


\begin{figure*}[htbp]
\begin{minipage}[c]{0.98\textwidth}
\small{(a)}\includegraphics[width=6.7cm,clip]{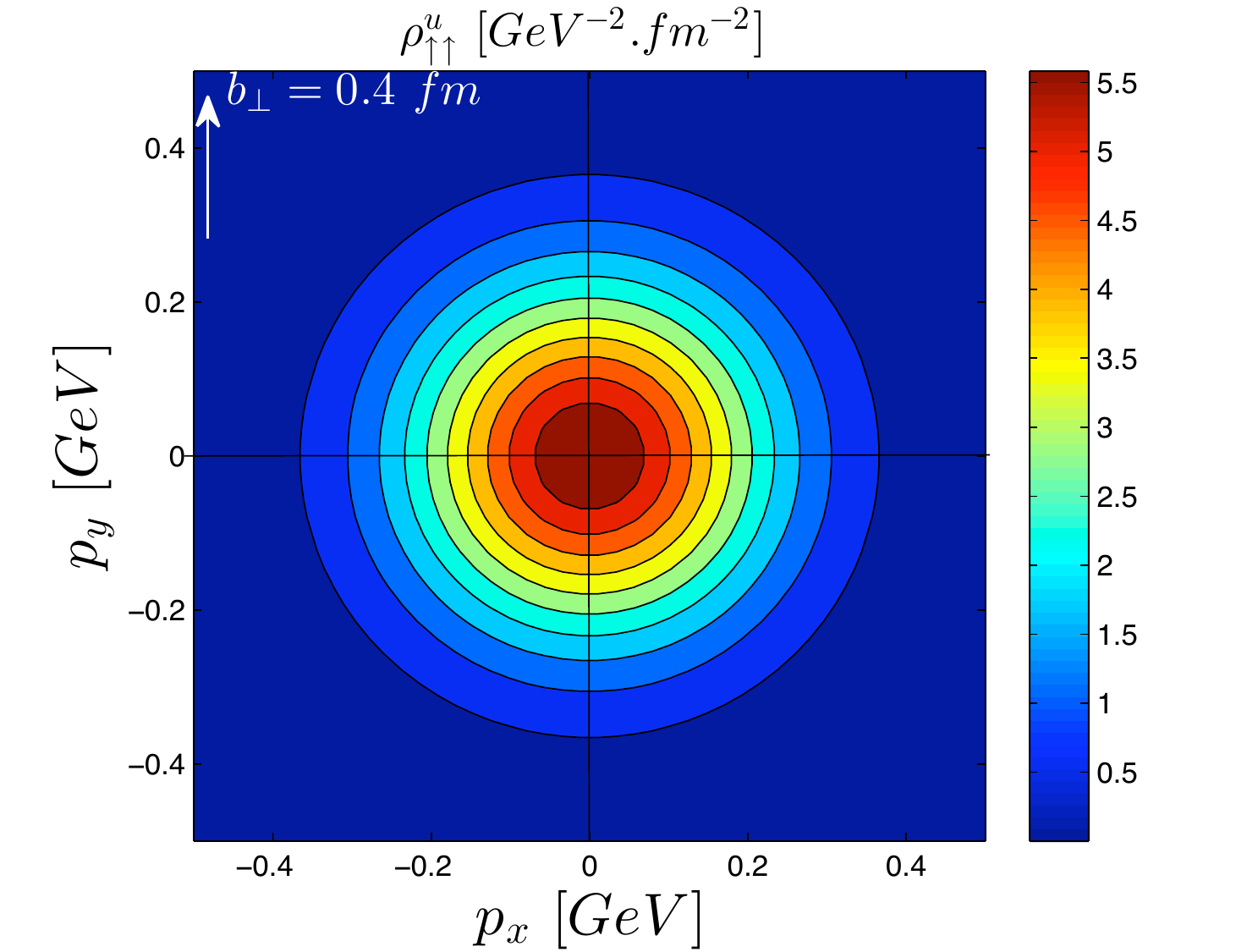}
\small{(b)}\includegraphics[width=6.7cm,clip]{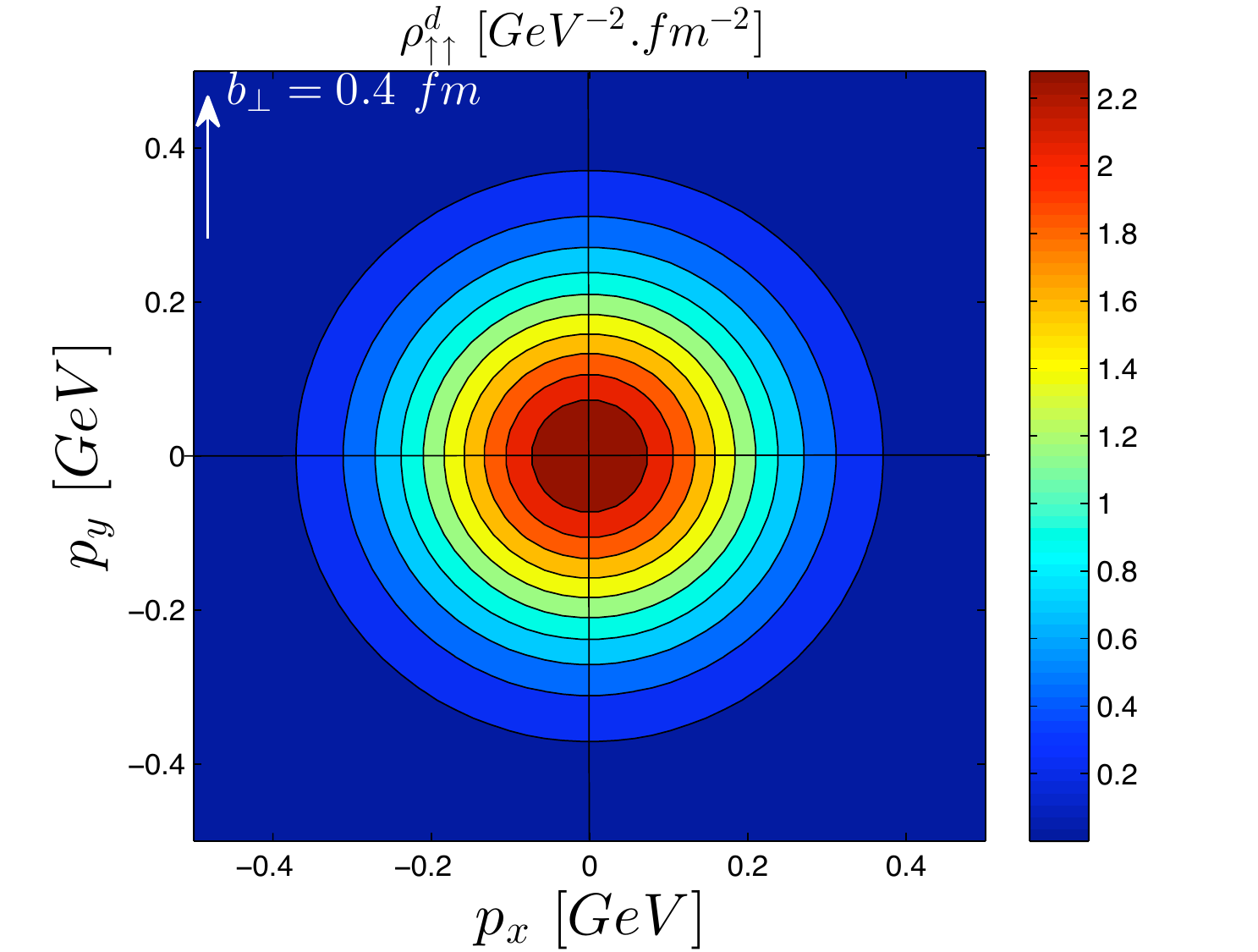}
\end{minipage}
\begin{minipage}[c]{0.98\textwidth}
\small{(c)}\includegraphics[width=6.7cm,clip]{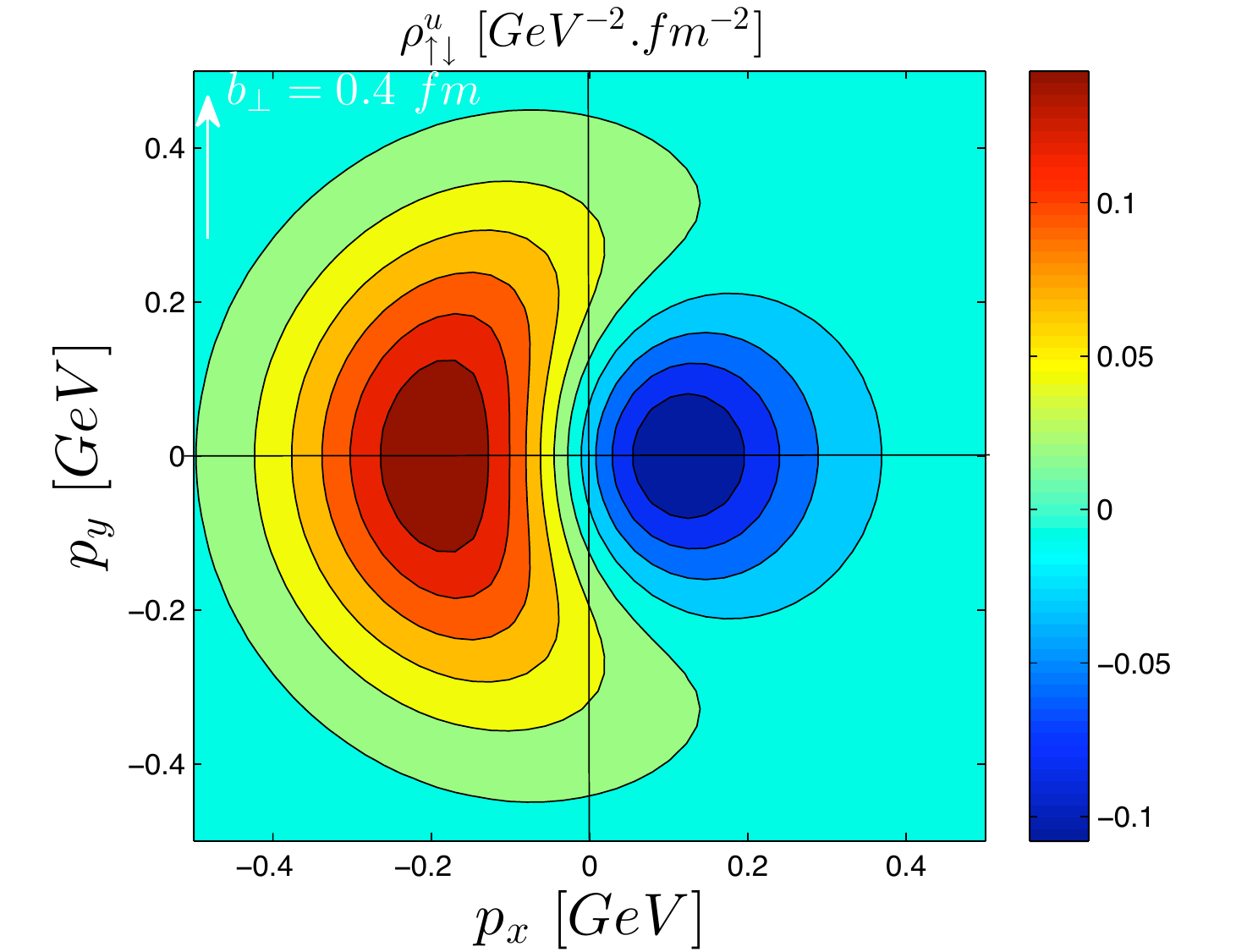}
\small{(d)}\includegraphics[width=6.7cm,clip]{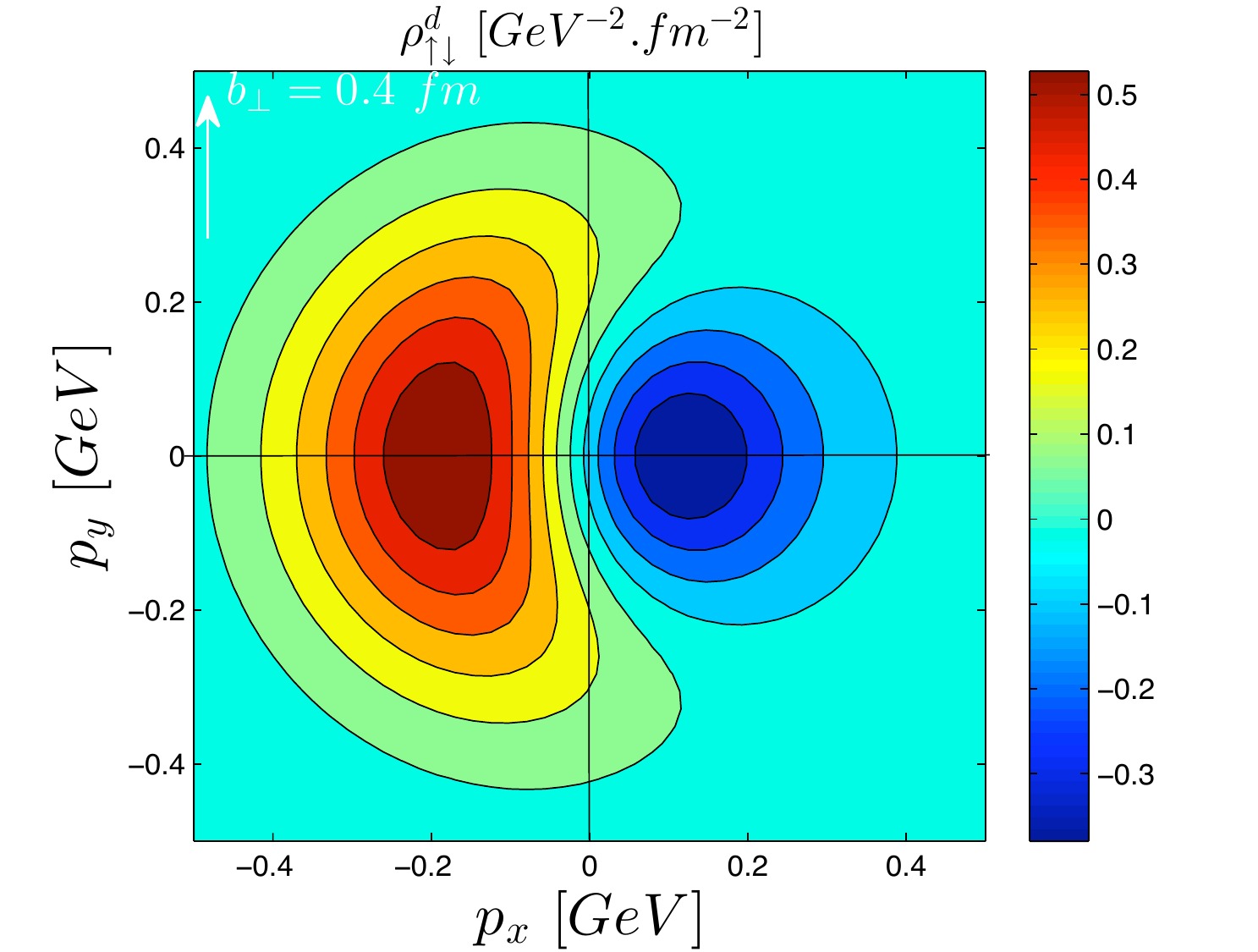} 
\end{minipage}
\begin{minipage}[c]{0.98\textwidth}
\small{(e)}\includegraphics[width=6.7cm,clip]{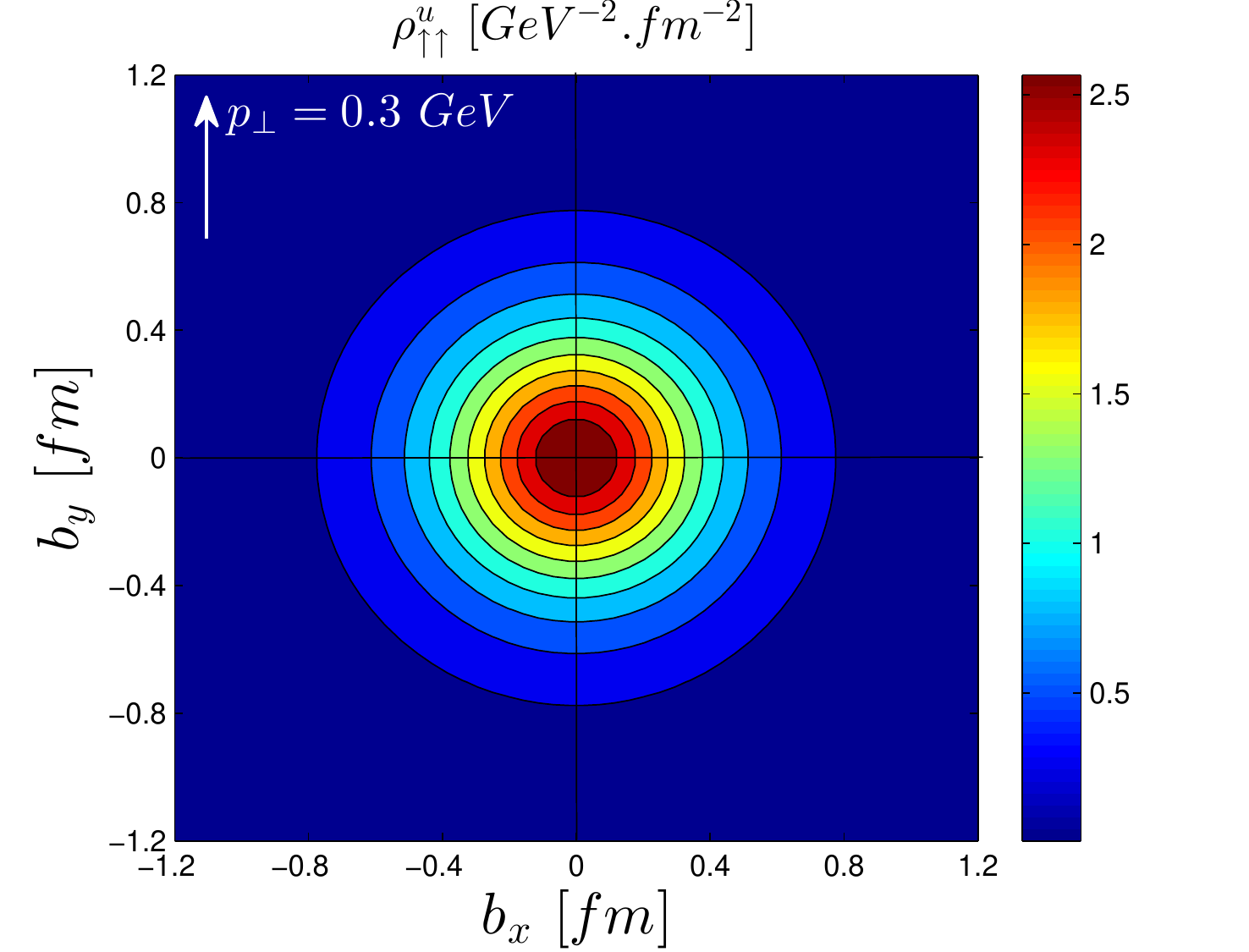}
\small{(f)}\includegraphics[width=6.7cm,clip]{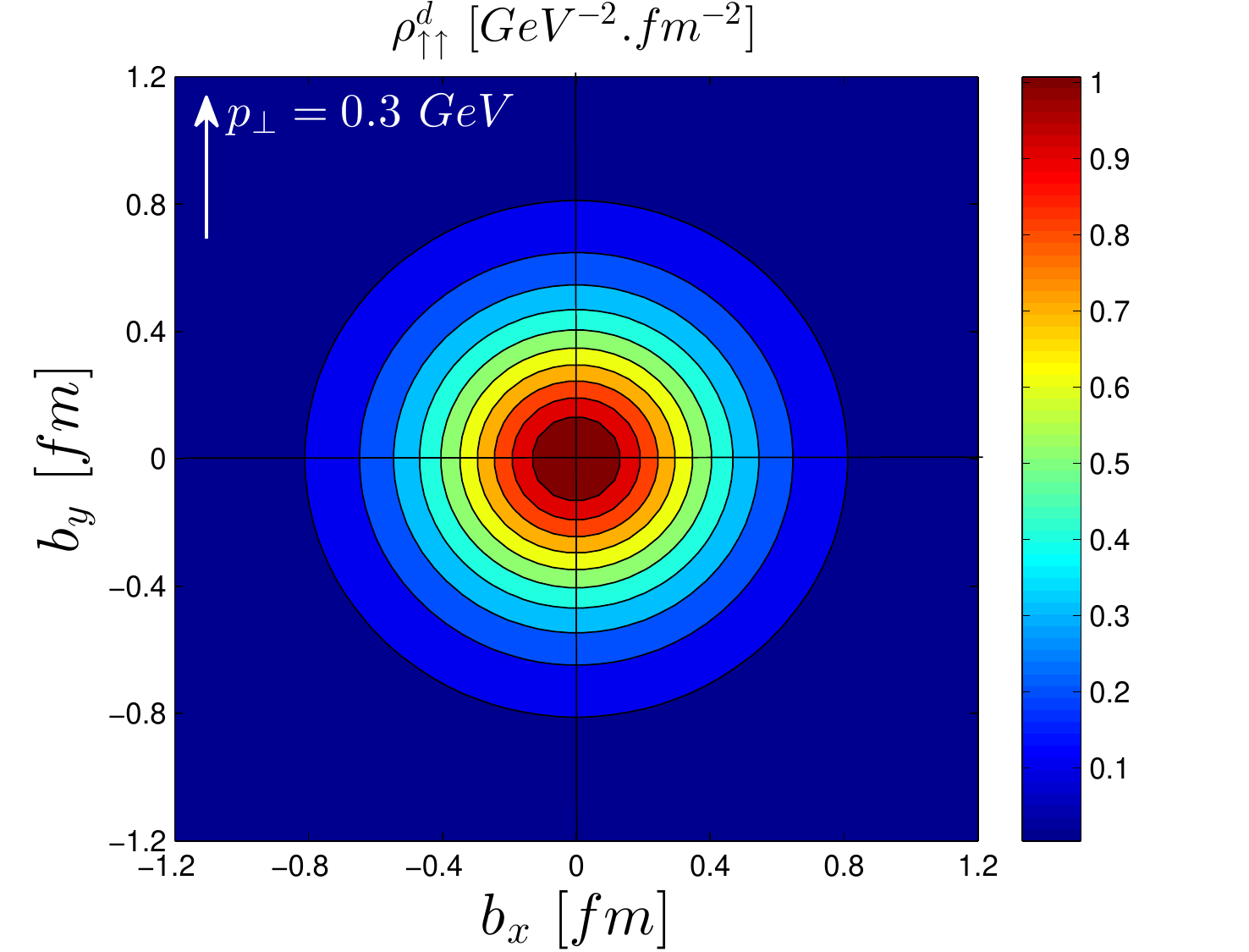}
\end{minipage}
\begin{minipage}[c]{0.98\textwidth}
\small{(g)}\includegraphics[width=6.7cm,clip]{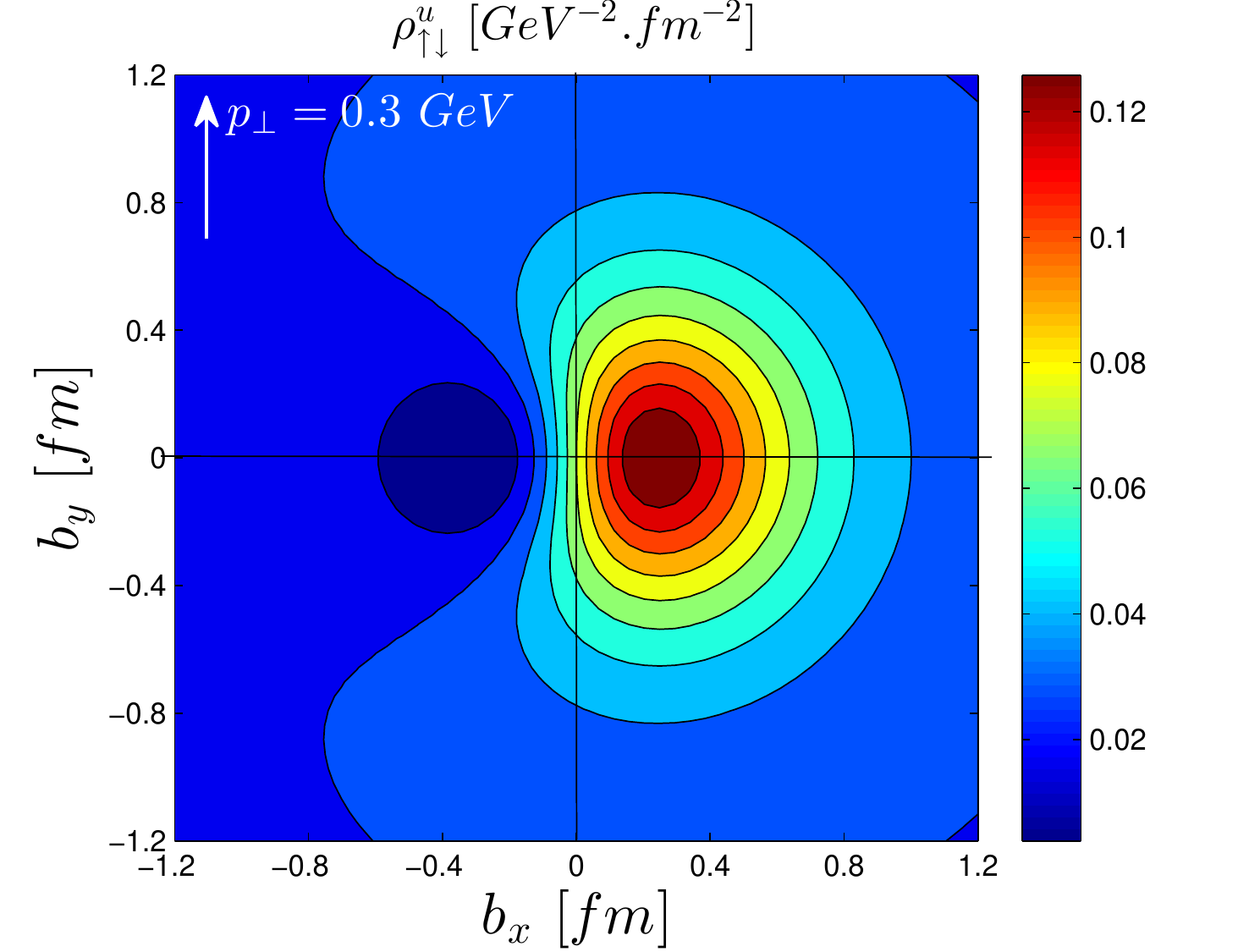}
\small{(h)}\includegraphics[width=6.7cm,clip]{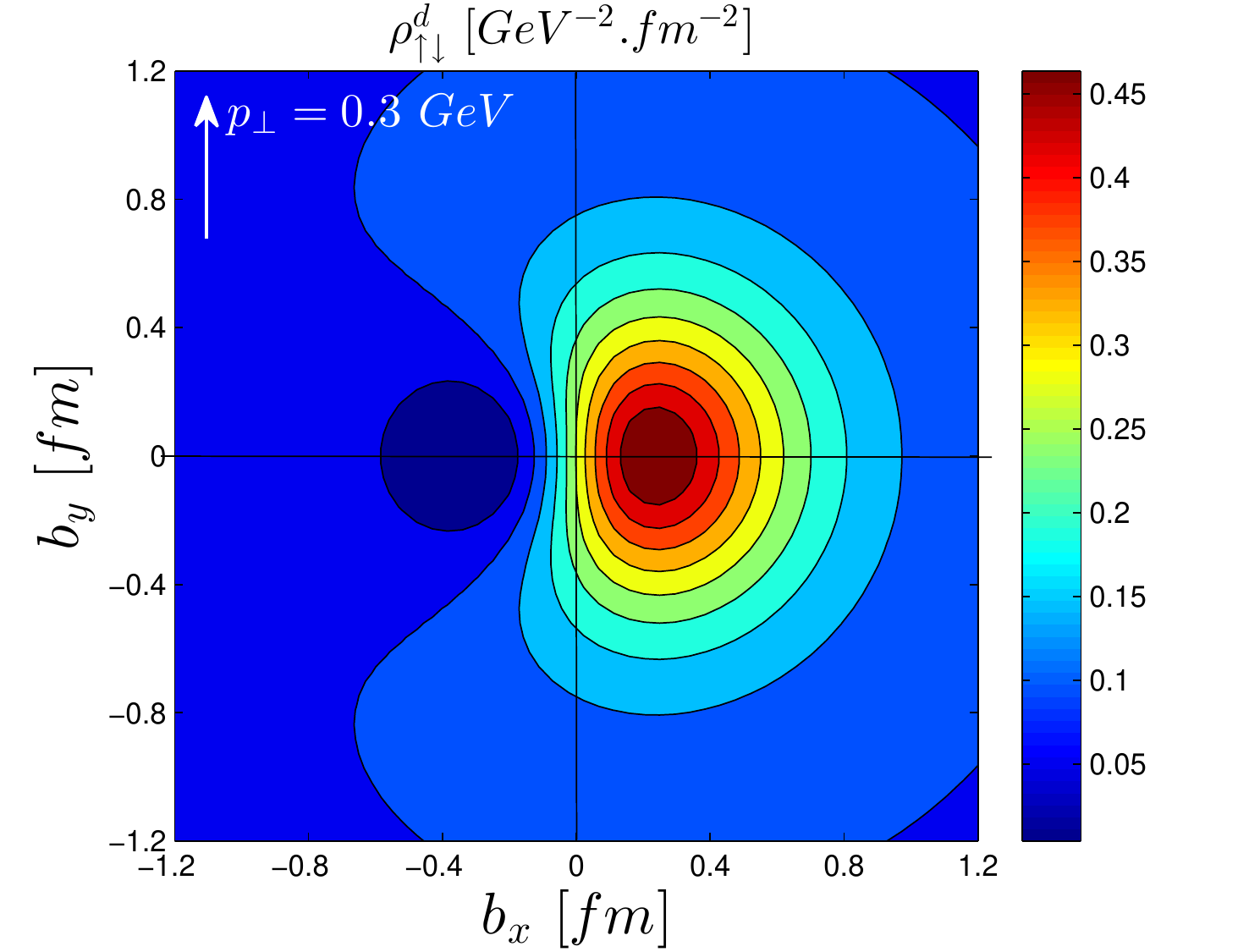}
\end{minipage}
\caption{\label{plot_rhoT}The $\rho^q_{\Lambda\lambda}(\bfb,\bfp)$ for $\Lambda=\uparrow$ and $\lambda=\uparrow,\downarrow$ in transverse momentum plane(a-d)with $\bfb=0.4\hat{y}~ fm$ and in transverse impact parameter plane(e-h)with $\bfp=0.3\hat{y}~ GeV$ for $u$ and $d$ quarks.}
\end{figure*} 
\begin{figure*}[htbp]
\begin{minipage}[c]{0.98\textwidth}
\small{(a)}\includegraphics[width=7.0cm,clip]{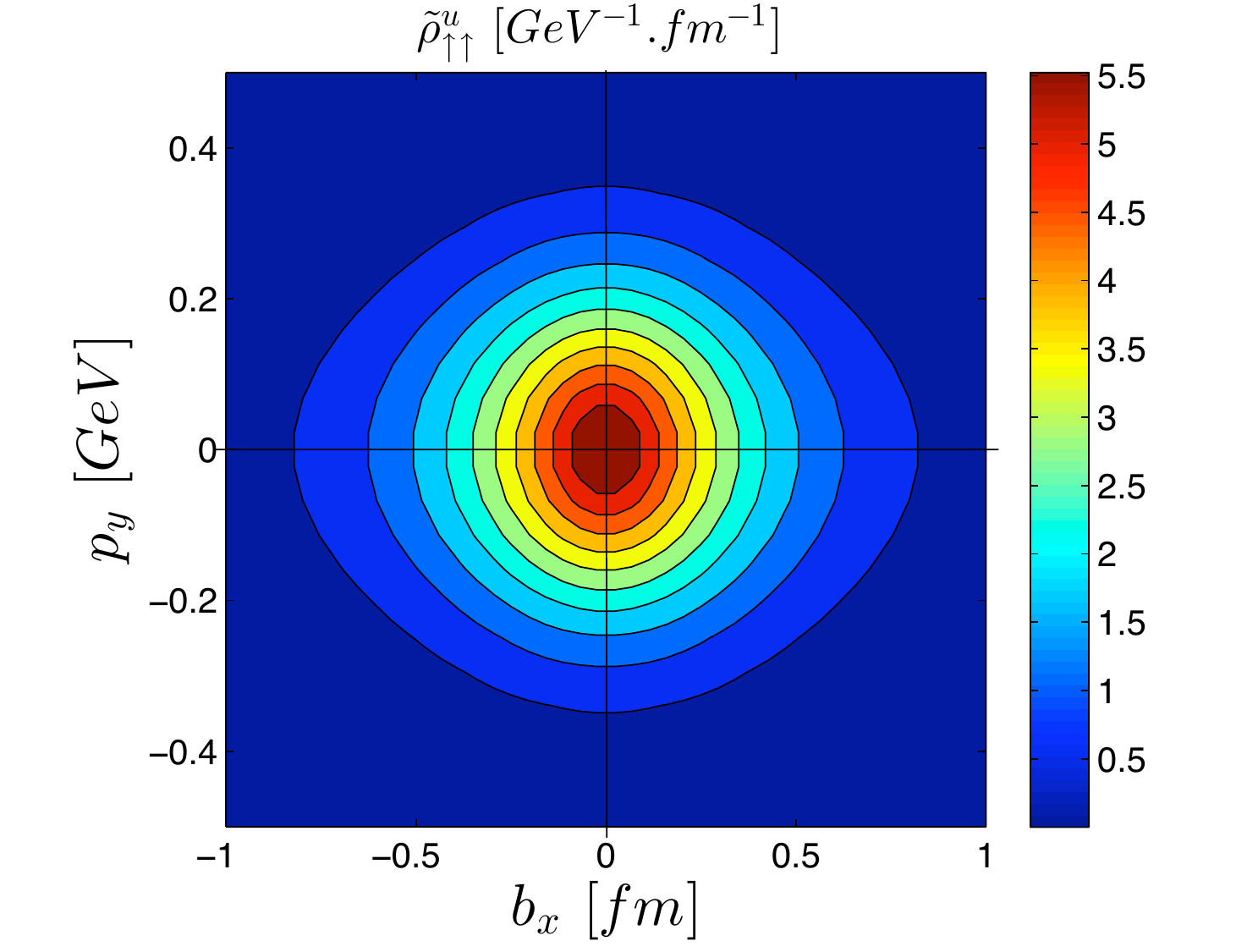}
\small{(b)}\includegraphics[width=7.0cm,clip]{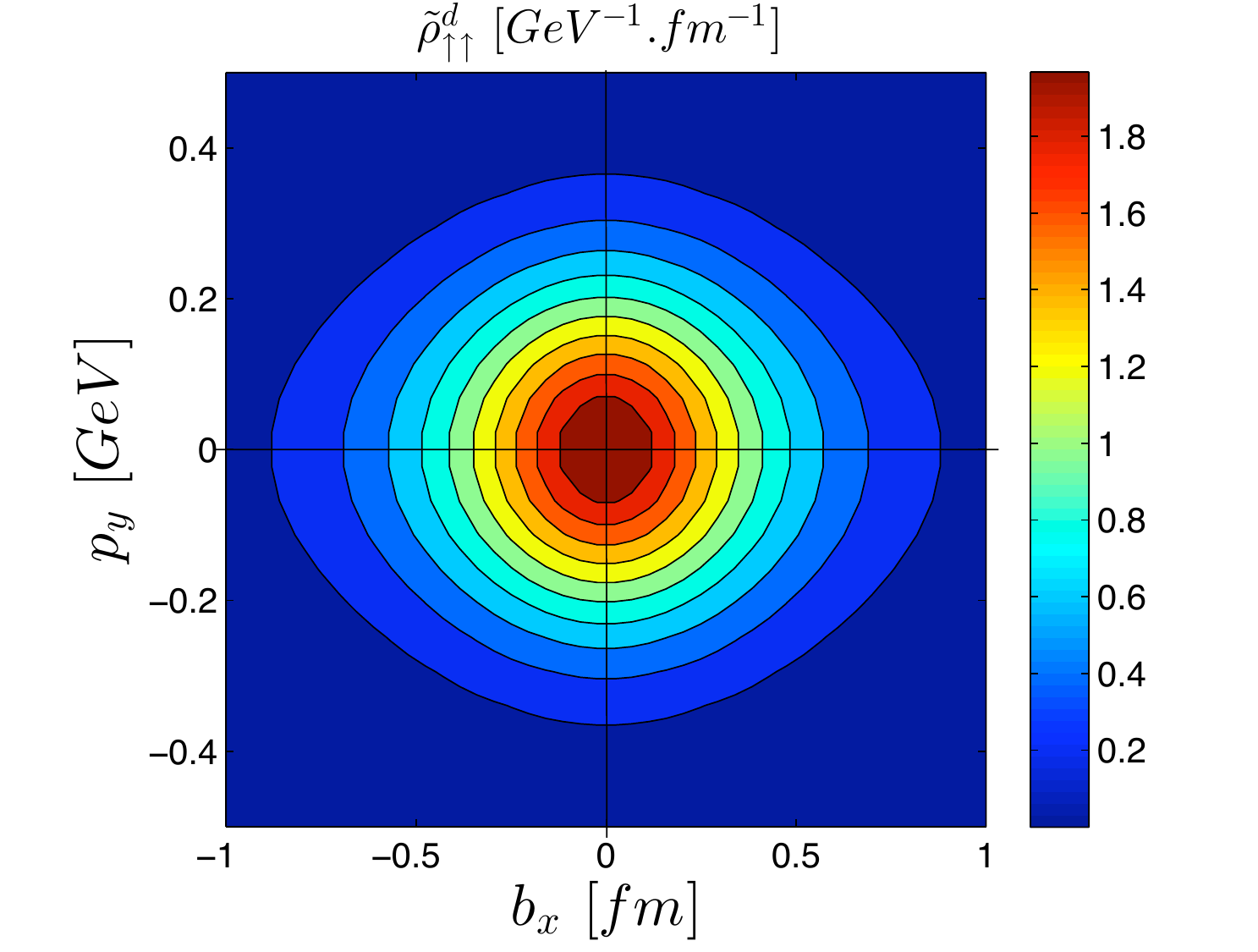}
\end{minipage}
\begin{minipage}[c]{0.98\textwidth}
\small{(c)}\includegraphics[width=7.0cm,clip]{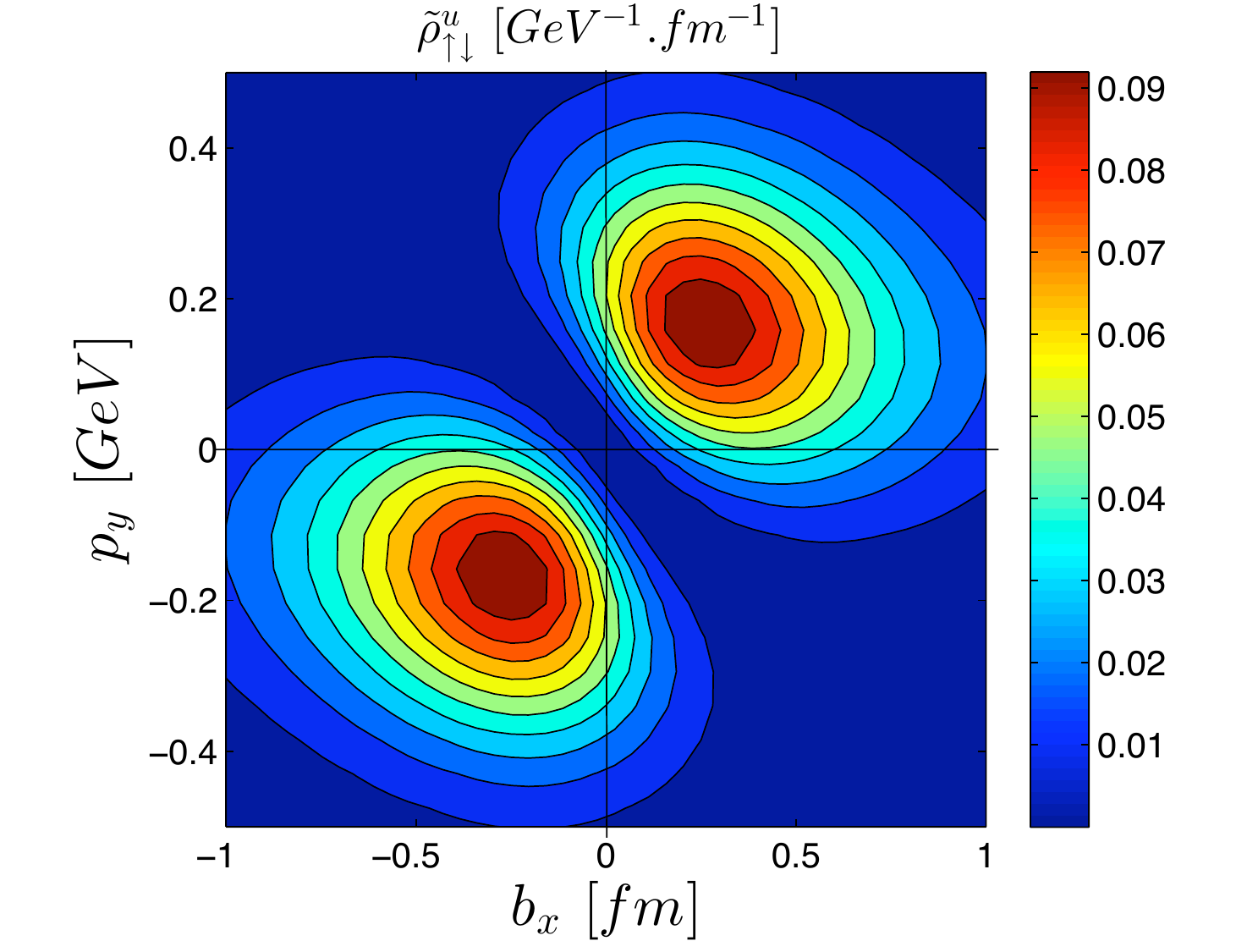}
\small{(d)}\includegraphics[width=7.0cm,clip]{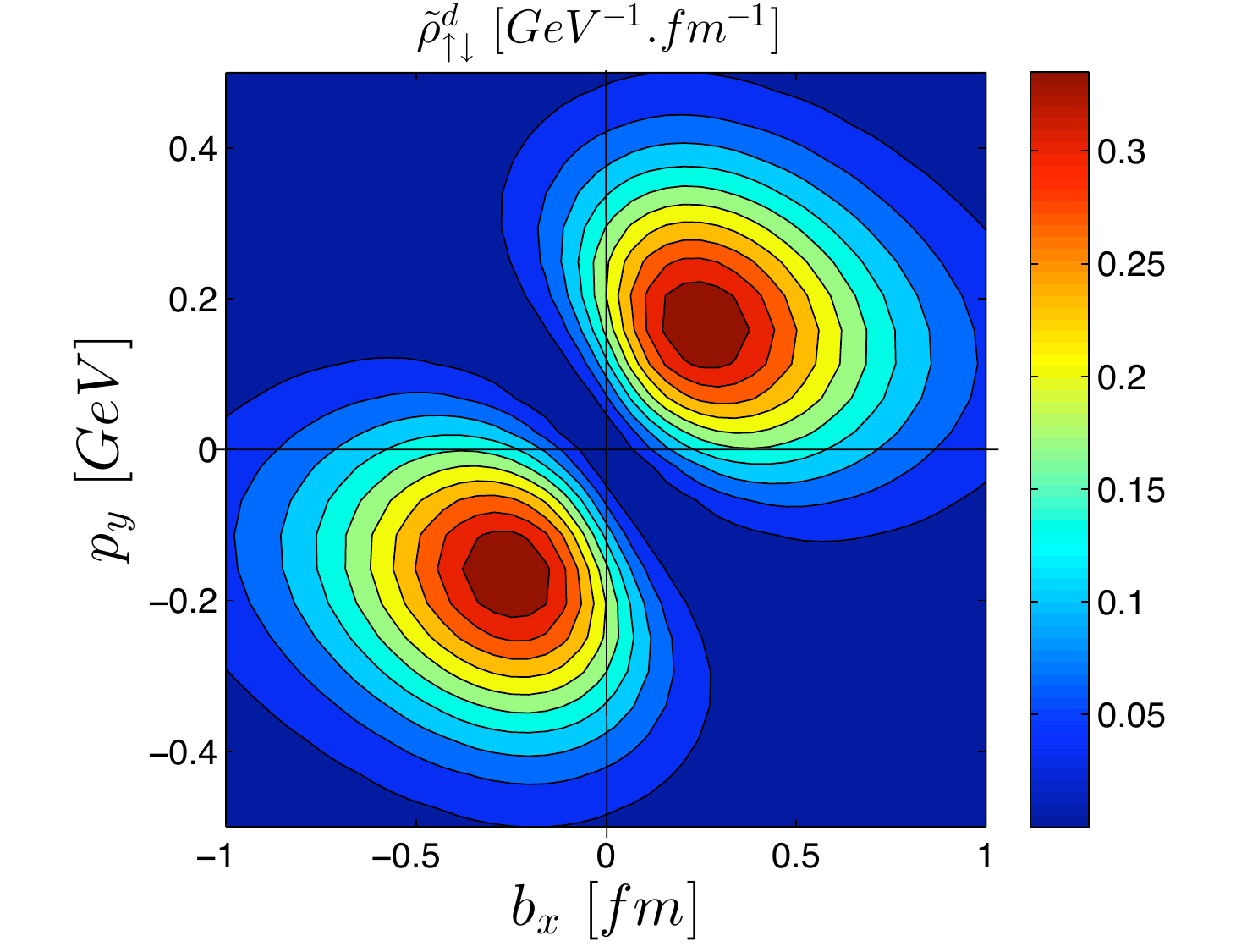}
\end{minipage}
\caption{\label{plot_rhoT_bxpy}The $\tilde{\rho}^q_{\Lambda\lambda}(b_x,p_y)$ for $\Lambda=\uparrow$ and $\lambda=\uparrow,\downarrow$ in mixed transverse plane for $u$ and $d$ quarks.}
\end{figure*}

The distributions $\rho^q_{\Lambda\lambda}(\bfb,\bfp)$ are shown in Fig.\ref{plot_rhoT} and Fig.\ref{plot_rhoT_bxpy} with the polarization of proton $\Lambda=\uparrow$ and quark polarization $\lambda=\uparrow,\downarrow$( Eq.(\ref{rho_long})). 
 Figs.\ref{plot_rhoT}(a-d) represent the variation of $\rho^q_{\Lambda\lambda}(\bfb,\bfp)$ in the transverse momentum plane for $u$ and $d$ quarks. We observe a circular symmetry for $\Lambda = \lambda $ but for $\Lambda \neq \lambda $ the distributions get distorted along $p_x$ for both $u$ and $d$ quarks. This is because, in Eq(\ref{rho_Lamlam}), the contributions from $\rho_{LU}$ and $\rho_{UL}$($\rho_{LU}=-\rho_{UL}$) get cancelled for $\Lambda=\lambda$, whereas for $\Lambda\neq\lambda$, the contributions add up and causes the distortion.
We have shown the distributions for $\Lambda=\uparrow$, 
the other possible spin combinations in transverse momentum plane can be found from $ \rho^q_{\downarrow\lambda^\prime}(\bfb,p_x,p_y)=\rho^q_{\uparrow\lambda}(\bfb,-p_x,p_y)$, where  $\lambda^\prime\neq\lambda$. 
 Figs.\ref{plot_rhoT}(e-h) show the variation of $\rho^q_{\Lambda\lambda}(\bfb,\bfp)$ in transverse impact parameter plane for $u$ and $d$ quarks. 
The distributions  are circularly symmetric  in transverse impact parameter space for $\Lambda=\lambda$ but the distributions get distorted for $\Lambda \neq\lambda$, due to the same reason as described in case of transverse momentum plane.
Similar to the momentum space,  the other possible spin combinations in the transverse impact parameter plane are found as $\rho^q_{\downarrow\lambda^\prime}(b_x,b_y,\bfp)=\rho^q_{\uparrow\lambda}(-b_x,b_y,\bfp)$, where  $\lambda^\prime\neq\lambda$. 
 The mixed transverse densities $\tilde{\rho}^q_{\Lambda\lambda}(b_x,p_y)$ are shown in Fig.\ref{plot_rhoT_bxpy} for $u$ and $d$ quarks. Again, for $\Lambda=\lambda$ the contribution from quadrupole distortions (Fig.(\ref{plot_rhoUL_bxpy},\ref{plot_rhoLU_bxpy}))  $\tilde{\rho}_{UL}$ and $\tilde{\rho}_{LU}$ get cancelled resulting the axial symmetry but for $\Lambda\neq\lambda$ the contributions add up. The maxima of $\tilde{\rho}_{UU} $ and $\tilde{\rho}_{LL}$  are nearly equal ( Figs. \ref{plot_rhoUU_bxpy} and \ref{plot_rhoLL_bxpy}). As a result, for $\Lambda\neq\lambda$, the destructive interference  of  these two distributions give almost zero  at the centre($b_x=0,p_y=0$) in Fig.\ref{plot_rhoT_bxpy}(c,d). 

\section{Spin-Spin and Spin-OAM Correlation}\label{corr}
In  Fig. \ref{plot_rhoUL} and Fig. \ref{plot_rhoLU}, we observe that  the quark OAM tends to be anti-aligned with quark spin and aligned to the proton spin for both $u$ and $d$ quarks. The correlation strength between proton spin and quark OAM is equal to the correlation between quark spin and quark OAM. Therefore, if the quark spin is parallel to the proton spin, i,e. $\Lambda=\uparrow, \lambda=\uparrow $ the contributions of $\rho_{UL}$ and $\rho_{LU}$ interfere destructively resulting the circular symmetry for $u$  and $d$ quarks, see Fig \ref{plot_rhoT}(a,b,e,f).
If the quark spin is anti-parallel to the proton spin, i,e. $\Lambda=\uparrow, \lambda= \downarrow$ the contributions of $\rho_{UL}$ and $\rho_{LU}$ interfere constructively resulting a significant shift for $u$  and $d$ quarks, see Fig \ref{plot_rhoT}(c,d,g,h). One can notice that from Fig.\ref{plot_rhoT}, the direction of shift flips with the polarization flip when 
 $\Lambda\neq\lambda$.
\begin{table*}[ht]
\centering 
\begin{tabular}{c | c| c | c || c| c | c |c}
    $\rho_{UL}(\bfp)$ & Our Model &   Ref.\cite{pasquini11} & Ref.\cite{liu_ma_WD}  & $\rho_{UL}(\bfb)$ & Our Model &   Ref.\cite{pasquini11} &  Ref.\cite{liu_ma_WD} \\ \hline
     $u$ &~ \textbf{- +} (0.1) &  ~\textbf{+ -} (0.6) &~ \textbf{- +} (0.010) & $ u$ &~ \textbf{+ -} (0.06) &  ~\textbf{- +} (0.5) & ~\textbf{+ -} (0.010)\\
     $ d$ & ~\textbf{- +} (0.3) &` \textbf{+ -} (0.6) &~ \textbf{- +} (0.005) & $ d$ &~ \textbf{+ -} (0.2) & ~\textbf{- +} (0.5) &~ \textbf{+ -} (0.005) \\
\end{tabular}
\caption{ Comparison of $\rho_{UL}$ in different models in momentum space(left panel) and in impact parameter space(right panel).  $\textbf{+ -}/\textbf{- +}$ represent the polarity of the dipolar distributions and the maxima of the distributions are given within the bracket.}
\label{tab_comp_UL} 
\end{table*}
\begin{table*}[ht]
\centering 
\begin{tabular}{c | c | c | c || c | c | c | c}
     $\rho_{LU}(\bfp)$ & Our Model &  ~ Ref.\cite{pasquini11} & ~ Ref.\cite{liu_ma_WD}  & $\rho_{LU}(\bfb)$ & Our Model &  ~ Ref.\cite{pasquini11} & ~ Ref.\cite{liu_ma_WD} \\ \hline
     $~~~ u$ & ~\textbf{+ -} (0.1) & ~ \textbf{+ -} (0.35) & ~\textbf{+ -} (0.010) & $~~~ u$ & ~\textbf{- +} (0.06) & ~ \textbf{- +} (0.3) & ~\textbf{- +} (0.005)\\
     $~~~ d$ & ~\textbf{+ -} (0.3) &~ \textbf{- +} (0.03) & ~\textbf{- +} (0.002) & $ ~~~ d$ & ~\textbf{- +} (0.2) & \textbf{+ -} (0.015) & ~\textbf{+ -} (0.0005) \\
\end{tabular}
\caption{ Comparison of $\rho_{LU}$ in different models in momentum space(left panel) and in impact parameter space(right panel). $\textbf{+ -}/\textbf{- +}$ represent the polarity of the dipolar distributions and the maxima of  the distributions are given within the bracket.}
\label{tab_comp_LU} 
\end{table*}

We compare our results  
with the  light cone constituent quark model (LCCQM) \cite{pasquini11} and light cone spectator model \cite{liu_ma_WD} in the tables \ref{tab_comp_UL} and \ref{tab_comp_LU}.  The polarities of $\rho_{UL}$ distributions are opposite to LCCQM but similar to the spectator model, whereas for $\rho_{LU}$, all the three models agree for $u$ quark, but the agreement is lost for $d$-quark.  
 In our model, the average quadrupole distortion  $Q^{ij}_b(\bfp)$ and $Q^{ij}_p(\bfb)$, in both the transverse momentum plane and transverse impact parameter plane, are found to be zero, whereas a nonzero small quadrupole distortion is found in \cite{pasquini11}.  
  This may be due the simple scalar diquark model considered here, inclusion of axial vector diquark might improve the result.
The quark OAM tends to be anti-aligned($C^u_z<0, C^d_z<0$) to quark spin for both $u$ and $d$ quarks in our model,  in LCCQM the quark OAM and quark spin tend to be aligned for both $u$ and $d$ quarks($C^u_z>0,~C^d_z>0$). In our model, the quark OAM tends to be aligned to proton spin for both $u$ and $d$ quarks($\ell^u_z>0,~\ell^d_z>0$). Whereas in \cite{pasquini11}, the quark OAM tends to be aligned($\ell^u_z>0$) to proton spin for $u$ quark and anti-aligned($\ell^d_z<0$) for $d$ quark.
For proton spin anti-aligned with quark spin, the distributions $\rho^q_{\uparrow\downarrow}$ for both $u$ and $d$ quarks show stronger dipolar structure in our model compared to the LCCQM. 
QCD or some model independent calculations  are required to resolve the differences.

\section{GTMDs}\label{gtmds}
\begin{figure*}[htbp]
\begin{minipage}[c]{0.98\textwidth}
\small{(a)}\includegraphics[width=7.9cm,,clip]{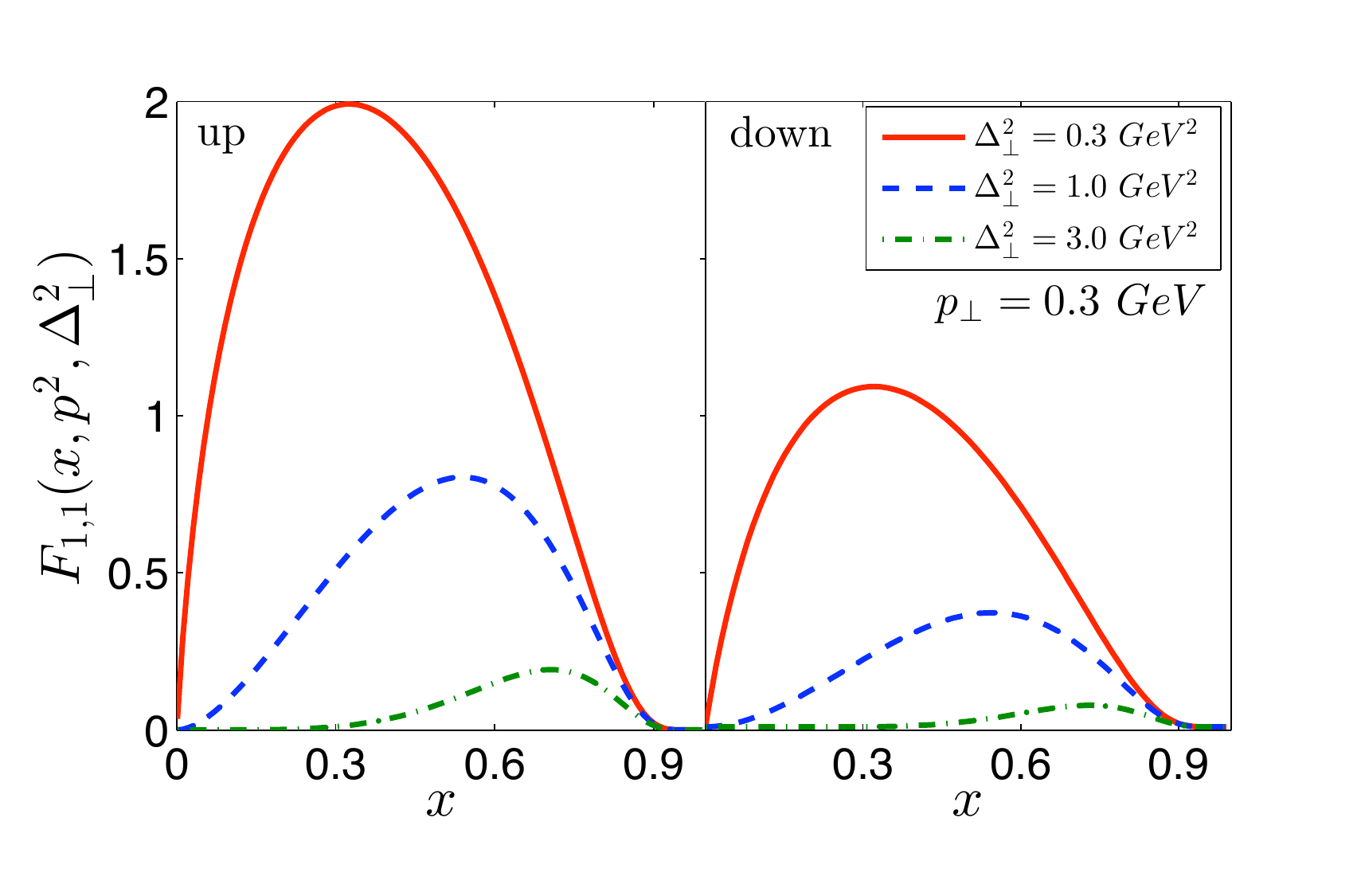}
\small{(b)}\includegraphics[width=7.9cm,,clip]{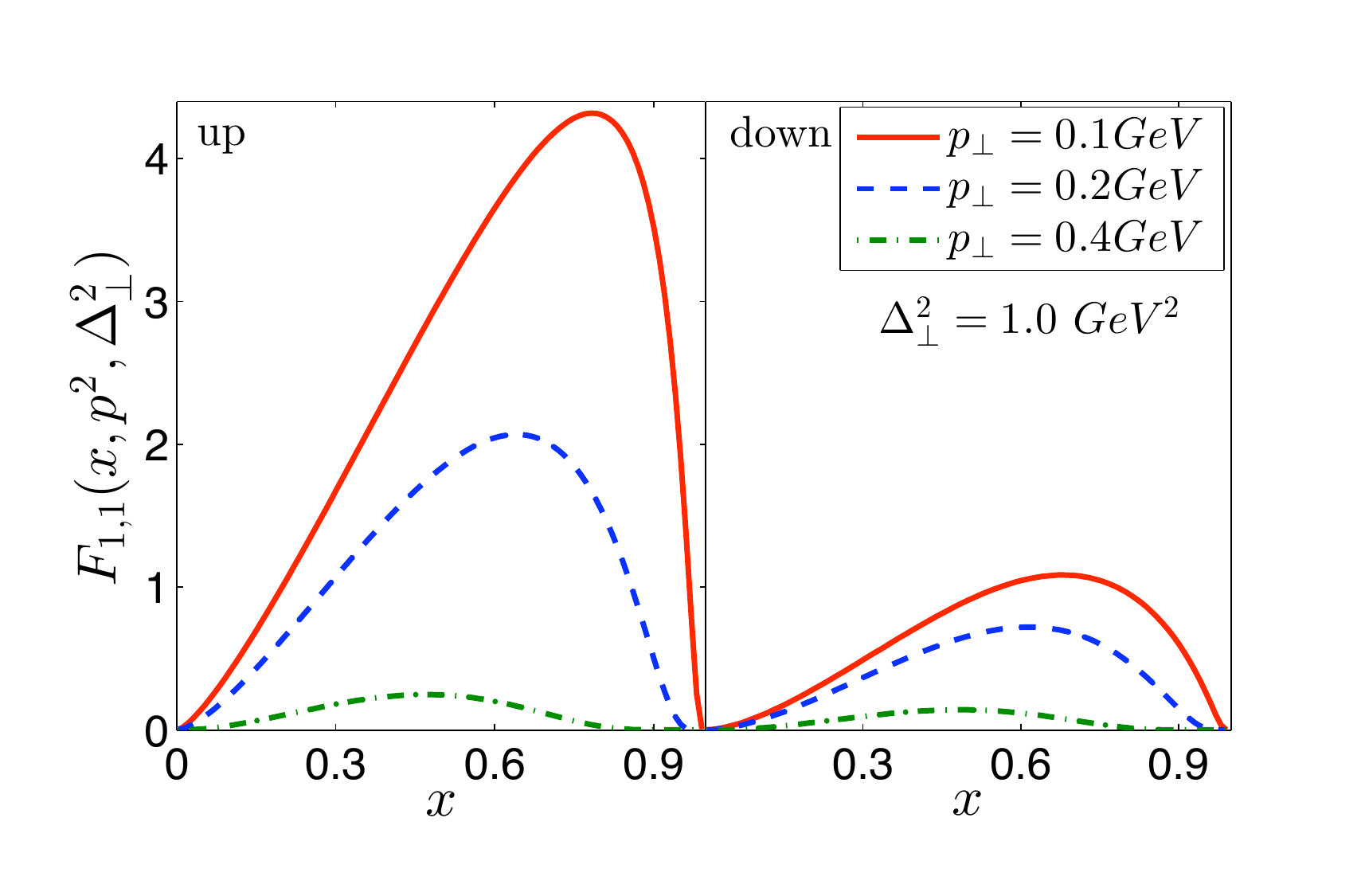}
\end{minipage}
\begin{minipage}[c]{0.98\textwidth}
\small{(c)}\includegraphics[width=7.9cm,clip]{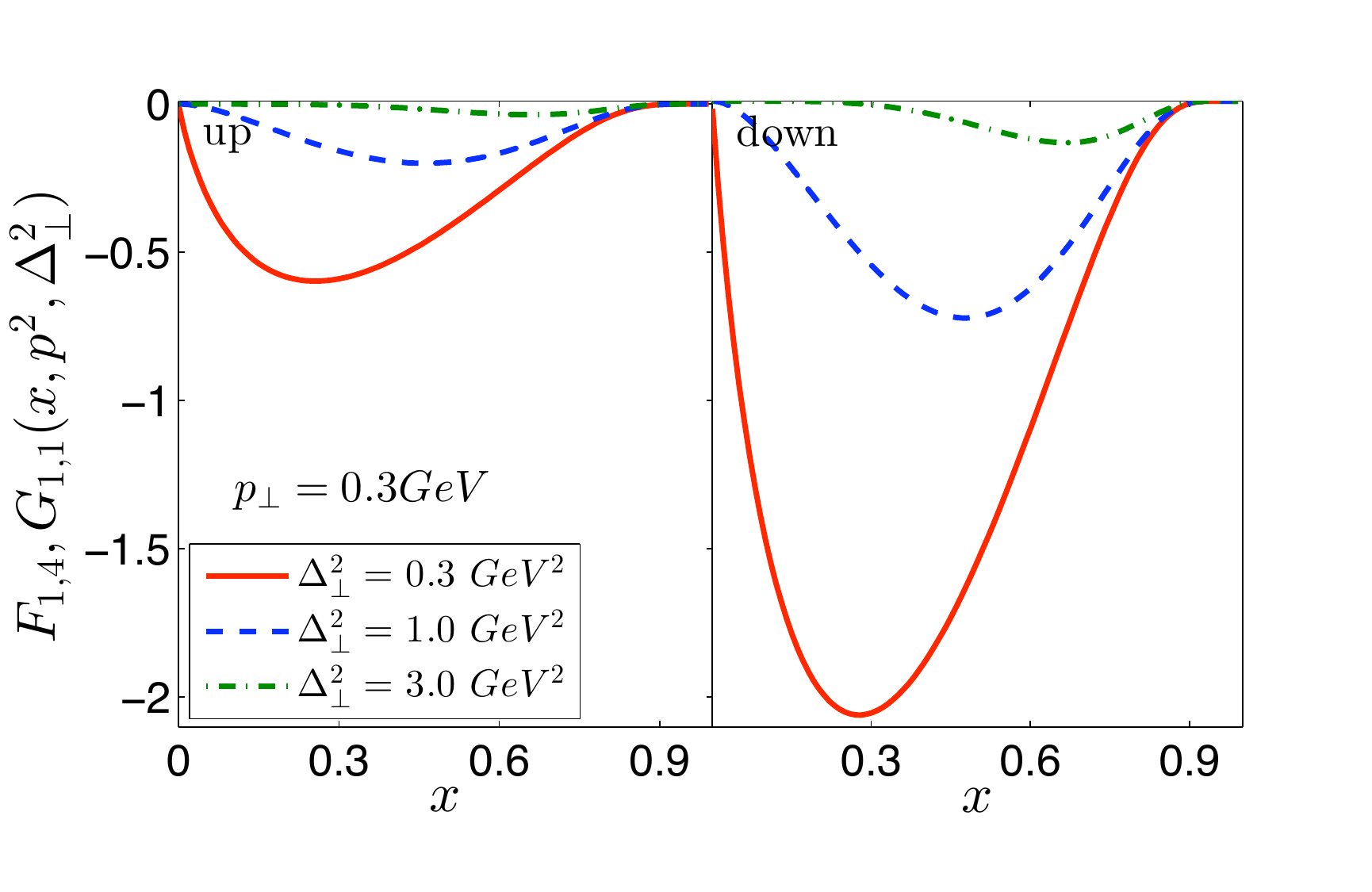}
\small{(d)}\includegraphics[width=7.9cm,clip]{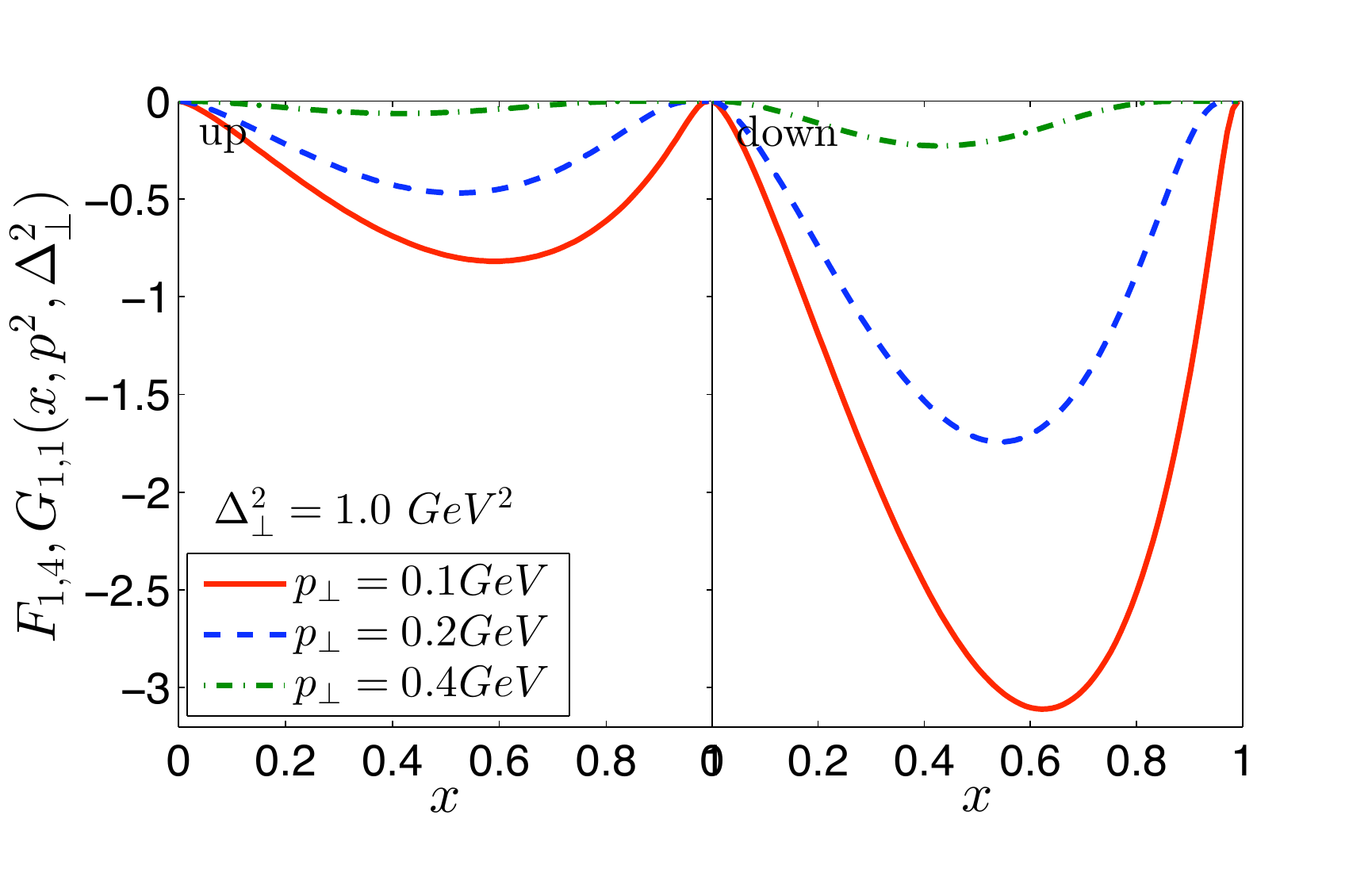}
\end{minipage}
\begin{minipage}[c]{0.98\textwidth}
\small{(e)}\includegraphics[width=7.9cm,clip]{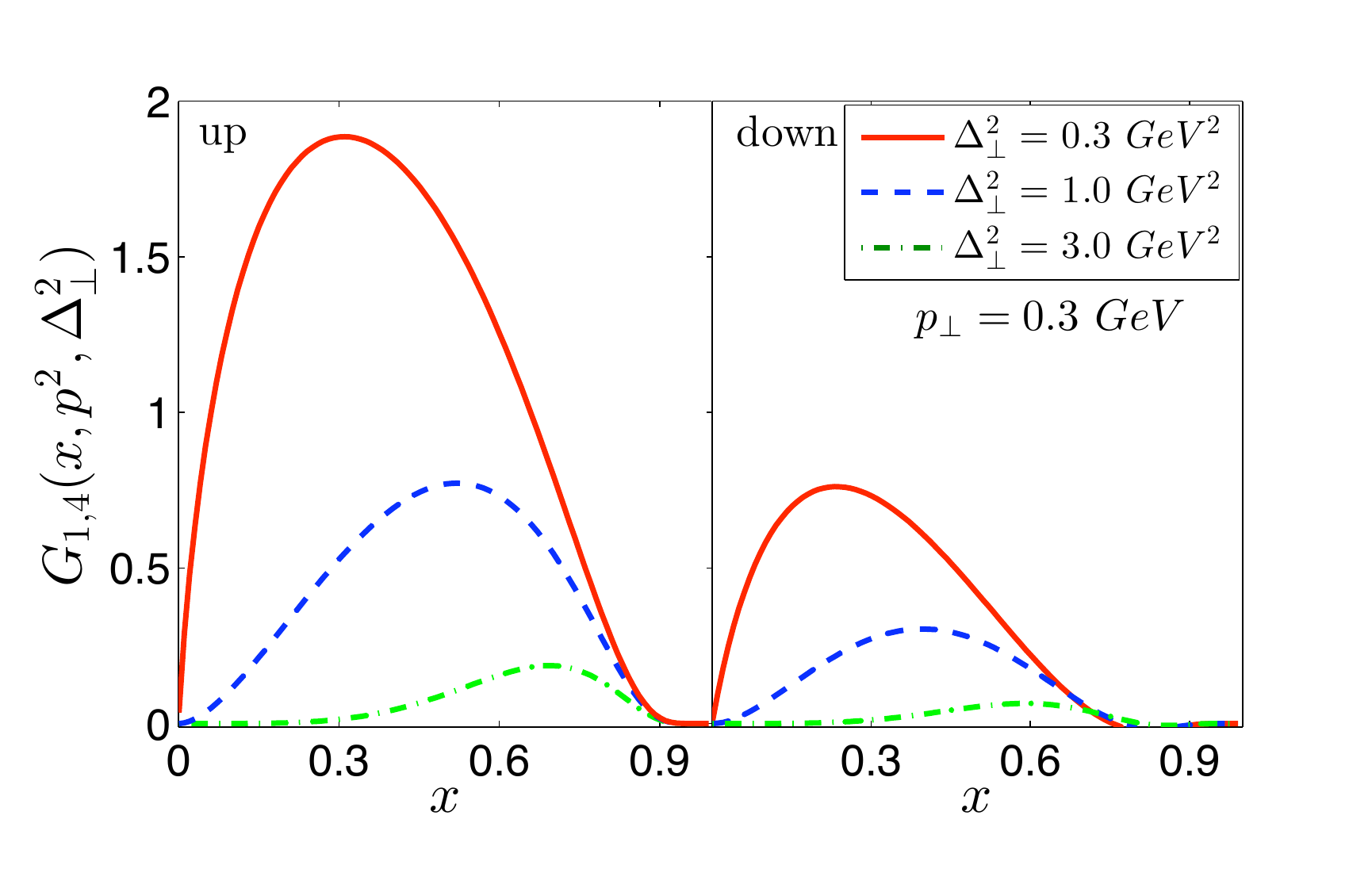}
\small{(f)}\includegraphics[width=7.9cm,clip]{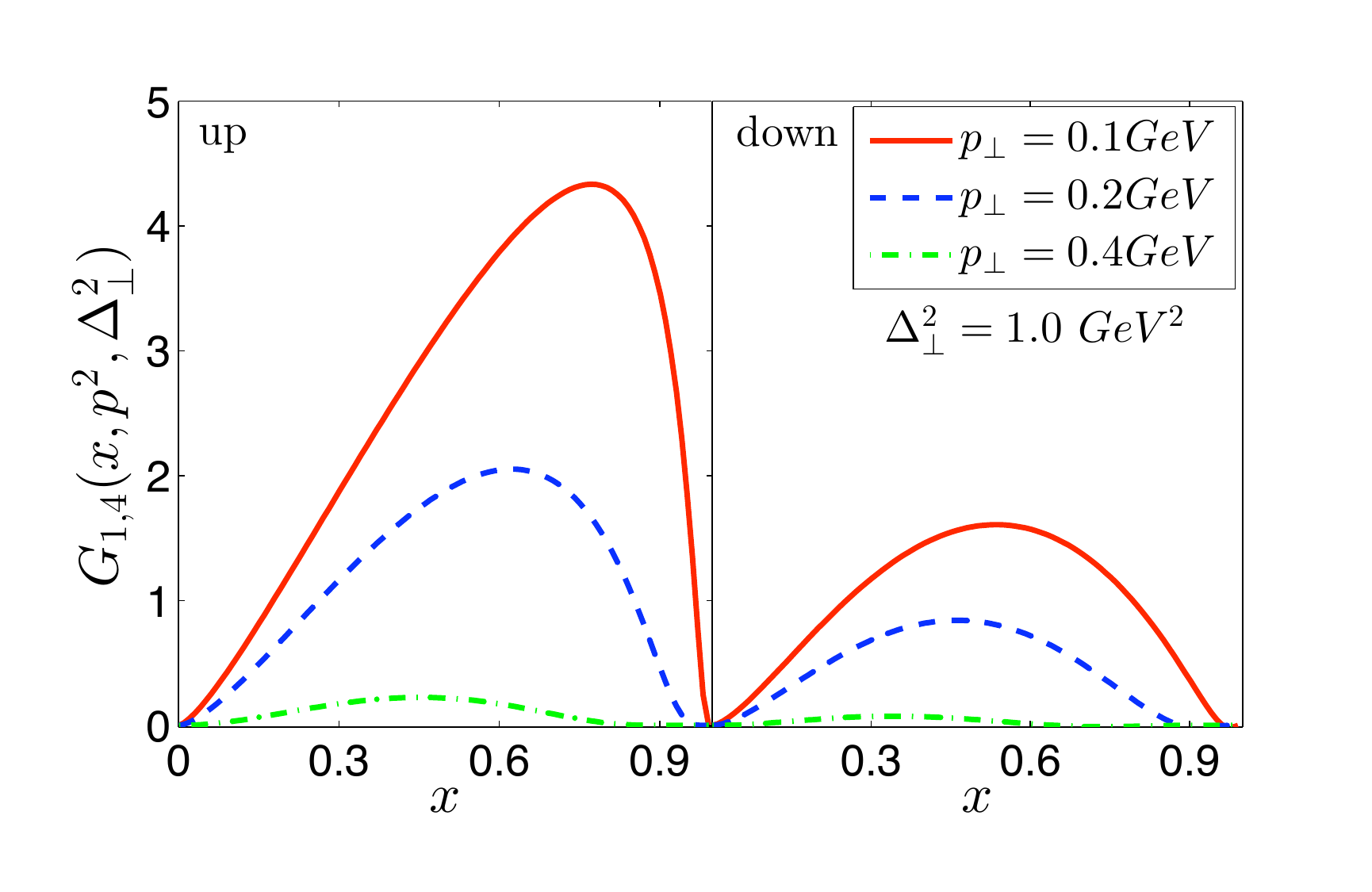}
\end{minipage}
\caption{\label{GTMD_x}  GTMDs as functions of x  for both $u$ and $d$ quarks at different fixed values of $\bfp$ and $\Dp$.}
\end{figure*}

At leading twist, there are sixteen GMDs. The variation of GTMDs (Eqs.(\ref{F11}-\ref{G14})) for $u$ and $d$ quarks are shown in Fig.(\ref{GTMD_x}). The left column is for different values of $\Delta^2_\perp$ with a fixed $\bfp=0.3~GeV$ and the right column is for different values of $p_\perp$ with a fixed $\Dp^2=1.0~GeV^2$. We observe that the peak of the distributions decrease with increasing $\Dp$ and shift towards higher $x$ . Thus, the distributions $F^q_{1,1},F^q_{1,4},G^q_{1,1},G^q_{1,4}$, having a quark with fixed transverse $\bfp$, highly depends on the momentum transfer $\Dp$ between initial and final proton. The behavior of $F_{1,1}$ for $u$ and $d$ quarks are almost same except in magnitude which is larger for $u$ quark than $d$ quark. In $F_{1,4}(=G_{1,1})$, the maxima for $d$ quark is greater than  the maxima for $u$ quark and opposite to $F_{1,1}$ and $G_{1,4}$. 
The GTMDs as functions of $x$ are shown in the right column of Fig.\ref{GTMD_x} for the different values of $\bfp$ with a fixed value of $\Delta^2_\perp=1.0~GeV^2$. In this case, the peak of the distributions shift towards lower $x$ and decreases as $\bfp$ increases.

\section{conclusions}\label{concl}

We have calculated the Wigner distributions in a quark-scalar diquark model
of the proton. We have used the light-front wave functions for the state
that are predicted by the soft wall ADS/QCD. The Wigner distributions of
both unpolarized quark in unpolarized proton as well as the distortions in
momentum and position space due to the polarization of the quark/proton are
calculated. The results are compared and contrasted with other model
estimates, in particular with those models that assume a confining
potential. Wigner functions are related to GTMDs that give information on
the canonical OAM as well as the spin-orbit correlation of the quarks. The
kinetic OAM can be calculated in terms of the GPDs in this model. We have
calculated both the canonical and kinetic OAM and compared with other model
calculations. In our case the proton state consists of an active quark
which can be either  a $u$ or a $d$ quark,  and a
scalar diquark. So the sum of the OAM of the $u$ and and the $d$ quark is not
expected to be the same. In fact the kinetic and canonical OAM of the $u$
quark are positive in this model whereas that of the $d$ quark are negative.
We have also calculated the pretzelosity in this model using a
model-dependent relation. As $x \rightarrow 1$ the difference  between
kinetic and canonical OAM vanishes as all the momentum is carried by the
active quark.  Further work would involve calculation of Wigner distributions
incorporating transverse polarization.


\end{document}